%% file: HIG-23-016_temp.tex
\pdfoutput=1
\documentclass[11pt,twoside,a4paper,cmspaper,final,collab]{cms-tdr}
\def\svnVersion{95ff461}\def\svnDate{2025/03/18}\def\cmsCernNoTag{CERN-EP-2024-294}\def\cmsCernDate{\today}\def\cmsMessage{Published in the Journal of High Energy Physics as \href{http://dx.doi.org/10.1007/JHEP03(2025)114}{\doi{10.1007/JHEP03(2025)114}.}}
\begin{document}\cmsNoteHeader{HIG-23-016}

\newcommand{\ZnnH}{\ensuremath{\PZ(\PGn\PGn)\PH}\xspace}
\newcommand{\VHbb}{\ensuremath{\PV\PH(\bbbar)}\xspace}
\newcommand{\ZtoNN}{\ensuremath{\PZ\to\PGn\PGn}\xspace}
\newcommand{\ZtoLL}{\ensuremath{\PZ\to\Pell\Pell}\xspace}
\newcommand{\WtoLN}{\ensuremath{\PW\to\Pell\PGn}\xspace}
\newcommand{\HBB}{\ensuremath{\PH\to\bbbar}\xspace}
\newcommand{\ptV}{\ensuremath{\pt(\PV)}\xspace}
\newcommand{\jj}{\ensuremath{\text{jj}}\xspace}
\newcommand{\dRJJ}{\ensuremath{\Delta R(\jj)}}
\newcommand{\twol}{\text{2-lepton}}
\newcommand{\onel}{\text{1-lepton}}
\newcommand{\zerol}{\text{0-lepton}}
\newcommand{\twom}{\text{2-muon}}
\newcommand{\twoe}{\text{2-electron}}
\newcommand{\onem}{\text{1-muon}}
\newcommand{\onee}{\text{1-electron}}
\newcommand{\DeepJet}{\textsc{DeepJet}\xspace}
\newcommand{\ParticleNet}{\textsc{ParticleNet}\xspace}
\newcommand{\msd}{\ensuremath{m_{\mathrm{SD}}}\xspace}
\newcommand{\cHqo}{\ensuremath{{c}^{(1)}_{\PH\PQq}}\xspace}
\newcommand{\cHqt}{\ensuremath{{c}^{(3)}_{\PH\PQq}}\xspace}
\newcommand{\cHu}{\ensuremath{{c}_{\PH\PQu}}\xspace}
\newcommand{\cHd}{\ensuremath{{c}_{\PH\PQd}}\xspace}
\newcommand{\cHW}{\ensuremath{{c}_{\PH\PW}}\xspace}
\newcommand{\cHWB}{\ensuremath{{c}_{\PH\PW\textrm{B}}}\xspace}
\newcommand{\cHB}{\ensuremath{{c}_{\PH\mathrm{B}}}\xspace}
\newcommand{\cHWtil}{\ensuremath{{c}_{\PH\widetilde{\PW}}}\xspace}
\newcommand{\cHWBtil}{\ensuremath{{c}_{\PH\widetilde{\PW}\textrm{B}}}\xspace}
\newcommand{\cHBtil}{\ensuremath{{c}_{\PH\widetilde{\textrm{B}}}}\xspace}
\newcommand{\cHD}{\ensuremath{{c}_{\PH\textrm{D}}}\xspace} 
\newcommand{\cHbox}{\ensuremath{{c}_{\PH\square}}\xspace} 
\newcommand{\gtZZ}{\ensuremath{{g}^{\PZ\PZ}_{2}}\xspace}
\newcommand{\gtZZtil}{\ensuremath{\widetilde{g}^{\PZ\PZ}_{2}}\xspace}
\newcommand{\gfZZ}{\ensuremath{{g}^{\PZ\PZ}_{4}}\xspace}
\newcommand{\gtZa}{\ensuremath{{g}^{\PZ\PGg}_{2}}\xspace}
\newcommand{\gtZatil}{\ensuremath{\widetilde{g}^{\PZ\PGg}_{2}}\xspace}
\newcommand{\gfZa}{\ensuremath{{g}^{\PZ\PGg}_{4}}\xspace}
\newcommand{\gtaa}{\ensuremath{{g}^{\PGg\PGg}_{2}}\xspace}
\newcommand{\gtaatil} {\ensuremath{\widetilde{g}^{\PGg\PGg}_{2}}\xspace}
\newcommand{\gfaa}{\ensuremath{{g}^{\PGg\PGg}_{4}}\xspace}
\newcommand{\gtWW}{\ensuremath{{g}^{\PW\PW}_{2}}\xspace}
\newcommand{\gfWW}{\ensuremath{{g}^{\PW\PW}_{4}}\xspace}

\providecommand{\cmsTable}[1]{\resizebox{\textwidth}{!}{#1}}

\cmsNoteHeader{HIG-23-016}
\title{Constraints on standard model effective field theory for a Higgs boson produced in association with \texorpdfstring{\PW or \PZ}{W or Z} bosons in the \texorpdfstring{$\PH\to \bbbar$}{H to \bbar} decay channel in proton-proton collisions at \texorpdfstring{$\sqrt{s} = 13\TeV$}{sqrt(s) = 13 TeV}}

\date{\today}

\abstract{
A standard model effective field theory (SMEFT) analysis with dimension-six operators probing nonresonant new physics effects is performed in the Higgs-strahlung process, where the Higgs boson is produced in association with a \PW or \PZ boson, in proton-proton collisions at a center-of-mass energy of 13\TeV. The final states in which the \PW or \PZ boson decays leptonically and the Higgs boson decays to a pair of bottom quarks are considered. The analyzed data were collected by the CMS experiment between 2016 and 2018 and correspond to an integrated luminosity of 138\fbinv. An approach designed to simultaneously optimize the sensitivity to Wilson coefficients of multiple SMEFT operators is employed. Likelihood scans as functions of the Wilson coefficients that carry SMEFT sensitivity in this final state are performed for different expansions in SMEFT. The results are consistent with the predictions of the standard model. 
}

\hypersetup{
pdfauthor={CMS Collaboration},
pdftitle={Constraints on standard model effective field theory for a Higgs boson produced in association with W or Z bosons in the H to bbar decay channel in proton-proton collisions at sqrt(s) = 13 TeV},
pdfsubject={CMS},
pdfkeywords={CMS, Higgs boson, Effective field theory, Machine learning}}

\maketitle 

\section{Introduction}
\label{sec:intro}

In the standard model (SM) of particle physics, the Brout--Englert--Higgs mechanism~\cite{Englert:1964et,Higgs:1964ia,Higgs:1964pj,Guralnik:1964eu} explains the electroweak symmetry breaking and allows the weak gauge bosons to acquire mass. The mechanism predicts the existence of a Higgs scalar field and the observation of its quantum, the Higgs boson, in 2012 by both the ATLAS~\cite{ATLAS:2012yve} and CMS~\cite{CMS:2012qbp,CMS:2013btf} Collaborations at the CERN LHC was one of the main goals of the LHC physics program.

In the SM, the Higgs boson, with a mass of 125\GeV, has the largest Yukawa coupling to \PQb quarks among the fermions to which its decay is kinematically allowed.
While the majority of Higgs bosons decay to \PQb quarks, the large rates of background from SM events composed uniquely of jets produced through the strong interaction, referred to as quantum chromodynamics (QCD) multijet production, and the lack of secondary vertex discrimination in the hardware-level trigger, make selecting events based solely on {\HBB} decays very difficult. The Higgs boson production in association with a vector boson $\PV$ ($\PV=\PW,\PZ$), referred to as $\PV\PH$ production, has the advantage of high efficiency via easily identified leptonic decays of the \PW and \PZ bosons, making it the most sensitive channel in the measurement of the {\HBB} decay. The presence of an energetic vector boson in the final state highly suppresses the QCD multijet events, while also providing an efficient trigger requirement when the \PW or \PZ boson decay to charged leptons and/or neutrinos. Requiring a large boost of the \PW or \PZ boson provides several additional advantages: it further reduces the large backgrounds from \PW and \PZ boson production in association with jets, it helps in reducing the large background from top quark production in the signal channels including neutrinos, it makes the \ZnnH\ channel accessible via large missing transverse momentum (\ptmiss), and it generally improves the mass resolution of the reconstructed Higgs boson candidates.
The $\PV\PH$ production mode was the dominant contributor to the observation of the {\HBB} decay by the ATLAS and CMS experiments using the data from proton-proton ($\Pp\Pp$) collisions at $\sqrt{s}=7$, 8, and 13\TeV collected during 2011--2017~\cite{ATLAS:2018kot,CMS:2018nsn}.
Both ATLAS and CMS experiments have recently performed cross section measurements for the $\PV\PH$ production mode where the vector boson decays leptonically in the phase space defined by the simplified template cross section (STXS) framework with $\Pp\Pp$ collision data at $\sqrt{s}=13\TeV$~\cite{ATLAS:2024yzu,CMS:2023vzh}.

The recent application of effective field theories~(EFTs)~\cite{Buchmuller:1985jz,Grinstein:1991cd,Chiu:2007dg} demonstrates that subtle deviations in observable distributions from SM expectations can give hints of new physics at energy scales beyond those directly accessible at the LHC. The underlying theoretical framework considered in this paper is the SM effective field theory~(SMEFT)~\cite{Degrande:2012wf,Jenkins:2013zja,Alonso:2013hga,Jenkins:2013wua,Englert:2014cva,Brivio:2017vri,Isidori:2023pyp}, which extends the SM with operators of higher mass dimensions, thereby smoothly modifying the kinematic spectra of the observables used for the extraction of the EFT effects. Several recent CMS results using the LHC Run 2 data set, targeting $\PH\to\PV\PV\to 4\,\mathrm{leptons}$~\cite{CMS:2021nnc}, $\PH\to\PW{\PW}\to \Pe \PGm \Pgne \Pgngm$~\cite{CMS:2024bua}, and $\PH\to\tau\tau$~\cite{MELAcmsHTT} decay modes and exploiting the matrix element likelihood approach (MELA)~\cite{Gao:2010qx,Bolognesi:2012mm,Anderson:2013afp}, report constraints on anomalous couplings of the Higgs boson, sometimes also interpreted in terms of bounds on SMEFT operator coefficients.

Notably, several dimension-six operators within the SMEFT framework affect the associated production of a Higgs boson with a \PW or \PZ boson and change the distributions of the kinematic variables of the final-state particles, so that the $\PV\PH$ production is an extremely important discovery tool for phenomena beyond the SM. 
So far, the $\PV\PH$ cross sections measured in the STXS framework have been used to probe the magnitude of coefficients corresponding to the relevant SMEFT operators by the ATLAS and CMS experiments~\cite{ATLAS:2019yhn,ATLAS:2020fcp,CMS-PAS-HIG-19-005}. 

Further constraints on SMEFT operators from global analyses are available outside the experimental collaborations~\cite{Ellis:2020unq,Ethier:2021bye}, with partial or complete data sets from LHC Run 2. While the $\PV\PH$ measurements performed within the STXS framework are sensitive to SMEFT effects, they encounter limitations in probing the degeneracy between the effects of different SMEFT operators because of a limited number of variables and phase space regions measured.
Moreover, there is no observable sensitive to the charge-parity ($CP$) nature of the SMEFT operators in the set of bins defined in the STXS framework, referred to as STXS stage 1.2. In this paper, we report a measurement designed to probe the size of several dimension-six SMEFT operator coefficients relevant in the $\PV\PH$ production process, targeting the $\HBB$ decay and leptonic decay modes of the W or Z boson: $\ZtoNN$, $\WtoLN$, and $\ZtoLL$, referred to as $\zerol$, $\onel$, and $\twol$ channels, respectively. 
Here, {\Pell} refers to either an electron or a muon, including the leptonic decays of the {\PGt} lepton.
A multivariate analysis is performed exploiting a rich set of kinematic information to separate the SMEFT effects from the SM, and also to disentangle the effects of different SMEFT operators. The measurements are performed using $\Pp\Pp$ collision data at $\sqrt{s}=13\TeV$ collected by the CMS experiment from 2016 to 2018, corresponding to an integrated luminosity of 138\fbinv.

This paper is organized as follows. Section~\ref{sec:EFTinVH} describes the EFT effects in the {\VHbb} channel. Section~\ref{sec:CMS} presents the CMS detector, while Section~\ref{sec:DataMC} describes the data and the simulated samples used in the analysis including the discussion of the simulation of SMEFT effects. Section~\ref{sec:reco} describes the object reconstruction employed in the analysis. Section~\ref{sec:selection} discusses the event selection for the different event topologies targeting different energy regimes, as well as the corrections applied to the simulation to improve the modeling. Section~\ref{sec:strategy} presents the analysis strategy employed to access and constrain SMEFT effects via operator coefficients with a dedicated focus on the definition of the observables which are sensitive to SMEFT effects. 
Section~\ref{sec:sys} describes the systematic uncertainties affecting the results and how they are included in the measurements. 
Results, with tabulated versions provided in HEPData~\cite{hepdata}, and conclusions are discussed in Sections~\ref{sec:result} and~\ref{sec:conclusion}, respectively.

\section{Effective field theory effects in Higgs-strahlung}
\label{sec:EFTinVH}

Low-energy manifestations of theories satisfying ultraviolet completion~\cite{Englert:2014cva} can be described by SMEFT at the {\TeV} energy scale, accessible at the LHC. It progressively includes operators of mass dimension greater than four that respect the SM gauge symmetries~\cite{Jenkins:2013zja,Alonso:2013hga,Jenkins:2013wua,Englert:2014cva,Brivio:2017vri}, and is defined by the Lagrangian
\begin{equation}
\mathcal{L}_{\text{SMEFT}} = \mathcal{L}_{\text{SM}} +  {\sum}_{i} \frac{\text{c}_i^{\left(5\right)}}{\Lambda} \mathcal{O}_{5,i} + {\sum}_{i} \frac{\text{c}_i^{\left(6\right)}}{{\Lambda}^{2}} \mathcal{O}_{6,i} + ... \;,
\label{Eq:SMEFT}
\end{equation}
where the sum runs over all possible operators with a particular mass dimension. 
This expansion captures all nonresonant phenomena beyond the SM below an arbitrarily chosen energy scale $\Lambda$.
The dimensionless Wilson coefficients $\text{c}^{\left(n\right)}$ are used to parameterize the effects on observables, while the terms $\mathcal{O}_5$ and $\mathcal{O}_6$ are the operators at mass dimensions five and six, respectively. The only possible dimension-five operator is the Weinberg operator~\cite{PhysRevLett.43.1566}, which predicts lepton flavor violation and is not relevant for the phenomenology of $\PV\PH$ production. The dimension-six SMEFT operators that affect the $\PV\PH$ production at the tree level are listed in Table~\ref{tab:smeft_operators}. The operators are written in the so-called Warsaw basis~\cite{Grzadkowski:2010es}. The interference term between the SM and dimension-eight SMEFT operators, which contributes to the overall cross section with the same order of magnitude ($\Lambda^{-4}$) as the squared terms of dimension-six SMEFT operators, is neglected in this analysis.  

\begin{table*}[!hbt]
\topcaption{
The dimension-six operators in the Warsaw basis affecting $\PV\PH$  production at leading order. Here ${{\PQq}_{\textrm{L}}}$ refers to a left-handed quark field and is a representation of an SU(2) quark doublets. ${\PQu}_{\textrm{R}}$ refers to a right-handed up quark singlet, and ${\PQd}_{\textrm{R}}$ a right-handed down quark singlet.}
\centering
\begin{tabular}{lcl lcl}
\multicolumn{1}{l}{Operator} & Definition & {Wilson} & Operator & Definition & {Wilson} \\ 
 &  & coefficient & & & coefficient \\
\hline
${\cal O}^{(1)}_{\PH\PQq}$&$i \PH^\dagger \overleftrightarrow{D}_\mu \PH {\bar{\PQq}_{\textrm{L}}}   \gamma^\mu {{\PQq}_{\textrm{L}}}$& $\cHqo$ &${\cal O}_{\PH\PW\textrm{B}}$&$ {\PH}^\dagger \sigma^a \PH {\PW}^a_{\mu\nu}\textrm{B}^{\mu\nu}$ & $\cHWB$\\
\rule{0pt}{4ex} ${\cal O}^{(3)}_{\PH\PQq}$&$i {\PH}^\dagger \sigma^a \overleftrightarrow{D}_\mu \PH {\bar{\PQq}_{\textrm{L}}}   \sigma^a \gamma^\mu {{\PQq}_{\textrm{L}}}$& $\cHqt$ &${\cal O}_{\PH\widetilde{\PW}\textrm{B}}$&$ {\PH}^\dagger \sigma^a {\PH} {\PW}^a_{\mu\nu}\widetilde{\textrm{B}}^{\mu\nu}$&$\cHWBtil$\\
\rule{0pt}{4ex} ${\cal O}_{\PH\PQu}$&$i {\PH}^\dagger \overleftrightarrow{D}_\mu \PH \bar{\PQu}_{\textrm{R}}  \gamma^\mu {\PQu}_{\textrm{R}}$& $\cHu$ &${\cal O}_{\PH{\PW}}$&$({\PH}^\dagger {\PH})\,{\PW}^{a}_{\mu\nu}{\PW}^{a \mu\nu}$&$\cHW$\\
\rule{0pt}{4ex} ${\cal O}_{\PH\PQd}$&$i {\PH}^\dagger \overleftrightarrow{D}_\mu {\PH} \bar{\PQd}_{\textrm{R}}  \gamma^\mu {\PQd}_{\textrm{R}}$& $\cHd$ & ${\cal O}_{\PH\widetilde{\PW}}$ & $ ({\PH}^\dagger \PH)\, {\PW}^a_{\mu\nu}\widetilde{\PW}^{a \mu\nu}$ & $\cHWtil$\\
\rule{0pt}{4ex} ${\cal O}_{\PH \textrm{D}}$&$({\PH}^\dagger  {D}^\mu \PH)^*({\PH}^\dagger  {D}_\mu \PH)$& $\cHD$ &${\cal O}_{\PH\textrm{B}}$ & $({\PH}^\dagger \PH)\,\textrm{B}_{\mu\nu}\textrm{B}^{\mu\nu}$&$\cHB$\\
\rule{0pt}{4ex} ${\cal O}_{\PH\square}$&$({\PH}^\dagger \PH) \square ({\PH}^\dagger \PH)$& $\cHbox$ & ${\cal O}_{\PH\widetilde{\textrm{B}}}$ & $({\PH}^\dagger \PH)\,\textrm{B}_{\mu\nu}\widetilde{\textrm{B}}^{\mu\nu}$ & $\cHBtil$
\end{tabular}
\label{tab:smeft_operators}
\end{table*}

\begin{figure*}[!hbt]
\centering
\includegraphics[width=0.42\textwidth]{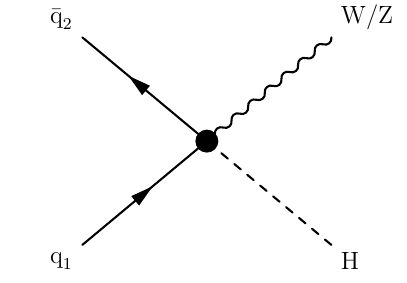}
\includegraphics[width=0.28\textwidth]{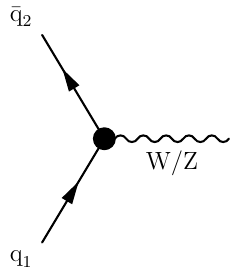}
\includegraphics[width=0.28\textwidth]{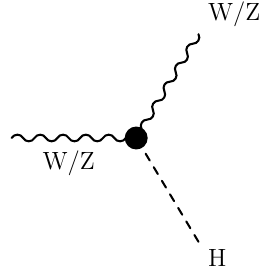}
\caption{
Representative Feynman diagrams for $\PV\PH$ production sensitive to different dimension-six operators. The EFT effects contribute in vertices highlighted with a black dot. The diagram on the left shows effects due to ${\cal O}^{(1)}_{\PH\PQq}$, ${\cal O}^{(3)}_{\PH\PQq}$, ${\cal O}_{\PH\PQu}$, and ${\cal O}_{\PH\PQd}$. The diagram at the center also includes contributions due to ${\cal O}_{\PH\textrm{D}}$ and ${\cal O}_{\PH\PW\textrm{B}}$. The diagram on the right displays effects from $\mathcal{O}_{\PH\PW}$, $\mathcal{O}_{\PH\PW\textrm{B}}$, $\mathcal{O}_{\PH\textrm{B}}$, and their $CP$ conjugates.
}
\label{fig:Feynman_digarams}
\end{figure*}

The operators ${\cal O}^{(1)}_{\PH\PQq}$, ${\cal O}^{(3)}_{\PH\PQq}$, ${\cal O}_{\PH\PQu}$, and ${\cal O}_{\PH\PQd}$, commonly referred to as current operators, introduce four-point interactions, depicted in Fig.~\ref{fig:Feynman_digarams}~(left), and result in an increase of the production cross section of the {\PV}{\PH} signal process with energy compared to the SM prediction. Together with ${\cal O}_{\PH\textrm{D}}$ and ${\cal O}_{\PH\PW\textrm{B}}$, the current operators describe all relevant {\PV}-fermion coupling modifications (Fig.~\ref{fig:Feynman_digarams}, middle). The remaining operators, $\mathcal{O}_{\PH\PW}$, $\mathcal{O}_{\PH\PW\textrm{B}}$, $\mathcal{O}_{\PH\textrm{B}}$, and their $CP$ conjugates, referred to as gauge coupling operators, together with ${\cal O}_{\PH\textrm{D}}$ and ${\cal O}_{\PH\square}$, also involving only gauge fields,  modify the \PH-\PV coupling, shown in Fig.~\ref{fig:Feynman_digarams} (right). Interference with SM interactions gives rise to a linear increase of the cross section with energy.

The operators ${\cal O}^{(1)}_{\PH\PQq}$, ${\cal O}_{\PH\PQu}$, ${\cal O}_{\PH\PQd}$, ${\cal O}_{\PH\textrm{B}}$, and ${\cal O}_{\PH\widetilde{\textrm{B}}}$ affect only the $\PZ\PH$ production, because both left- and right-handed initial-state quarks of the same flavor take part in $\PZ\PH$ production, while only the left-handed quarks of different flavors are involved in $\PW\PH$ production~\cite{Falkowski:2014tna,Banerjee:2018bio}.

\begin{figure*}[!hbt]
\centering
\includegraphics[width=0.9\textwidth]{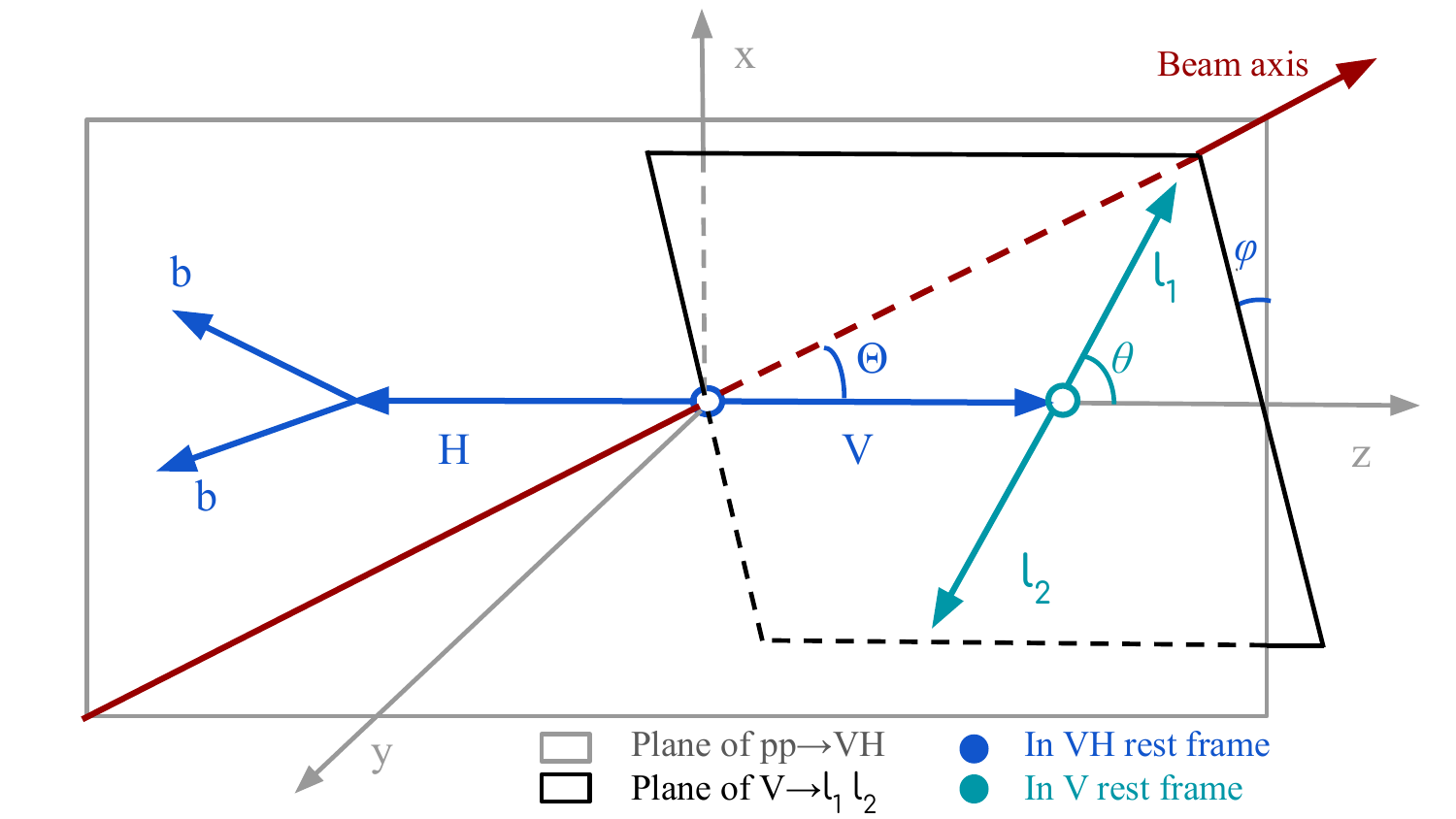}
\caption{Decay planes and angles in the $\PV(\to \Pell_1 \Pell_2){\PH}(\to \PQb \PAQb)$ production. The $\Theta$ angle is defined in the $\PV\PH$ rest frame, while $\theta$ is defined in the \PV rest frame. Figure modified from Ref.~\cite{Banerjee:2019twi}. The coordinate system used in the sketch of the decay plane is independent of the general CMS coordinate system that is used for the analysis.}
\label{fig:HelicityFrame}
\end{figure*}

The measurement reported in this paper probes the operators in Table~\ref{tab:smeft_operators} using kinematic variables exhibiting cross section growth with energy due to the presence of EFT operators and, for the first time in this channel, the angular structure that is sensitive to the interference effects. For the angular information, we exploit the techniques of Ref.~\cite{Banerjee:2019twi} and define the angles and decay planes of the $\PV\PH$ process as shown in Fig.~\ref{fig:HelicityFrame}. In this Figure, $\theta$ is the polar angle of the negatively charged lepton for $\PV={\PW}^{-},\PZ$ and the neutrino for $\PV=\PW^{+}$, in the vector boson rest frame coordinate system, $\varphi$ is the angle between the vector boson decay plane and the plane of the vector boson and the beam axis, and $\Theta$ is the angle between the beam axis and the vector boson direction in the center of mass frame of the $\PV\PH$ system. The squared amplitude of $\PV(\to \Pell_1 \Pell_2)\PH(\to \PQb \PAQb)$ production, when summing over the lepton helicities, has the form
\begin{equation}
	\abs{\mathcal{M} \left(\hat{s}, \Theta, \theta, \varphi \right)}^{2} = {\sum}_{i}\; a_i \left(\hat{s}\right) f_i \left(\Theta, \theta, \varphi \right),
\label{Eq:Amplitude}
\end{equation}
where index $i$ represents the lepton helicity, $a_i\left(\hat{s}\right)$ are functions of the Wilson coefficients and the partonic center-of-mass energy $\sqrt{\hat{s}}$ involved in the process.

The functions $f_i$ depend on the three angles in Fig.~\ref{fig:HelicityFrame} and are summarized in Eq.~(\ref{Eq:angular_moments}) following Ref.~\cite{Banerjee:2019twi}:
\begin{equation}
\begin{aligned}
f_{1} &= f_{LL} = \sin^{2}\Theta \sin^{2}\theta  \\ 
f_{2} &= f^1_{TT} = \cos\Theta \cos\theta  \\  
f_{3} &= f^2_{TT} = (1+\cos^{2}\Theta)(1+ \cos^{2}\theta)  \\ 
f_{4} &= f^1_{LT} = \cos\varphi \sin\Theta \sin\theta  \\ 
f_{5} &= f^2_{LT} = \cos\varphi \sin\Theta \sin\theta \cos\Theta \cos\theta  \\
f_{6} &= \tilde{f}^{\,1}_{LT} = \sin\varphi \sin\Theta \sin\theta  \\
f_{7} &= \tilde{f}^{\,2}_{LT} = \sin\varphi \sin\Theta \sin\theta \cos\Theta \cos\theta  \\
f_{8} &= f_{TT'} = \cos^{2}\varphi \sin^{2}{\Theta} \sin^{2}{\theta}  \\
f_{9} &= \tilde{f}_{TT'} =  \sin^{2}{\varphi} \sin^{2}{\Theta} \sin^{2}{\theta}, 
\label{Eq:angular_moments}
\end{aligned}
\end{equation}
where the indices $T$ and $T'$ refer to two transverse polarizations of the intermediate vector boson and index $L$ refers to its longitudinal polarization.

Angular observables are useful to probe the $CP$ structure of the \PH-\PV interaction. A traditional inclusive measurement implicitly integrates over $\Theta, \theta, \varphi$, thus removing all the terms in Eq.~(\ref{Eq:angular_moments}), except $f_{LL}$ and $f^2_{TT}$. This leads to the loss of important information due to the removal of most of the angular moment dependencies, which can be recovered for $f^2_{LT}$, $\tilde f^{\,2}_{LT}$, $f_{TT'}$, and $\tilde f_{TT'}$  by using a triple-differential analysis with respect to all three angles. Because the incoming quark and antiquark directions are not known at the LHC, the terms proportional to $f^1_{TT}$, $f^1_{LT}$, and $\tilde{f}^1_{LT}$ cannot be extracted.

In this analysis, the complete event information is exploited to build optimal observables with likelihood-free inference methods~\cite{Brehmer:2018kdj,Brehmer:2018hga,Chen:2020mev,Chatterjee:2021nms,Chatterjee:2022oco,GomezAmbrosio:2022mpm}, extracting the SMEFT effects to be described in Section~\ref{sec:strategy}.

Finally, a rotation to the mass eigenstate basis~\cite{deFlorian:2016spz} is performed for the coefficients $\cHW$, $\cHB$, and $\cHWB$ to reduce the number of parameters probed in the measurement and avoid unconstrained directions in the space of Wilson coefficients~\cite{Davis:2021tiv}:
\begin{equation}\begin{aligned}
{\gtZZ} & =  -2\frac{v^2}{\Lambda^2} \left( s_w^2 {\cHB} + c_w^2 {\cHW} + s_wc_w {\cHWB} \right), \\
{\gtZa} & =  -2\frac{v^2}{\Lambda^2} \left( s_wc_w({\cHW} - {\cHB}) + \frac{1}{2}(s_w^2 - c_w^2){\cHWB} \right), \\
{\gtaa} & =  -2\frac{v^2}{\Lambda^2} \left( c_w^2 {\cHB} + s_w^2 {\cHW} - s_wc_w {\cHWB} \right),
\label{eq:mass_eigenstates}
\end{aligned}\end{equation}
where $s_w$ and $c_w$ represent the sine and cosine functions of the weak mixing angle, respectively, and $v$ is the SM vacuum expectation value of the Higgs field.
We also perform a similar translation for the Wilson coefficients of the corresponding complex conjugate operators:
\begin{equation}\begin{aligned}
{\gfZZ} = {\gtZZtil} & =  -2\frac{v^2}{\Lambda^2} \left( s_w^2 {\cHBtil} + c_w^2  {\cHWtil} + s_wc_w  {\cHWBtil} \right),\\
{\gfZa} = {\gtZatil} & =  -2\frac{v^2}{\Lambda^2} \left( s_wc_w({\cHWtil} -  {\cHBtil}) + \frac{1}{2}(s_w^2 - c_w^2) {\cHWBtil} \right),\\
{\gfaa} = {\gtaatil} & =  -2\frac{v^2}{\Lambda^2} \left( c_w^2  {\cHBtil} + s_w^2  {\cHWtil} - s_wc_w  {\cHWBtil} \right).
\label{eq:mass_eigenstates_odd}
\end{aligned}\end{equation}
The  anomalous couplings $\gtZa$, $\gtaa$ and their $CP$-odd counterparts are not expected to be constrained from $\PW\PH$ and $\PZ\PH$ production and are therefore not considered for this analysis. The prefactor $-2\frac{v^2}{\Lambda^2}$ is dropped in the rest of the paper. 
The Wilson coefficients $\cHD$ and $\cHbox$ do not introduce new Lorentz structure and are dropped due to the smallness of their effects. 
Therefore, the final set of Wilson coefficients probed in this paper are $\cHqo$, $\cHqt$, $\cHu$, $\cHd$, $\gtZZ$ and $\gfZZ$. 
It is important to note that the corresponding degrees of freedom in the $\PW\PH$ production, namely $\gtWW$ and $\gfWW$, are fully defined through the constraints of the previously listed coefficients after removing the unconstrained directions in Wilson coefficient space~\cite{Davis:2021tiv}. While the structure of the EFT effects in the $\PW\PH$ channel is different, this does not lead to additional degrees of freedom. Obtaining constraints on $\gtWW$ and $\gfWW$ from the ones on $\gtZZ$ and $\gfZZ$ is equivalent to a scaling by the cosine squared of the weak mixing angle. 

The EFT effects in the gluon-induced $\PZ\PH$ production are not considered. The vector couplings do not affect the gluon-induced $\PZ\PH$ process~\cite{Rossia:2023hen}. It is also checked that the normalization of this process does not affect the sensitivity to the Wilson coefficients considered in the analysis.

\section{The CMS detector}
\label{sec:CMS}

The central feature of the CMS apparatus is a superconducting solenoid of 6\unit{m} internal diameter, providing a magnetic field of 3.8\unit{T}. Within the solenoid volume are a silicon pixel and strip tracker, a lead tungstate crystal electromagnetic calorimeter (ECAL), and a brass and scintillator hadron calorimeter (HCAL), each composed of a barrel and two endcap sections. Forward calorimeters extend the pseudorapidity coverage provided by the barrel and endcap detectors. Muons are reconstructed in gas-ionization detectors embedded in the steel flux-return yoke outside the solenoid. More detailed descriptions of the CMS detector, together with a definition of the coordinate system used and the relevant kinematic variables, can be found in Refs.~\cite{CMS:2008xjf,CMS:2023gfb}. 

The silicon tracker used in 2016 measured charged particles within the range $\abs{\eta} < 2.5$. For nonisolated particles of $1 < \pt < 10\GeV$ and $\abs{\eta} < 1.4$, the track resolutions were typically 1.5\% in \pt and 25--90 (45--150)\mum in the transverse (longitudinal) impact parameter~\cite{CMS:2014pgm}. At the start of 2017, a new pixel detector was installed~\cite{Phase1Pixel}; the upgraded tracker measures particles up to $\abs{\eta} = 3.0$ with typical resolutions of 1.5\% in \pt and 20--75\mum in the transverse impact parameter~\cite{DP-2020-049} for nonisolated particles of $1 < \pt < 10\GeV$. The CMS particle-flow (PF) algorithm~\cite{CMS:2017yfk} aims to reconstruct and identify each individual particle in an event, with an optimized combination of information from the various elements of the CMS detector.

Events of interest are selected using a two-tiered trigger system. The first level (L1), composed of custom hardware processors, uses information from the calorimeters and muon detectors to select events at a rate of around 100\unit{kHz} within a fixed latency of about 4\mus~\cite{CMS:2020cmk}. The second level, known as the high-level trigger, consists of a farm of processors running a version of the full event reconstruction software optimized for fast processing, and reduces the event rate to around 1\unit{kHz} before data storage~\cite{CMS:2016ngn}. 

\section{Data and simulated samples}
\label{sec:DataMC}

The data used in this analysis are from $\Pp\Pp$ collisions at $\sqrt{s}=13\TeV$ collected by the CMS experiment from 2016 to 2018, corresponding to an integrated luminosity of 138\fbinv. Background estimates are obtained primarily from simulation, but also include corrections derived from data for the main background processes, as discussed in Sections~\ref{sec:reco} and \ref{sec:result}.

Several triggers are used to collect events containing final-state particles consistent with the signal processes considered. The triggers used to select events in the $\zerol$ channel make use of \ptmiss and missing hadronic transverse momentum, \mht. These quantities are derived from the reconstructed objects as identified by the PF algorithm. The $\mht$ is defined as the magnitude of the negative vector $\ptvec$ sum of all reconstructed jets with $\pt>20\GeV$ and ${\abs{\eta}<2.5}$. The main triggers used in each of the data-taking periods require the same threshold on $\ptmiss$ and $\mht$. This threshold is 110\GeV in 2016, and 120\GeV in the 2017 and 2018 data-taking periods. In the $\onel$ channel, single-lepton triggers are used. The \pt threshold for electrons is 27\GeV in the 2016 data-taking period, rising to 32\GeV in 2017--2018. For muons, the \pt threshold is 24\GeV in the 2016 and 2018 data-taking periods, and is increased to 27\GeV in 2017~\cite{muontrigger}. Dilepton triggers are used to select events in the $\twol$ channel. The $\pt$ thresholds for electrons are 23 and 12\GeV in all data-taking periods. For muons, the \pt thresholds are 17 and 8\GeV in all data-taking periods; the triggers used in 2017 and 2018 differ from those used in 2016 by the additional requirement that the dimuon invariant mass must be greater than 3.8\GeV. In addition to the \pt thresholds, the triggers require the leptons to pass stringent identification criteria. The trigger-level leptons are also required to be isolated from other tracks and energy deposits in the calorimeters.

The \POWHEG~2.0 generator~\cite{Nason:2004rx,Frixione:2007vw,Alioli:2010xd} is employed to generate \ttbar events at next-to-leading order (NLO) in perturbative QCD~\cite{Frixione:2007nw}. For the normalization of the \ttbar sample, the production cross section is calculated at next-to-next-to-leading order (NNLO) with the resummation of soft gluons at next-to-next-to-leading logarithmic precision with the \textsc{Top++}2.0 program~\cite{Czakon:2011xx}. Event generation for the production of a single top quark in the $t$ channel~\cite{Frederix:2012dh} and in association with a \PW boson~\cite{Re:2010bp} is performed at NLO with the \POWHEG generator as well. Events with the production of a single top quark in the $s$ channels are generated at NLO using the \MGvATNLO~2.4.2 generator~\cite{Alwall:2014hca} and top quark decays are simulated with {\sc madspin}~\cite{Artoisenet:2012st}. The production of a \PW or \PZ boson in association with jets is simulated at NLO with {\MGvATNLO}. These samples are categorized based on jet multiplicity and \pt of the \PW or \PZ boson. Event samples overlapping in phase space regions are reweighted so that the total cross section of a given process is conserved. Additionally, the simulated {\PW}+jets and {\PZ}+jets samples are corrected as a function of the angular momentum separation between jets ({\dRJJ}), in the region $\dRJJ < 1$. This correction is needed to account for a shape mismodeling observed in the region $\dRJJ < 1$, as discussed in Section~\ref{sec:selection}. The correction improves the agreement between data and the NLO $\PV$+jets predictions. The reweighting is derived for $\PV$+jets events and is flavor-agnostic, hence one single correction is derived for the $\PV$+jets processes, separately in each analysis channel. It has been verified that the shape corrections are consistent across flavors of the jets. The simulation-to-data reweighting is propagated to $\PV$+jets events in all analysis regions. Shape-altering uncertainties associated with this correction of the $\PV$+jets simulation are extracted as the statistical uncertainty of the correction factors and are small in size. These uncertainties are uncorrelated across analysis channels to account for residual channel-specific data-to-simulation non-closures after the application of the corrections. The differences in these corrections originate from variations in background composition across channels, e.g., the larger contribution of backgrounds due to top quark production in the {\zerol} and {\onel} channels compared to the {\twol} channel. The diboson processes, $\PW\PW$, $\PW\PZ$, and $\PZ\PZ$ are also simulated at NLO in QCD with the {\POWHEG} or {\MGvATNLO} generators. The QCD multijet events are produced at LO using \MGvATNLO with up to four outgoing partons in the final state. The merging schemes employed for the {\MGvATNLO} LO and NLO samples with the parton shower model are MLM~\cite{Alwall:2007fs} and FxFx~\cite{Frederix:2012ps}, respectively.

The quark-induced Higgs boson production in association with a \PW or \PZ boson, including the effects of SMEFT operators, is simulated using \MGvATNLO at LO plus one additional jet (LO+1j). This follows the prescription discussed in the LHC EFT working group note~\cite{Belvedere:2024nzh}. The production cross sections used to normalize the simulated signal samples at LO+1j are rescaled to the NLO cross-section predictions using the {\POWHEG}~\cite{Nason:2004rx,Frixione:2007vw,Alioli:2010xd} event generator extended with the \textsc{MiNLO} procedure~\cite{Hamilton:2012np,Luisoni:2013cuh}. 
The NLO electroweak corrections are applied to the $\PW\PH$ and $\PZ\PH$ processes as functions of \PW or \PZ bosons \pt, respectively~\cite{deFlorian:2016spz}. Additionally, a reweighting of the $\ptV$ observable for the nominal \MGvATNLO LO+1j signal sample is performed to match NLO {\POWHEG} accuracy. The SMEFT effects in the $\PV\PH$ production are simulated using the \textsc{SMEFTsim}~v3.0~\cite{Brivio:2017btx,Brivio:2020onw} software package with the \texttt{topU3l} flavor scheme at LO considering single-operator insertions without SMEFT corrections to the propagators. The effect of the SMEFT corrections to the propagators is negligible for the operators considered in this analysis, as described in Ref.~\cite{Brivio:2020onw}. The SMEFT cutoff scale, $\Lambda$, is set to 1{\TeV}. In the \textsc{SMEFTsim}~model, the correction to the total Higgs decay width is computed using separate $k$ factors for each decay channel following Ref.~\cite{Brivio:2019myy}. The LO+1j samples are generated under the SM hypothesis and, for each simulated event, a number of weights, computed by matrix-element reweighting using \textsc{MadWeight}~\cite{Artoisenet:2008zz}, are stored. The stored weights encode the EFT effects that arise when setting the Wilson coefficients to nominal values (1 and 2) and are interpolated using a second-order polynomial. A sufficient number of nominal values are simulated which allows the interpolation to recover the full polynomial EFT dependency, which is then used to compute event weights for any combination of Wilson coefficients. 

The gluon-induced $\PZ\PH$ process in the SM is simulated at LO with {\POWHEG} and its cross section is calculated at NLO. 

The NNPDF3.1~\cite{Ball:2017nwa} NNLO parton distribution functions (PDFs), with the strong coupling constant $\alpS (M_{\mathrm{\PZ}})$ set to 0.118, are used in all simulated samples to model the momentum distribution of partons inside the colliding protons. Generated partons undergo parton showering and hadronization using \PYTHIA~{8.212}. The underlying event activity in each sample is simulated using the CP5 tune, derived by tuning the model parameters for multiple parton interactions in \PYTHIA using minimum bias data collected by the CMS experiment~\cite{CMS:2019csb}. For all samples, in order to match the pileup conditions in data and simulation, a weighting is performed in simulation based on the value of the total inelastic $\Pp\Pp$ cross section, which is taken to be 69.2\unit{mb}~\cite{CMS:2018mlc}. The generated samples are processed through the CMS detector simulation based on \GEANTfour~\cite{Agostinelli:2002hh}, using the same reconstruction algorithms used on data.

\section{Event reconstruction}
\label{sec:reco}

The CMS PF algorithm~\cite{CMS:2017yfk} aims to reconstruct and identify each individual particle in an event, with an optimized combination of information from the various elements of the CMS detector. The energy of photons is obtained from the ECAL measurement. The energy of electrons is determined by using the PF algorithm and from a combination of the electron momentum at the primary interaction vertex as determined by the tracker, the energy of the corresponding ECAL cluster, and the energy sum of all bremsstrahlung photons spatially compatible with originating from the electron track. The energy of muons is obtained from the curvature of the corresponding track. The energy of charged hadrons is determined from a combination of their momentum measured in the tracker and the matching ECAL and HCAL energy deposits, corrected for the response function of the calorimeters to hadronic showers. Finally, the energy of neutral hadrons is obtained from the corresponding corrected ECAL and HCAL energies. 

The primary vertex (PV) is taken to be the vertex corresponding to the hardest scattering in the event, evaluated using tracking information alone, as described in Section 9.4.1 of Ref.~\cite{CMS-TDR-15-02}. The PV must be within 24\cm of the nominal interaction point along the beam axis and within 2\cm in the transverse plane.

The electron reconstruction is performed with the Gaussian sum filter algorithm~\cite{CMS:2020uim}. The momentum resolution for electrons with $\pt \approx 45\GeV$ from $\PZ \to \Pe \Pe$ decays ranges from 1.6 to 5\%. It is generally better in the barrel region than in the endcaps, and also depends on the bremsstrahlung energy emitted by the electron as it traverses the material in front of the ECAL~\cite{CMS:2020uim,CMS-DP-2020-021}. Electrons are preselected by requiring $\pt>7\GeV$, $\abs{\eta}<2.4$, $\abs{d_{xy}}<0.05\cm$, and $\abs{d_z}<0.2\cm$, where $d_{xy}$ and $d_z$ are the transverse and longitudinal impact parameters associated with the electron tracks, respectively.

A multivariate approach is used to identify prompt electrons arising from the decays of \PW or \PZ bosons. Two identification criteria corresponding to the expected electron identification efficiency of either 90\% (loose working point) or 80\% (tight working point) are used in this analysis. The loose working point is used when counting the number of additional leptons beyond the selected electrons in each event, as well as for the event selection of the {\twol} channel. The tight working point is required to select events in the {\onel} channel. The electron \pt threshold in the {\onel} channel is 30\GeV. For the {\twol} channel, the thresholds are 25 and 17\GeV for the two electrons. This stricter requirement is needed in the {\onel} channel to reduce the multijet background.

Muons are measured in the pseudorapidity range $\abs{\eta} < 2.4$, with detection planes made using three technologies: drift tubes, cathode strip chambers, and resistive plate chambers. The efficiency to reconstruct and identify muons is greater than 96\%. Matching muons to tracks measured in the silicon tracker results in a relative transverse momentum resolution, for muons with \pt up to 100\GeV, of 1\% in the barrel and 3\% in the endcaps. The \pt resolution in the barrel is better than 7\% for muons with \pt up to 1\TeV~\cite{CMS:2018rym}. Muons are preselected by requiring the following conditions: $\pt>5\GeV$, $\abs{\eta}<2.4$, $\abs{d_{xy}}<0.5\cm$, and $\abs{d_z}<1.0\cm$. 

Two sets of selection conditions, corresponding to 96--98\% (tight working point) and $\approx$99\% (loose working point) efficiency for the identification of prompt muons, are used in the analysis. The use of loose and tight working points across different final states is similar to the approach used for electrons. The muon \pt threshold in the {\onel} channel is 25\GeV, and 25 and 15\GeV in the {\twol} channel.

For each event, hadronic jets are clustered from these reconstructed particles using the infrared- and collinear-safe anti-\kt algorithm~\cite{Cacciari:2008gp} with distance parameters 0.4 (AK4 jets) and 0.8 (AK8 jets), as implemented in the \FASTJET package~\cite{Cacciari:2011ma}.
The distance between two particles in the $\eta$-$\phi$ plane, where $\phi$ is azimuthal angle in radians, is defined as $\Delta R = \sqrt{\smash[b]{(\Delta\eta)^2+(\Delta\phi)^2}}$. Jet momentum is determined as the vectorial sum of all particle momenta in the jet, and is found from simulation to be, on average, within 5 to 10\% of the momentum of the particle-level jets reconstructed using stable particles (lifetime $>30\unit{ps}$), excluding neutrinos, 
over the whole $\pt$ spectrum and detector acceptance~\cite{CMS-DP-2021-033}. Additional $\Pp\Pp$ interactions within the same or nearby bunch crossings (pileup) can contribute with additional tracks and calorimetric energy depositions to the jet momentum. To mitigate this effect, tracks identified as originating from pileup vertices are discarded before jet reconstruction~\cite{CMS:2017yfk}.  For AK4 jets, an offset correction~\cite{Cacciari:2007fd} is applied to correct for remaining pileup contributions~\cite{CMS:2016lmd}. For AK8 jets, the pileup-per-particle identification (PUPPI) algorithm~\cite{Bertolini:2014bba} is used to mitigate the effect of pileup at the reconstructed particle level. It has been shown that the PUPPI algorithm improves the resilience of jet substructure observables against pileup~\cite{CMS:2020ebo}. Jet energy corrections are derived from simulation to bring the measured response of jets to that of the corresponding particle-level jets on average. In situ measurements of the momentum balance in dijet, $\PGg + \text{jet}$, $\PZ + \text{jet}$, and multijet events are used to account for any residual differences in the jet energy scale (JES) between data and simulation~\cite{CMS:2016lmd, CMS-DP-2021-033}. The jet energy resolution (JER) amounts typically to 15--20\% at 30\GeV, 10\% at 100\GeV, and 5\% at 1\TeV~\cite{CMS:2016lmd, CMS-DP-2021-033}.  Additional selection criteria are applied to each jet to remove jets potentially dominated by anomalous contributions from various subdetector components or reconstruction failures~\cite{CMS-PAS-JME-16-003}. Jets overlapping with the selected muons and electrons, satisfying $\Delta R < 0.4$ for AK4 jets and $\Delta R < 0.8$ for AK8 jets, are removed. Jets contained within the tracker volume, satisfying $\abs{\eta}<2.5$, are considered in the analysis. Thresholds used on \pt of AK4 jets are 25\GeV in 0- and {\onel} final states, whereas it is 20\GeV in the {\twol} final state. AK8 jets with $\pt > 250\GeV$ are considered in the analysis. 

When the \PH boson has a large \pt, its decay products are merged into a single large-area jet. In this scenario, referred to as the boosted topology, AK8 jets are used to reconstruct the $\PH\to\PQb\PAQb$ decay. In a large fraction of events with $\PV\PH$ production, the \PH boson has relatively small \pt, and two \PQb quarks from the $\PH\to\PQb\PAQb$  give rise to two separate small-area jets. In such cases, referred to as the resolved topology, AK4 jets are used to reconstruct particles from \PQb quark fragmentation.
In the resolved topology, jets from final-state radiation, exceeding 20\GeV in \pt, are recovered by adding the momenta of jets close to the Higgs boson candidate in the dijet mass calculation.

In CMS, a deep neural network (DNN) based algorithm, \DeepJet~\cite{Bols:2020bkb}, is used for \PQb jet identification from AK4 jets. Both charged and neutral particles, as well as secondary vertices within an AK4 jet, are exploited to check whether the jet contains at least one \PQb hadron. To match the efficiency of \PQb tagging in data, a correction factor depending on \pt, $\eta$, and flavor of the initiator quark or gluon is applied to the simulation. The correction factor is derived using samples enriched in dileptonic and single-lepton \ttbar events for jets initiated by \PQb and \PQc quarks, and $\PZ+$jets events for the jets initiated by light quarks and gluons. Performance of the {\DeepJet} tagger and its calibration results using Run 2 data are summarized in Ref.~\cite{CMS-DP-2023-005}. For an AK4 jet to be declared as $\PQb$ tagged, its \DeepJet \PQb tagging discriminator value is required to be larger than a specific threshold. The threshold values, corresponding to mistag rates for jets initiated by light quarks or gluons of approximately 10, 1, and 0.1\% are referred to as loose, medium, and tight working points, respectively. The selection of AK4 jets based on \PQb tagging discriminator values for different analysis regions is discussed in Section~\ref{sec:selection}. Given that the \DeepJet score is correlated to the vector boson kinematics recoiling against the jet system in the $\PV\PH$ topology, it is observed that $\ptV$ and the additional jet (\ie, additional to the $\Pb$-tagged jets associated with the reconstructed Higgs boson candidate) observables are modified in shape and normalization when applying the \DeepJet calibration corrections discussed earlier in this Section. These vector-boson-related observables are corrected with a simulation-to-simulation reweighting employed after the application of the \DeepJet calibration corrections in order to match their shape and normalization before the application of the {\DeepJet} calibration corrections. These reweighting factors have been derived by employing a one-dimensional ($\ptV$) or two-dimensional ($\ptV$ and additional jet multiplicity) reweighting, respectively, depending on the analysis regions. This reweighting is applied for all analysis channels in $\ptV$, for lepton flavor bins, per signal and background processes, and separately among data-taking eras.

In the resolved topology, the precision of the four-momenta of the {\PQb}-tagged jets is improved with a multivariate regression analysis utilizing a DNN trained on simulated \PQb jets from \ttbar events~\cite{CMS:2019uxx}, which, in turn, improves the dijet invariant mass resolution by 10--15\%. The momenta of the Higgs boson candidate jets are corrected by the application of the \PQb jet energy regression described above. After the application of the regression, dedicated scale and smearing corrections are applied to the \PQb jet energy to match the dijet invariant mass resolution in simulation to that in data. The scale and smearing parameters are extracted using events containing two jets recoiling against a {\PZ} boson that decays into leptons,  where the transverse momentum of the second jet is small compared to the first jet and to the vector boson. The parameters are extracted by extrapolating the transverse momentum of the second jet to 0. Because the \PZ boson \pt is balanced with the jet \pt, and given that the lepton momentum measurement is more precise, the ratio of the reconstructed jet momentum ($\pt^{j}$) to the \PZ boson momentum ($\pt^{\Pell\Pell}$) enables a precise measurement of the jet momentum and energy. The distribution of the energy difference between the \pt balance procedure, exploiting the ratio of the reconstructed jet momentum ($\pt^{j}$) to the \PZ boson momentum ($\pt^{\Pell\Pell}$), and the \PQb jet energy regression is used to estimate the \PQb jet energy regression scale and smearing corrections and uncertainties. 

To identify AK8 jets that are consistent with being initiated by two b quarks from the Higgs boson decay, a dynamic graph convolutional neural network based tagger, \ParticleNet~\cite{Qu:2019gqs}, and the groomed jet mass are used.  The soft drop algorithm, a generalization of the modified mass drop algorithm~\cite{Butterworth:2008iy,Dasgupta:2013ihk},
with angular exponent $\beta = 0$, soft cutoff threshold $z_{\text{cut}} = 0.1$, and characteristic radius $R_{0} = 0.8$~\cite{Larkoski:2014wba} is used to groom the AK8 jets, and the corresponding groomed mass, known as the soft-drop mass~(\msd), is required to be within a window of 90--150\GeV for a jet to be \PH tagged. In this algorithm, the constituents of the AK8 jets are reclustered using the Cambridge--Aachen algorithm~\cite{Dokshitzer:1997in,Wobisch:1998wt} and  the relative \pt between the successive clusters of particles merged during the jet clustering is checked to remove soft, wide-angle particles from the jet. The \ParticleNet algorithm uses jet constituents, as well as the secondary vertices associated with the jet and has been studied extensively in CMS~\cite{CMS-DP-2020-002}. The relative score of the \ParticleNet tagger, specifying how likely the jet is originated from a heavy particle decaying to a pair of \PQb quarks ($\PX\to\PQb\PQb$), as compared to the same from a light quark or gluon, has been used in the analysis. The \ParticleNet tagger performance is decorrelated from the mass of the parent particle by using simulated samples containing Lorentz-boosted spin-0 particles with a flat mass spectrum between 15 and 250\GeV and subsequently decaying to a pair of quarks as the signal sample, and the multijet QCD events as the background sample in training. In the analysis, the threshold used on the \ParticleNet discriminator corresponds to a misidentification rate of approximately 1\% for light-quark and gluon jets and signal efficiency of 80--85\%  in simulation. The calibration of the \ParticleNet tagger is performed using an event sample enriched in jets originating from the $\Pg\to\PQb\PAQb$ decay, which have substructure similar to that for jets arising from the $\PH\to\PQb\PAQb$ decay~\cite{CMS-DP-2022-005}, and correction factors depending on the jet \pt are applied in simulation to match the efficiency of the {\ParticleNet} tagger in data~\cite{CMS-PAS-BTV-22-001}.

The missing transverse momentum vector \ptvecmiss is computed as the negative vector sum of the transverse momenta of all the PF candidates in an event, and its magnitude is denoted as \ptmiss~\cite{CMS:2019ctu}. The \ptvecmiss is modified to account for corrections to the energy scale of the reconstructed jets in the event. Track-based missing transverse momentum ${\vec{p}_{\mathrm{T}}^{\mkern3mu\text{trk,miss}}}$~\cite{CMS-PAS-JME-09-010} is also used in the analysis. In this case, only tracks with \pt above a minimum momentum threshold and with an impact parameter consistent with the PV are considered in the vectorial sum. The tracks are also required to pass quality requirements designed to limit the contribution from misreconstructed tracks. Corrections are applied to ensure that the azimuthal angle of \ptvecmiss has a flat distribution. Anomalous high-\ptmiss events can be due to a variety of reconstruction failures, detector malfunctions or noncollision backgrounds. Such events are rejected by event selection conditions that are designed to identify more than 85--90\% of the spurious high-\ptmiss events with a mistagging rate less than 0.1\%~\cite{CMS:2019ctu}. 

\section{Event selection and background estimation}
\label{sec:selection}

The basic selection criteria closely follow those used in the \VHbb cross section measurement in the STXS phase space~\cite{CMS:2023vzh}. Some key differences are the usage of the updated CMS flavor tagging algorithms for resolved and boosted topologies, \ie, \DeepJet and \ParticleNet, for the identification of AK4 and AK8 jets stemming from the \HBB decays, respectively.

In the resolved topology, the two AK4 jets with the highest \PQb tagging scores, referred to as $\textrm{b}_{1,2}$, form the \PH boson candidate. In the boosted topology, the AK8 jet with the highest \ParticleNet $\PX\to\PQb\PQb$/QCD score ratio provides the \PH boson candidate. 

For all three final states considered in the analysis, $\zerol$, $\onel$, and $\twol$, the signal region (SR) is constructed to capture most of the $\PV(\to \Pgn\Pgn / \Pell \Pgn / \Pell\Pell)\PH(\to \PQb \PAQb)$ signal events. A condition that the reconstructed \PH candidate mass is within 90--150\GeV is imposed in the SR for all channels. Several control regions (CRs) are defined targeting events from different processes, vector boson production in association with heavy-flavor ({\PV}+HF) and light-flavor ({\PV}+LF) jets, and top quark pair production (\ttbar) -- by changing the conditions on different variables. 

Events that are selected in the SR (CRs) of both resolved and boosted categories are assigned to the SR (CRs) in the boosted category. If an event is selected in the SR of the resolved category and any CR of the boosted category, it is considered only in the former category. The treatment of the overlap has been optimized using the sensitivity of the analysis under the SM hypothesis as a figure of merit, as also discussed in Ref.~\cite{CMS:2023vzh}. It was checked that the signal acceptance in the EFT-sensitive phase space does not change in the presence of the SMEFT operators considered in this analysis. 

In the following, selection conditions for different CRs and SR in the resolved and boosted categories of different final states are discussed. The kinematic selections on electrons, muons, jets, and missing transverse momentum are already discussed in Section~\ref{sec:reco}.

\subsection{Selection criteria in the \texorpdfstring{$\zerol$}{zero-lepton} channel}

Events in the $\zerol$ final state are selected by requiring a significant amount of \ptmiss. Selected events must not contain any isolated lepton with $\pt > 15\GeV$ and $|\eta|<2.5$. The value of \ptmiss in the resolved and boosted categories is required to be higher than 190 and 250\GeV, respectively.
In the resolved category, the minimum of \ptmiss and {\mht}, defined as the magnitude of the negative vectorial sum of all AK4 jets in the event, is required to be higher than 150\GeV, ensuring that the trigger efficiency reaches the plateau, and the presence of at least two AK4 jets is required.
To reduce the QCD multijet background, a requirement of $\Delta\phi > 0.5$ between $\ptvecmiss$ and any jet with $\pt > 30\GeV$ is applied. 
In the resolved category, among two \PQb candidate AK4 jets, the highest \pt jet is required to satisfy $\pt > 60\GeV$, whereas the other jet is required to have $\pt > 30\GeV$, and the \PH candidate \pt, \ie, the \pt of the dijet system, is required to exceed 120\GeV to reduce the background contribution. A requirement of  $\Delta\phi > 2.5$ between \ptvecmiss and the \PH candidate dijet system is imposed to further reduce the QCD multijet contamination. 
A similar requirement is also used in the boosted category, where a looser condition of $\Delta\phi > \frac{\pi}{2}$ is used between \ptvecmiss and the \PH candidate AK8 jet since there are few signal events at small $\Delta\phi$ values.
 
To separate the \ttbar CR from other regions, the number of additional jets, defined as the AK4 jets with $\pt>30\GeV$ satisfying $\DR>0.4$ with respect to the $\PH\to\PQb\PAQb$ candidate AK4 jets, is used. In the boosted category, at least one \PQb-tagged AK4 jet with $\pt > 20\GeV$ outside the \PH boson candidate AK8 jet is required to separate this CR from other regions. The additional set of main event selection criteria in the $\zerol$ channel is reported in Table~\ref{tab:selection_0l_resolved} for the resolved category. The main event selection criteria used in the boosted category of the $\zerol$ channel are listed in Table~\ref{tab:selection_0l_boosted}.

\begin{table*}[!htb]
\topcaption{Selection criteria for the resolved category in the $\zerol$ final state. Momenta and masses have units of \GeV. }
\centering
\renewcommand{\arraystretch}{1.2}
\cmsTable{
  \begin{tabular}{lllll}
  Variable & SR & {\PV}+HF CR & {\PV}+LF CR & \ttbar CR \\
  \hline
  {Max (\PQb tag score of $\textrm{b}_{1}$ and $\textrm{b}_{2}$)}  & $\geq$ medium & $\geq$ medium  & $<$ medium  & $\geq$ medium \\
  {Min (\PQb tag score of $\textrm{b}_{1}$ and $\textrm{b}_{2}$)} & $\geq$ loose  & $\geq$ loose   & $\geq$ loose   & $\geq$ loose  \\
  No. of additional jets & $<2$ &  $<2$ & $<2$ & $\geq 2$ \\
  $\Delta\phi$ (${{\vec p}_{\mathrm{T}}^{\mkern3mu\text{trk,miss}}}$, {\ptvecmiss}) & $<0.5$ & $<0.5$   & $<0.5$  & \NA \\
  $M$ (b$_1$, b$_2$) & $\in [90,150]$ & $\in [50,250]$ $\cup$ $\notin [90, 150]$ & $\in [50,250]$ & $\in [50,250]$  \\
  \end{tabular}
}
\label{tab:selection_0l_resolved}
\end{table*}

\begin{table*}[!htb]
\topcaption{Selection criteria for the boosted category in the $\zerol$ final state. Momenta and masses have units of \GeV.}
\centering
\renewcommand{\arraystretch}{1.2}
\begin{tabular}{lllll}
Variable & SR & {\PV}+HF CR & {\PV}+LF CR & \ttbar CR \\
\hline
{\PH} {\ParticleNet} score & $\geq 0.94$ & $\geq 0.94$ & $\in [0.1,0.94)$ & $\geq 0.94$ \\
$\text{N}_{\text{{\PQb}-tagged}}$ jets outside \PH cand. & $=0$ &  $=0$ & $=0$ & $> 0$ \\
${\msd}^{\PH}$ & $\in [90,150]$ & $\in [50,250]$ $\cup$ $\notin [90, 150]$ & $>50$ & $>50$ \\
\end{tabular}
\label{tab:selection_0l_boosted}
\end{table*}

\subsection{Selection criteria in the \texorpdfstring{$\onel$}{one-lepton} channel}

The $\onel$ channel is characterized by the presence of exactly one isolated lepton and missing transverse momentum. 
The \PW boson momentum is reconstructed using the lepton, $\ptvecmiss$, and by fixing the \PW boson mass to be 80.4{\GeV}~\cite{PDG2022} to obtain the $z$ component of the momentum. When two physical solutions exist for the $z$ component of the momentum of the neutrino from the {\PW} boson decay, both are retained for use in the multivariate analysis discussed in Section~\ref{sec:strategy}. A requirement of the reconstructed \PW boson $\pt$ to be higher than 170 and 250\GeV is applied in the resolved and boosted categories, respectively, in order to reduce the background contribution. 
The reconstructed \PW boson is also required to satisfy $\Delta\phi < 2$ with respect to  $\ptvecmiss$ in both categories.
Since the \PW and \PH candidates are expected to be produced in a back-to-back topology, $\Delta\phi$ between those is required to be higher than $\frac{\pi}{2}$ and 2.5 in the boosted and resolved categories, respectively. A tighter condition in the resolved category helps to remove background contributions, especially in the $\ttbar$ CR, which is small in the boosted category. The additional set of main event selection criteria for the resolved category in the $\onel$ final state are described in Table~\ref{tab:selection_1l_resolved}. In the boosted category, a condition on the number of {\PQb}-tagged AK4 jets with $\pt > 20\GeV$ outside the \PH boson candidate AK8 jet is applied, which reduces its overlap with the resolved category and separates the \ttbar CR from other regions. 
Table~\ref{tab:selection_1l_boosted} lists the main event selection criteria used in the boosted category of the $\onel$ channel.

\begin{table*}[!htb]
\topcaption{Selection conditions for the resolved category in the $\onel$ final state. Momenta and masses have units of \GeV. }
\centering
\cmsTable{
  \begin{tabular}{lllll}
  Variable & SR & {\PV}+HF CR & {\PV}+LF CR & \ttbar CR \\
  \hline
  {Max (\PQb tag score of $\textrm{b}_{1}$,$\textrm{b}_{2}$)} & $\geq$ medium & $\geq$ medium  & $\geq$ loose and $<$ medium   & $\geq$ tight \\
  {Min (\PQb tag score of $\textrm{b}_{1}$,$\textrm{b}_{2}$)} & $\geq$ loose & $\geq$ loose  & \NA  & \NA \\
  No. of additional jets & $<2$ &  $<2$ & \NA & $\geq2$ \\
  $M$ (b$_1$, b$_2$) & $\in [90,150]$ & {$\in [50,250]$ $\cup$ $\notin [90, 150]$} & $\in [50,250]$ &  $\in [50,250]$ \\
  \end{tabular}
}
\label{tab:selection_1l_resolved}
\end{table*}

\begin{table*}[!htb]
\topcaption{Selection conditions for the boosted category in the $\onel$ final state. Momenta and masses have units of \GeV.}
\centering
\renewcommand{\arraystretch}{1.2}
\cmsTable{
  \begin{tabular}{lllll}
  Variable & SR & {\PV}+HF CR & {\PV}+LF CR & $\ttbar$ CR \\
  \hline
  {\PH} {\ParticleNet} score & $\geq 0.94$ & $\geq 0.94$ & $\in [0.1,0.94)$ & $\geq 0.94$ \\
  $\text{N}_{\text{{\PQb}-tagged}}$ jets outside \PH cand. & $=0$ &  $=0$ & $=0$ & $> 0$ \\
  ${\msd}^{\PH}$ & $\in [90,150]$ & $\in [50,250]$ $\cup$ $\notin [90, 150]$& $>50$ & $>50$ \\
  \end{tabular}
}
\label{tab:selection_1l_boosted}
\end{table*}

\subsection{Selection criteria in the \texorpdfstring{\twol}{two-lepton} channel}

In the $\twol$ channel, two oppositely charged isolated leptons with an invariant mass closest to the nominal {\PZ} boson mass form the \PZ candidate. 
A \pt threshold of 75\GeV is used on the {\PZ} candidate in the resolved category, whereas it is 250\GeV in the boosted category. In the resolved category, the dilepton invariant mass is used to separate the \ttbar CR, while the boosted category also uses the number of {\PQb}-tagged AK4 jets similar to the $\onel$ channel. The vector boson mass requirements for the definition of the V+HF and \ttbar CRs are different for the resolved and boosted analysis selections, being looser for the latter to increase the data statistics and to improve the precision on the constraints of the background processes extracted in the boosted phase-space. The vector boson mass requirements around the \PZ mass pole is tighter for the resolved V+HF CR selection compared to the V+LF CR selection to increase the V+HF background purity inside the \PZ mass window. The additional set of main event selection criteria used for the resolved and boosted categories in the $\twol$ final state is given in Tables~\ref{tab:selection_2l_resolved} and ~\ref{tab:selection_2l_boosted}, respectively.

\begin{table*}[!htb]
\topcaption{Selection conditions for the resolved category in the $\twol$ final state. Momenta and masses have units of \GeV.}
\centering
\renewcommand{\arraystretch}{1.2}
\cmsTable{
\begin{tabular}{lllll}
Variable & SR & {\PV}+HF CR & {\PV}+LF CR & $\ttbar$ CR \\
\hline
{Max (\PQb tag score of $\textrm{b}_{1}$,$\textrm{b}_{2}$)} & $\geq$ medium & $\geq$ medium  & $<$ loose  & $\geq$ tight \\
{Min (\PQb tag score of $\textrm{b}_{1}$,$\textrm{b}_{2}$)} & $\geq$ loose & $\geq$ loose    & $<$ loose  & $\geq$ loose \\
$m^{\PV}$  & $\in [75, 105]$ & $\in [85, 97]$   & $\in [75, 105]$  & $\in [10, 75]$ or $\geq 120$ \\
$M$ (b$_1$, b$_2$) & $\in [90,150]$ & $\in [50,250]$ $\cup$  $\notin [90, 150]$ & $\in [90,150]$ & $\in [50,250]$ \\
\end{tabular}
}
\label{tab:selection_2l_resolved}
\end{table*}

\begin{table*}[!htb]
\topcaption{Selection conditions for the boosted category in the $\twol$ final state. Momenta and masses have units of \GeV.}
\centering
\renewcommand{\arraystretch}{1.2}
\begin{tabular}{lllll}
Variable & SR & {\PV}+HF CR & {\PV}+LF CR & $\ttbar$ CR \\
\hline
{\PH} {\ParticleNet} score & $\geq 0.94$ & $\geq 0.94$ & $< 0.94$ & $\geq 0.94$ \\
$m^{\PV}$  & $\in [75, 105]$ & $\in [75, 105]$   & $\in [75, 105]$  &  $\leq 75$ or $\geq 105$  \\
${\msd}^{\PH}$ & $\in [90,150]$ & $\in [50,250]$ $\cup$  $\notin [90, 150]$ & $>50$ & $>50$ \\
\end{tabular}
\label{tab:selection_2l_boosted}
\end{table*}

\subsection{Application of process-specific corrections}
\label{sec:rw_corrections}

A mismodeling of the $\PW$+jets and $\PZ$+jets samples simulated at NLO is observed at small values of $\DR$ between the {\HBB} candidate AK4 jets, as discussed in Section~\ref{sec:DataMC}. A reweighting is performed in the $\DR(\text{b}_1,\text{b}_2)<1$ range of the {\PV}+HF CR and the correction factors extracted differentially in $\DR(\text{b}_1,\text{b}_2)$ are applied to the simulated {\PV}+jets samples in all resolved regions of the analysis. In the resolved category of the $\twol$ final state, a data-to-simulation mismodeling is observed at low-\pt values of the \PH candidate dijet system in both {\PV}+HF and {\PV}+LF CRs, while the \pt spectra of individual AK4 jets are well modeled. This also affects the modeling of other variables characterizing the dijet system. An additional reweighting is performed separately for electron and muon final states and different \pt ranges of the dilepton system. The weights are derived in the mismodeled $\pt(\text{b}_1,\text{b}_2)<100\GeV$ region using the {\PV}+LF CR and applied to the $\PZ$+jets process in all regions of the 2-lepton channel. After reweighting, a significant improvement is observed in the modeling of kinematic variables corresponding to the dijet system, e.g., $\pt(\text{b}_1,\text{b}_2)$, $\Delta\eta(\text{b}_1,\text{b}_2)$, and $\Delta\phi(\text{b}_1,\text{b}_2)$. Shape uncertainties associated to these corrections are propagated to the fit model as discussed in Section~\ref{sec:sys}.

\section{Analysis strategy for effective field theory parameter extraction}
\label{sec:strategy}

A multivariate approach using a number of kinematic variables characterizing the $\PV\PH$ system is used to construct optimal observables for probing Wilson coefficients corresponding to the SMEFT operators described in Section~\ref{sec:EFTinVH}. Specifically, the boosted information tree (BIT)~\cite{Chatterjee:2021nms,Chatterjee:2022oco}, a likelihood-free inference method using boosted decision trees, is employed in this analysis. The core idea of the BIT approach, which differs from the MELA method applied in previous CMS analyses~\cite{Gao:2010qx,Bolognesi:2012mm,Anderson:2013afp}, is briefly discussed below.  

\subsection{Likelihood ratio estimation}
\label{sec:Likelihood_ratio_estimation}

An event, characterized by detector-level observables $\mathbf{x}$, can be produced by a parton-level configuration $\mathbf{z}$. 
The relation between these two quantities and any theory parameter $\boldsymbol{\theta}$ affecting the event is: 
\begin{equation}
  p(\mathbf{x}\abs{\boldsymbol{\theta}) = \int \rd{z} \, p(\mathbf{x}}\mathbf{z}) p(\mathbf{z}|\boldsymbol{\theta}).
 \label{eq:convolution}
\end{equation}
For this analysis, $\boldsymbol{\theta}$ corresponds to the set of considered Wilson coefficients. 
In Eq.~(\ref{eq:convolution}), the transfer function $p(\mathbf{x}|\mathbf{z})$ models the parton showering and hadronization processes as well as all detector effects. The differential cross-section at a specific point $\boldsymbol{\theta}$ in the Wilson coefficient space with respect to the parton-level configuration is proportional to a quadratic sum of contributions at the matrix-element ($\mathcal{M}$) level:
\begin{equation}
\rd\sigma_{\boldsymbol{\theta}}(\mathbf{z}) \propto | \mathcal{M_{\text{SM}}}(\mathbf{z}) + \sum_{1\leq i \leq M} \theta_i \mathcal{M}_i(\mathbf{z}) |^2 \rd{z} ,
\end{equation}
where the sum in the second term runs over the $M$ Wilson coefficients probed in the analysis, \ie, $\boldsymbol{\theta} \left(= {\theta}_{i=1,\ldots,M} \right)$. 

The parton-level probability distribution function  $p(\mathbf{z}|\boldsymbol{\theta})$ is written as: 
\begin{equation}
p(\mathbf{z}|\boldsymbol{\theta}) = \frac{1}{\sigma(\boldsymbol{\theta})}\frac{\rd\sigma_{\boldsymbol{\theta}}(\mathbf{z})}{\rd{z}}
\end{equation}
and is simulated in the analysis as described in Section~\ref{sec:DataMC}. 

The parton-level dependence on the Wilson coefficients is at most quadratic. The factorization of this polynomial dependence on $\boldsymbol{\theta}$ can be extracted based on the integral of Eq.~(\ref{eq:convolution}). Therefore, the polynomial dependence also applies to the detector-level probability distribution function  $p(\mathbf{x}|\boldsymbol{\theta})$:
\begin{equation}
p(\mathbf{x} |\boldsymbol{\theta})  = p_0(\mathbf{x}) + \sum_{1\leq i \leq M} p_{l,i}(\mathbf{x}) \frac{\theta_i}{\Lambda^2} +  \sum_{1 \leq i \leq M} p_{q,i}(\mathbf{x}) \frac{\theta_i^2}{\Lambda^4} +  \sum_{1 \leq i < j \leq M} p_{m,i,j}(\mathbf{x}) \frac{\theta_i \theta_j}{\Lambda^4},
\label{eq:factorization}
\end{equation}
where $p_0(\mathbf{x})$ describes the probability distribution function of the event $\mathbf{x}$ under the SM hypothesis and the $i$ and $j$ indices run over the $M$ different Wilson coefficients considered in the analysis. The SMEFT effects in Eq.~(\ref{eq:factorization}) consist of three different types of contributions: 
\begin{itemize}
\item Linear part, labeled $p_{l,i}(\mathbf{x})$: describes the interference between SM and SMEFT operators. 
\item Quadratic part, labeled $p_{q,i}(\mathbf{x})$: describes the squared contribution from a single dimension-six operator.
\item Mixed part, labeled $p_{m,i,j}(\mathbf{x})$: describes the contribution from the interference between two dimension-six operators.
\end{itemize}

The likelihood ratio between two SMEFT hypotheses corresponding to different values of Wilson coefficients ($\boldsymbol{\theta}$ and $\boldsymbol{\theta_0}$) is used as a test statistic to extract the EFT effects in the analysis. Using notations and developments presented in Ref.~\cite{Chatterjee:2022oco}, the likelihood ratio can be written as:
\begin{equation}
q_\theta = - \log \frac{L(\mathcal{D}|\boldsymbol{\theta})}{L(\mathcal{D}|\boldsymbol{\theta_0})},
\label{eq:LLR}	
\end{equation}	
while the likelihood is given by: 
\begin{equation}
 L(\mathcal{D}|\boldsymbol{\theta}) = \frac{\re^{-\mathcal{L}\sigma(\boldsymbol{\theta})}}{N!} \prod_{1 \leq i \leq N} \mathcal{L}\sigma(\boldsymbol{\theta}) p(\mathbf{x}_i|\boldsymbol{\theta}),
\label{eq:LLH}	
\end{equation}	
where $\mathcal{L}$ is the integrated luminosity of the data set $\mathcal{D}$, consisting of $N$ events, and $\sigma(\boldsymbol{\theta})$ is the inclusive cross section of the considered process, the $\PV\PH$ production in this case, under the hypothesis of the presence of a given Wilson coefficient $\boldsymbol{\theta}$. Under these assumptions, the likelihood ratio in Eq.~(\ref{eq:LLR}) reads:	
\begin{equation}
 q_\theta = \mathcal{L}(\sigma(\boldsymbol{\theta}) - \sigma(\boldsymbol{\theta_0}) ) - \sum_{1 \leq i \leq N} \log R(\mathbf{x}_i|\boldsymbol{\theta}, \boldsymbol{\theta_0}),
\label{eq:LLH_simplified}	
\end{equation}	
where 
\begin{equation}
R(\mathbf{x}|\boldsymbol{\theta}, \boldsymbol{\theta_0}) = \frac{\sigma(\boldsymbol{\theta}) p(\mathbf{x}|\boldsymbol{\theta})}{\sigma(\boldsymbol{\theta_0}) p(\mathbf{x}|\boldsymbol{\theta_0})}.
\label{eq:R_function}	
\end{equation}	
The normalization factor $\sigma(\boldsymbol{\theta})/\sigma(\boldsymbol{\theta_0})$ is independent of event kinematics and is obtained directly from simulation. 
The remaining ratio of probabilities is intractable, \ie, not known from the parton-level configuration, but can be estimated using the methodology described in Refs.~\cite{Brehmer:2018eca,Brehmer:2018hga} that corresponds to the generic minimization of a functional with a target function given as a joint probability distribution of detector- and parton-level quantities. This makes it possible to estimate a detector-level ratio of probabilities. More precisely, using the following ratio as the target in the previously mentioned minimization:  
\begin{equation}
 R(\mathbf{x},\mathbf{z}| \boldsymbol{\theta}, \boldsymbol{\theta_0}) = \frac{p(\mathbf{x},\mathbf{z}|\boldsymbol{\theta})}{p(\mathbf{x},\mathbf{z}|\boldsymbol{\theta_0})} = \frac{p(\mathbf{z}|\boldsymbol{\theta})}{p(\mathbf{z}|\boldsymbol{\theta_0})},
\label{eq:this_target}
\end{equation}	
makes it converge to estimate $R(\mathbf{x}|\boldsymbol{\theta}, \boldsymbol{\theta_0})$. 
As discussed in Section~\ref{sec:DataMC}, the joint likelihood ratios in Eq.~(\ref{eq:this_target}) are simulated and can be used in a regression to estimate the detector-level likelihood ratio. 

As described in Ref.~\cite{Chatterjee:2022oco}, the likelihood ratio estimation is performed using decision trees.
Making use of the polynomial dependence of $p(\mathbf{x} |\boldsymbol{\theta})$ on $\boldsymbol{\theta}$, the likelihood ratio in Eq.~(\ref{eq:R_function}) can be expanded as:
\begin{equation}
R(\mathbf{x}|{\boldsymbol{\theta}}, {\boldsymbol{\theta}_0}) = 1 + \sum_{1 \leq i \leq M} (\theta_i - \theta_0)R_i(\mathbf{x}) + \sum_{1 \leq i \leq j \leq M}\frac{1}{2}(\theta_i - \theta_0)(\theta_j - \theta_0) R_{i,j}(\mathbf{x}),
 \label{eq:R_expanded_simple}
\end{equation}
where the indices $i$ and $j$ run over the considered Wilson coefficients and $R_i(\mathbf{x})$ and $R_{i,j}(\mathbf{x})$ correspond to the partial derivatives with respect to the Wilson coefficients, as defined in Eq.~(\ref{eq:partial_Rs}):
\begin{equation}
  R_{i}(\mathbf{x}) =  \left.\frac{\partial}{\partial{\theta}_{i}} R(\mathbf{x}|{\boldsymbol{\theta},\boldsymbol{\theta}_0}) \right\vert_{{\theta} = {\theta}_0}, \qquad
  R_{i,j}(\mathbf{x}) = \left.\frac{\partial}{\partial{\theta}_{i}}\frac{\partial}{\partial{\theta}_{j}} R(\mathbf{x}|{\boldsymbol{\theta}, \boldsymbol{\theta}_0}) \right\vert_{{\theta} = {\theta}_0}.
\label{eq:partial_Rs}
\end{equation}
The regression uses partial derivatives of the joint likelihood ratio in Eq.~(\ref{eq:this_target}), obtained using the SMEFT weights in the simulated $\PV\PH$ samples, as described in Section~\ref{sec:DataMC}, as target functions to make the minimization converge to $R_{i}$ and $R_{i,j}$. 
For the background processes, since no EFT effects are considered, $R_{i}(\mathbf{x})$ and $R_{i,j}(\mathbf{x})$ are always zero. The regression can be performed using boosted decision trees as implemented in the BIT framework described in Ref.~\cite{Chatterjee:2022oco}.
Each component $R_i$ or $R_{i,j}$ of the likelihood ratio Eq.~(\ref{eq:R_expanded_simple}) are then estimated independently as the output of the regression.

As an alternative to the Mean Squared Error (MSE) loss function used in Ref.~\cite{Chatterjee:2022oco}, the cross-entropy loss function is used for the trainings described in Section~\ref{sec:BIT}. The expression of the functional that is minimized during the training is given by Eq.~(8) of Ref.~\cite{GomezAmbrosio:2022mpm}:
\begin{equation}
  L\left[ g(\mathbf{x}|\boldsymbol{\theta}) \right] = - \left(\int \mathrm{d}x \mathrm{d}z \frac{\sigma(\boldsymbol{\theta})}{ \sigma(\boldsymbol{\theta}_0)} \frac{p(\mathbf{x,z}| \boldsymbol{\theta})}{p(\mathbf{x,z}| \boldsymbol{\theta}_0)}\log(1 - g(\mathbf{x}|\boldsymbol{\theta})) + \log(g(\mathbf{x}|\boldsymbol{\theta})) \right), 
  \label{eq:functional_CE}
\end{equation}
where the notation was adapted to match the definitions given earlier in this Section. 
The minimization of the loss function in Eq.~(\ref{eq:functional_CE}) with respect to $g(\mathbf{x}|\boldsymbol{\theta})$ leads to the minimizer $g^{*}(\mathbf{x}|\boldsymbol{\theta})$: 
\begin{equation}
g^{*}(\mathbf{x}|\boldsymbol{\theta}) = \frac{1}{1 + R(\mathbf{x}|\boldsymbol{\theta},\boldsymbol{\theta}_0)},
\label{eq:functional_min_CE}
\end{equation}	
which is a monotonic function of the likelihood ratio that can be inverted to recover it. 
The main difference compared to the MSE loss function lies in the fact that the target function (given as a ratio of probabilities) has to be positive. 
For the MSE loss function, terms in Eq.~(\ref{eq:partial_Rs}) can take negative values. 
While the MSE loss function could be used to train individual SMEFT components, e.g., linear or quadratic terms of SMEFT expansion separately, the cross-entropy loss function requires the full polynomial expression in the weights during training to guarantee positivity. Nonetheless, the number of degrees of freedom remains the same and the trained polynomials can be recombined to obtain the linear and quadratic components shown in Eq.~(\ref{eq:R_expanded_simple}). 
Since the implementation described above closely follows the one presented in Ref.~\cite{Chatterjee:2022oco} using per-event information of the simulated data sets encoding the SMEFT predictions, the same label of boosted information tree (BIT) is used to refer to it.

\subsection{The BIT training regions and input features}
\label{sec:BIT}

The regression discussed in Section~\ref{sec:Likelihood_ratio_estimation} is implemented using the light gradient boosting machine (LightGBM)~\cite{NIPS2017_6449f44a} tree boosting framework and following the BIT method with the cross-entropy loss function. The training is performed separately in the resolved and boosted categories defined in Section~\ref{sec:selection} and inclusively across data taking eras (2016--2018), given the same behavior of the input features of the BIT training across data-taking eras. The signal consists of the SM $\PV\PH$ process, while the SMEFT effects are encoded in the targets of the regression. 
As described in Eq.~(\ref{eq:R_expanded_simple}), each Wilson coefficient is associated with a linear component $R_i$ and to quadratic and mixed components $R_{i,j}$. Each component in this expansion corresponds to an individual BIT training. 

The training is performed  in the analysis SRs defined in Section~\ref{sec:selection} separately for each final state after combining event samples from all data-taking eras. Angular observables sensitive to SMEFT effects as presented in Section~\ref{sec:EFTinVH} are included in the training for the 1- and $\twol$ final states along with kinematic features of these final states, which carry separation power between signal (SM $\PV\PH$ and SMEFT effects) and backgrounds. It is relevant to note that, due to the nature of the $\zerol$ final state topology, it is not possible to construct angular variables, so only energy-sensitive kinematic variables and observables characterizing the {\HBB} system are used in the training. The {\DeepJet} discriminator values (full shape) of the {\PH} candidate {\PQb}-tagged jets and the {\ParticleNet} discriminator value (binned in the calibrated working points according to the descriptions in Section~\ref{sec:reco}) of the {\PH} candidate AK8 jet are used as inputs to the BIT trainings in the resolved and boosted categories, respectively. Since a boosted decision tree technique is used, the algorithm selects the group of features resulting in the largest improvement of the loss function for each iteration. Therefore, a single feature can be employed several times to define the splitting if it carries a significant signal vs. background discrimination power. On the other hand, some features characterized by little to no discrimination power are not employed in the algorithm split. Thus, the training defines itself the relevance of the input features. A fraction of $50\%$ of the events are used for the training, which are then removed from the analysis. 
The regression scores are evaluated for the remaining $50\%$ of the events.

In order to validate the BIT implementation, a background-free BIT training was performed. The target of the training $R(\mathbf{x},\mathbf{z}| \theta, \theta_0)$ is a joint quantity over the $(\mathbf{x},\mathbf{z})$ space, but the output of the training is only a function of $\mathbf{x}$, \ie, the detector-level configurations. 
Thus, the distributions of the target and output cannot be compared directly. 
To obtain comparable quantities, profiling over each of the features used during training is necessary. 
In each bin of the profile, the parton-level configurations will be averaged out and this allows the comparison of the regression output and target on detector-level configurations, hence the validation of the BIT regression training. An excellent agreement between the two detector-level quantities was found, demonstrating the ability of the model to learn the likelihood ratio in the absence of background. 

Nonetheless, the background events need to be considered in the training in order to apply the regression to the analysis. The BIT trainings need to perform two tasks, \ie, separate the background events with vanishing EFT weights, as mentioned in Section~\ref{sec:Likelihood_ratio_estimation}, from the signal events, as well as learning the likelihood ratio of the signal component. The hyperparameters of the BIT are optimized to achieve maximal separation between events that are not affected by EFT effects and the EFT-relevant signal. The BIT parameters are optimized by minimizing the loss function for each SMEFT component training individually. A grid search is performed over hyperparameters to find the optimal number of leaves in a tree, maximum depth of a tree, minimum number of events in a node to allow a split, learning rate, and number of boosting iterations. 
The training data are split into $K$-folds for cross-validation in such a way that the proportion of signal is conserved in each part of the split. In the resolved and boosted categories, the training data are split in $K = 3$ and $K = 5$ folds, respectively. The training was then performed $K$ times by leaving out one of the $K$ data sets which was then used as validation data. Out of the set of trainings for each hyperparameter point in the grid, the one set performing the best on average defines the optimal hyperparameters for that training.

The EFT-sensitive variables, such as the \pt of the vector boson, the invariant mass of the {\PV}{\PH} system, as well as the angular variables described in Section~\ref{sec:EFTinVH} play an important role in distinguishing the effects of EFT operators from the {\PV}{\PH} signal process as predicted by the SM, whereas the variables not relevant for EFT effects, e.g., {\PH} candidate mass, ratio of the {\PV} boson \pt to the {\PH} candidate \pt, are useful to separate the background contributions from the SM {\PV}{\PH} component and the EFT signal. Hence, both classes of input features are highly ranked by the BIT discriminant.

\subsection{Optimization of discriminator shapes}
\label{sec:opt_discriminator}

By performing the training described in Section~\ref{sec:BIT}, all the $R_i$ and $R_{i,j}$ components in the expansion of the likelihood ratio in Eq.~(\ref{eq:R_expanded_simple}) are estimated. 

An example case of $N = 3$ arbitrarily chosen Wilson coefficients affecting the $\PV\PH$ production (labeled generically c1, c2, and c3) is considered. Each point in this three-dimensional space corresponds to a reweighting of the SM events as described by Eq.~(\ref{eq:factorization}). Therefore, the likelihood ratio for a point $\boldsymbol{\theta}=(\theta_1,\theta_2,\theta_3)$ with respect to the SM, \ie, $(0,0,0)$, following Eq.~(\ref{eq:R_expanded_simple}) can be written as follows:
\begin{equation}
R(x|\boldsymbol{\theta}) = 1 + \theta_1  R_{\text{c1}}(x) +\theta_2 R_{\text{c2}}(x) + \theta_3 R_{\text{c3}}(x) + \frac{1}{2}\theta_1^2 R_{\text{c1, c1}}(x) + \theta_1\theta_2 R_{\text{c1, c2}}(x) + \cdots,
\label{eq:R_expanded_3_WC}
\end{equation} 
which is the optimal observable for the separation of the SMEFT signal at a non-SM point $\boldsymbol{\theta}$ from the prediction of the SM. 
However, $R(x|\boldsymbol{\theta})$ is not an optimal test statistic separating the SMEFT signal at a different point, e.g., $\boldsymbol{\theta}^{\prime}$ from the SM.
In Eq.~(\ref{eq:R_expanded_3_WC}), the coefficients of the polynomial are the components of $\boldsymbol{\theta}$, which are extracted in this analysis. In order to retain the maximum separation power of the likelihood ratio for the Wilson coefficient extractions based on the binned templates of $R_{\text{c1}}(x)$, $R_{\text{c2}}(x)$, and $R_{\text{c3}}(x)$, it would be necessary to compute different BIT template shapes for each point in the Wilson coefficient space. 
This is unfeasible for binned templates since up and down variations of the templates due to systematic uncertainties would need to be constructed at each Wilson coefficient point. To address this, a novel template optimization technique is developed. The overall objective of the template optimization is to define a template that retains as much information in signal vs. background separation as possible when all six Wilson coefficients under study in this analysis are profiled together in the fit model. Additionally, for the template optimization process, it is important to avoid regions in the Wilson coefficient space that deteriorate the sensitivity for some coefficients significantly because of anticorrelation effects. However, the SMEFT signal extraction in Section~\ref{sec:result} is performed in the full Wilson coefficient space.

Bayesian optimization is a flexible way of minimizing functions using Gaussian processes and it is used in this analysis to define the optimal template shape for the extraction of Wilson coefficient values that encapsulate the SMEFT effects. 
This model does not assume a specific functional form for the minimization and is typically used when the function to minimize is expensive to compute~\cite{frazier2018tutorial}. This allows global minimization in a limited number of iterations, without the need to compute derivatives of the function to minimize. The figure of merit used for this optimization is the product of the fully profiled 95\% confidence intervals on the Wilson coefficients that the analysis has the power to constrain. Therefore, the optimization consists of finding a candidate point $\boldsymbol{\theta}$ in the $N$-dimensional Wilson coefficient space, thus defining the observable of Eq.~(\ref{eq:R_expanded_3_WC}) that maximizes the expected sensitivity to the $N$ Wilson coefficients using a fully profiled $N$-dimensional fit. Correlations between Wilson coefficients are taken into account in the optimization process because the 95\% confidence intervals in the product are extracted from fully profiled fits in the $N$-dimensional volume of the Wilson coefficient space. 

The constraining power of the various channels depends on the Wilson coefficients that are being targeted. The inclusion of Wilson coefficients with little to no constraining power leads to unconstrained directions in the likelihood scans, which is undesirable as the variations of such coefficients in very large ranges impact the optimal working point of the shape optimization. While the optimization of the template shape is performed on subsets of Wilson coefficients, which are the sensitive Wilson coefficients for a given channel, all of the six Wilson coefficients will be profiled together for the EFT signal extraction.

For the $\zerol$ channel, the Wilson coefficients that are considered in the Bayesian optimization are: $\cHqo$, $\cHqt$, $\cHu$, and $\cHd$. Such coefficients are mostly constrained by kinematic properties and carry minimal dependence on angular variables, as in the $\zerol$ channel it is not possible to use angular correlation between the \PZ boson decay products. In the case of the $\onel$ channel, only three Wilson coefficients: $\cHqt$, $\gtWW$, and $\gfWW$ (which directly translate to $\gtZZ$ and  $\gfZZ$ in terms of numerical values of the coefficients as explained in Section~\ref{sec:EFTinVH}) are used in the optimization since those correspond to the operators that affect the $\PW\PH$ production. 
The optimization in the $\twol$ channel is performed over all six Wilson coefficients probed in this analysis: $\cHqo$, $\cHqt$, $\cHu$, $\cHd$, $\gtZZ$, and $\gfZZ$. The optimization process results in a BIT template, \ie, a binned BIT score, that is optimal only at one particular set of values of Wilson coefficients and is used for the EFT signal extraction. The process does not result in an optimal BIT template in the full Wilson coefficient space -- for any other values of Wilson coefficients, the template does not guarantee the optimal separation of SMEFT signals from the SM background.

\begin{figure*}[!htb]
\centering
\includegraphics[width=0.49\textwidth]{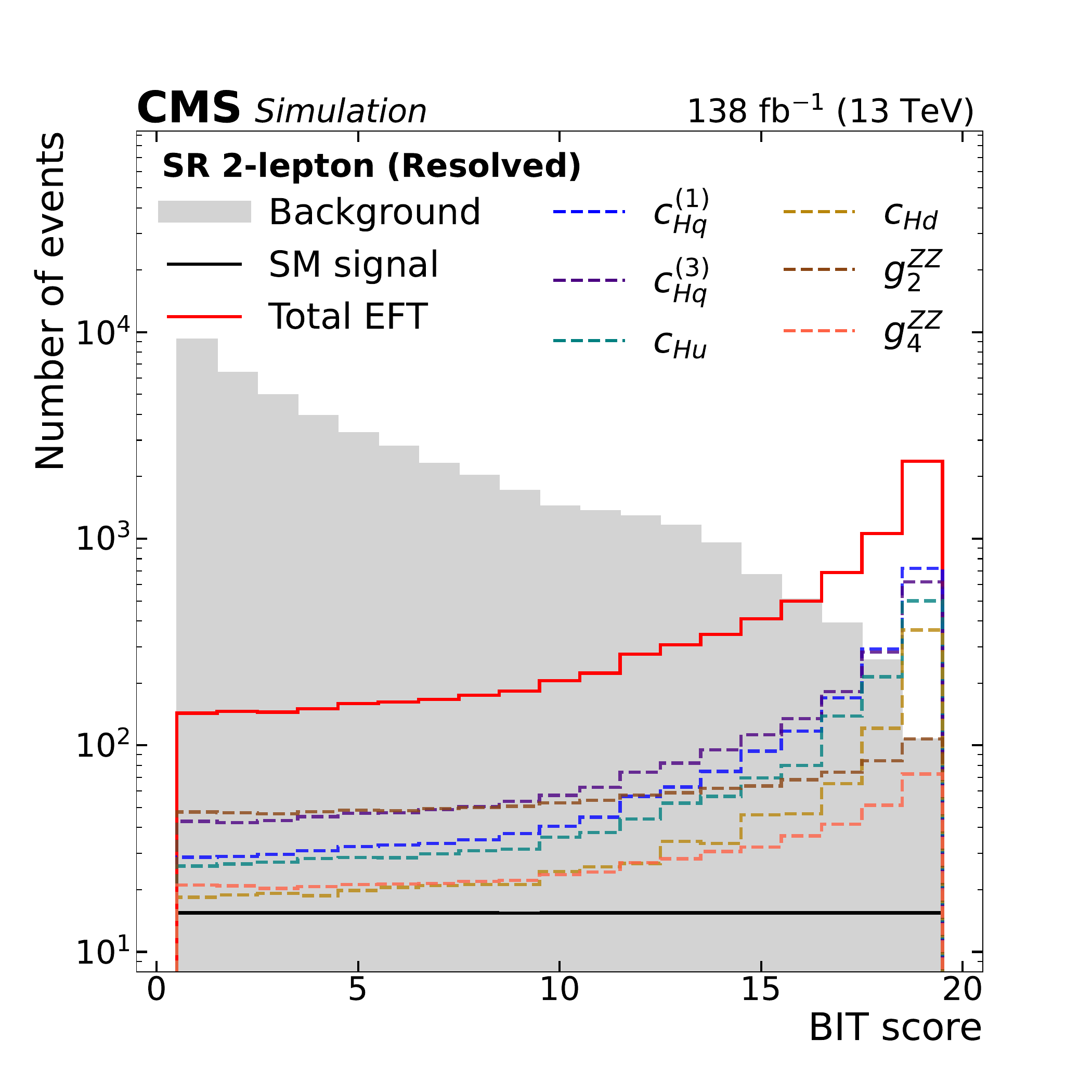}
\includegraphics[width=0.49\textwidth]{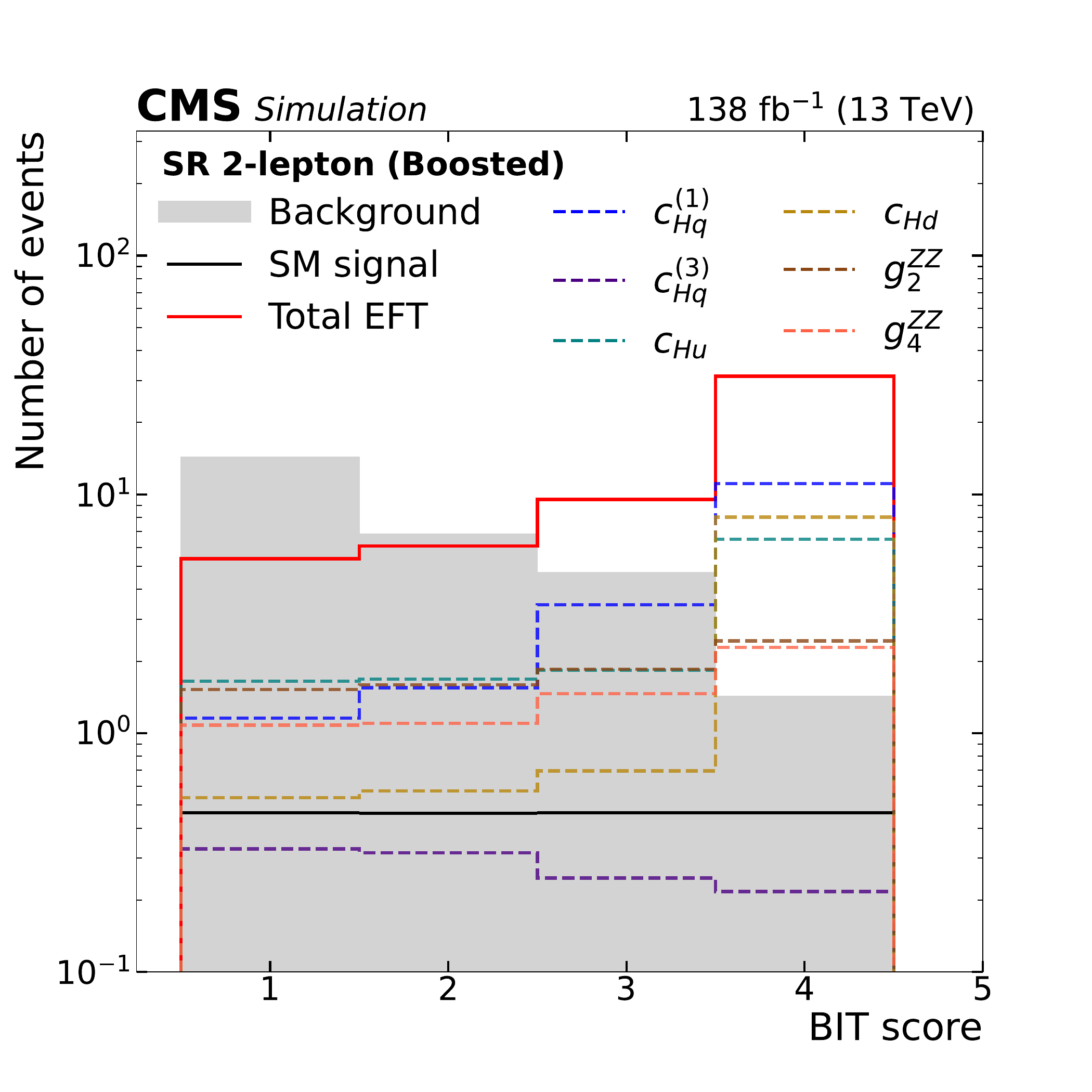}
\caption{Selected template shapes after the optimization process described in Section~\ref{sec:opt_discriminator} in the resolved (left) and boosted (right) categories of the $\twol$ channel. 
The template shapes of the EFT signal components  are shown for arbitrary values of the Wilson coefficients: ($\cHqo$, $\cHqt$, $\cHu$, $\cHd$,  $\gtZZ$,  $\gfZZ$) = (1, 0.8, 1, 1, 2, 2) and (0.2, -0.03, 0.2, 0.2, 1, 1) in the resolved and boosted categories, respectively, and are referred to as 'Total EFT'. 
Individual template shapes for each Wilson coefficient, where only one coefficient is non-zero (set to the same values as above) while the others are set to zero, are also displayed and labeled by their respective symbols.
The SM $\PV\PH$ signal is flat by construction. The background is shown as the grey histogram.}
\label{fig:BIT_templates_2lep}
\end{figure*}

Figure~\ref{fig:BIT_templates_2lep} shows the distribution of the BIT score for several EFT signal hypotheses for values of the Wilson coefficients in the $\twol$ channel (resolved and boosted categories) selected by the discriminant shape optimization procedure described in this section. The SR binning is defined in order to keep a constant SM signal contribution for each bin of the BIT observable and the number of bins is chosen by requiring a minimum number of raw background events: 20 bins are used in the resolved category, while three or four bins are used in the boosted category, depending on the channel. 

The templates, constructed from the sum of all background components, are also reported for comparison and are found to peak at low values of the BIT score, hence guaranteeing signal vs. background separation for the EFT signal extraction discussed in Section~\ref{sec:result}. 

\section{Systematic uncertainties}
\label{sec:sys}

Theoretical uncertainties arising from the choice of PDFs and the value of the strong coupling constant, 1.9\% for the quark-induced $\PZ\PH$ and $\PW\PH$ processes, and 2.4\% for the gluon-induced $\PZ\PH$ process, are derived for each signal and background process following the recommendations given in Ref.~\cite{pdfLHC} and are treated as correlated across data-taking eras. Uncertainties in the factorization and renormalization scales, $\mu_{\mathrm{F}}$ and $\mu_{\mathrm{R}}$, are considered for all background processes and correlated across data-taking eras. These theoretical scale uncertainties are derived by varying $\mu_{\mathrm{F}}$ and $\mu_{\mathrm{R}}$ individually up and down by a factor of two around their default values. The theoretical uncertainty in the $\PH\to\PQb\PAQb$ branching fraction is 0.5\%~\cite{deFlorian:2016spz}. The NLO electroweak corrections to the $\PZ\PH$ and $\PW\PH$ processes have a 2\% uncertainty~\cite{deFlorian:2016spz}.

Since simulated samples are employed to obtain shapes of observable distributions for the various background processes, dedicated modeling uncertainties impacting both shape and normalization of the processes are included in the fit model. These shape uncertainties originate from the reweighting procedure employed to improve the background modeling, e.g., the $\DR$($\PQb\PQb$) and \PH candidate dijet transverse momentum corrections discussed in Section~\ref{sec:rw_corrections}, and are propagated as shape uncertainties to the fit model. These uncertainties are treated as uncorrelated between data-taking eras and across background processes.

The corrections applied to account for differences in the electron and muon trigger, reconstruction, and identification efficiencies in data and simulation are impacted by systematic uncertainties stemming from several sources, \eg, the selection conditions applied to the leptons, the efficiency measurement method, and the limited size of the simulated samples used in the efficiency measurement. They depend on the lepton \pt and $\abs{\eta}$, and affect the process normalizations by 1--2\%. An uncertainty in the $\ptmiss$ trigger efficiency correction is included and has a $\approx 1$\% effect. 

Uncertainties in the $\PQb$ tagging efficiency and misidentification rate measurements for the {\DeepJet} algorithm used in the resolved topology of the analysis depend on the jet flavor, \pt, and $\eta$. These uncertainties are split into nine independent sources and fifteen \pt and $|\eta|$ ranges, and treated as uncorrelated between data-taking eras. For the tagging efficiencies of the {\ParticleNet} discriminator used in the boosted topology, uncertainties are extracted for the {\HBB} events at the particle level. These uncertainties are parameterized in regions of jet \pt (200--300, 300--400, 400--500, 500--600, $>$600\GeV) and are extracted for the medium (0.94--0.98) and tight ($>$0.98) {\ParticleNet} working points used in the analysis for the BIT training. 

Uncertainties in the jet energy scale (JES) and jet energy resolution (JER) depend on the \pt and $\abs{\eta}$ of the jets, and affect the kinematic properties of AK4 and AK8 jets, as well as the {\ptmiss} in the event. The uncertainties in the JES are split into independent sources accounting for different experimental effects~\cite{CMS:2016lmd}. Some of these uncertainty sources are uncorrelated between the different data-taking periods, \eg, when that uncertainty component depends on the size of the available data sample. For {\PQb}-tagged jets, to which the previously described \PQb jet energy regression is used, additional uncertainties in the JES and JER corrections are considered. 
To reduce the effects from statistical fluctuations on the alternative JES and JER template variations, a smoothing technique is applied to templates exhibiting large fluctuations with respect to the nominal templates. The normalization of the systematic variation is fixed, and the ratio of the template to the nominal one is smoothed. 
Additionally, a symmetrization step is also employed for all templates to construct the up and down-variation template histograms. Finally, constrained JES nuisance parameters are decorrelated across the SRs and CRs to release such constraints. The set of uncertainties relevant for the $\PV\PH$ samples applies to the templates of the EFT predictions in the same way those affect the SM prediction.

The uncertainty in the integrated luminosity measurement is 1.2, 2.3, and 2.5\% in the 2016, 2017, and 2018 data-taking periods, respectively~\cite{CMS:2021xjt, CMS:2018elu, CMS:2019jhq}, resulting in an overall uncertainty of 1.6\%. These uncertainties are treated as partially correlated between the three data-taking years. A 4.6\% uncertainty in the total inelastic $\Pp\Pp$ cross section, used to evaluate the pileup profile in data for reweighting to the simulated pileup profile, is applied.

The scale factors correcting the normalization of the main background processes extracted from simulation, \ie, \ttbar, {\PV}$+$LF ($\PV{+}\PQu\PQd\PQs\Pg\PQc$) and {\PV}$+$HF (split in $\PV{+}\PQb$, $\PV{+}\PQb\PQb$), are implemented as freely floating rate parameters and are adjusted by the fit. These scale factors are constrained mostly in the CRs because of their significant statistical power and extrapolated to the SRs. These parameters that scale the process normalizations are treated as uncorrelated between lepton flavors (\Pe, \PGm) for the light-flavor scale factors due to slightly different data-to-simulation modeling behavior of the electron and muon channels. Such a split is found to improve the overall goodness-of-fit due to additional degrees of freedom in the fit model. As we do not observe differences across lepton flavors for the \ttbar and {\PV}+HF processes, we do not employ the flavor-specific splitting for these processes. In addition to the process-specific scale factors, additional unconstrained parameters, used to measure flavor-tagging scale factors {\it in situ\/} in the boosted categories, are employed to account for the (mis)tagging efficiency difference between data and simulation for high-momentum jets initiated by \PQb, \PQc, or light quarks and gluons. These parameters are treated as fully correlated between channels, and are not correlated with the scale factors corresponding to background processes due to their different nature. 

To account for the finite sizes of the simulated samples, each bin of the simulated signal-plus-background template is allowed to vary within its statistical uncertainty, independently from the other bins in the distribution, following the Barlow--Beeston ``light" approach~\cite{bbb}.  Here, signal refers to the SM $\PV\PH$ signal production. 

\section{Results}
\label{sec:result}

Signal-enriched and background-enriched regions are used to extract the SMEFT signal and to constrain the main background processes (\ttbar and {\PV}+ light/heavy-flavor jets), respectively. The definition of the SR and CRs used in the analysis is reported in Section~\ref{sec:selection}. 

In the resolved category of the analysis, the background predictions in the heavy-flavor CRs are extracted by using a template with bins corresponding to exclusive combinations of \DeepJet \PQb tagging scores, binned by working point, of the leading and subleading {\PQb}-tagged AK4 jets. 
The contribution of the \ttbar process is constrained by its rate measured in \ttbar-enriched CRs.
As discussed in Section~\ref{sec:sys}, the process scale factors correcting the normalization of the main background processes extracted from simulation, \ie, \ttbar, {\PV}+LF, {\PV}+HF are considered inclusively in vector boson \pt and in other observables of the analysis. In the 0- and $\twol$ channels the {\PV}$+${\PQb} and {\PV}$+${$\PQb\PQb$} components are split, while in the $\onel$ channel, a freely floating parameter for the $\PV{+}\PQb\PQb$ process is used in addition to a prior constraint that governs the ratio of $\PV{+}\PQb$ to $\PV{+}\PQb\PQb$. This implementation is needed because the yields of $\PV{+}\PQb$ events in the $\onel$ channel are limited due to the tight \PQb tagging requirement applied in the selection and the suppression of the {\PW}+{\PQb} process due to the small size of the corresponding mixing parameters. This approach follows the strategy used in Ref.~\cite{CMS:2023vzh}. Furthermore, all other systematic uncertainties described in Section~\ref{sec:sys} are included in the fit model.

A statistical analysis is performed to probe for the potential presence of SMEFT operator effects in $\PV\PH$ production with a binned maximum likelihood fit using the CMS statistical analysis tool \textsc{Combine}~\cite{CMS:2024onh}, which is based on the \textsc{RooFit}~\cite{Verkerke:2003ir} and \textsc{RooStats}~\cite{Moneta:2010pm} frameworks. 
The BIT template shapes, which were defined as described in Section~\ref{sec:strategy}, are used in the SR for extracting the SMEFT effects. 
The modified frequentist approach~\cite{CMS-NOTE-2011-005,Thomas_1999,Read_2002} is used in this search to set intervals on the set of the Wilson coefficients specified in Section~\ref{sec:EFTinVH}. %with an asymptotic approximation to the profile likelihood test statistic~\cite{Cowan:2010js}.
Goodness-of-fit tests using the saturated model test statistic~\cite{saturated} show excellent closure of the systematics model to data in the SRs and CRs after the application of the maximum likelihood fit.

\begin{figure*}[!htb]
\centering
\includegraphics[width=0.425\textwidth]{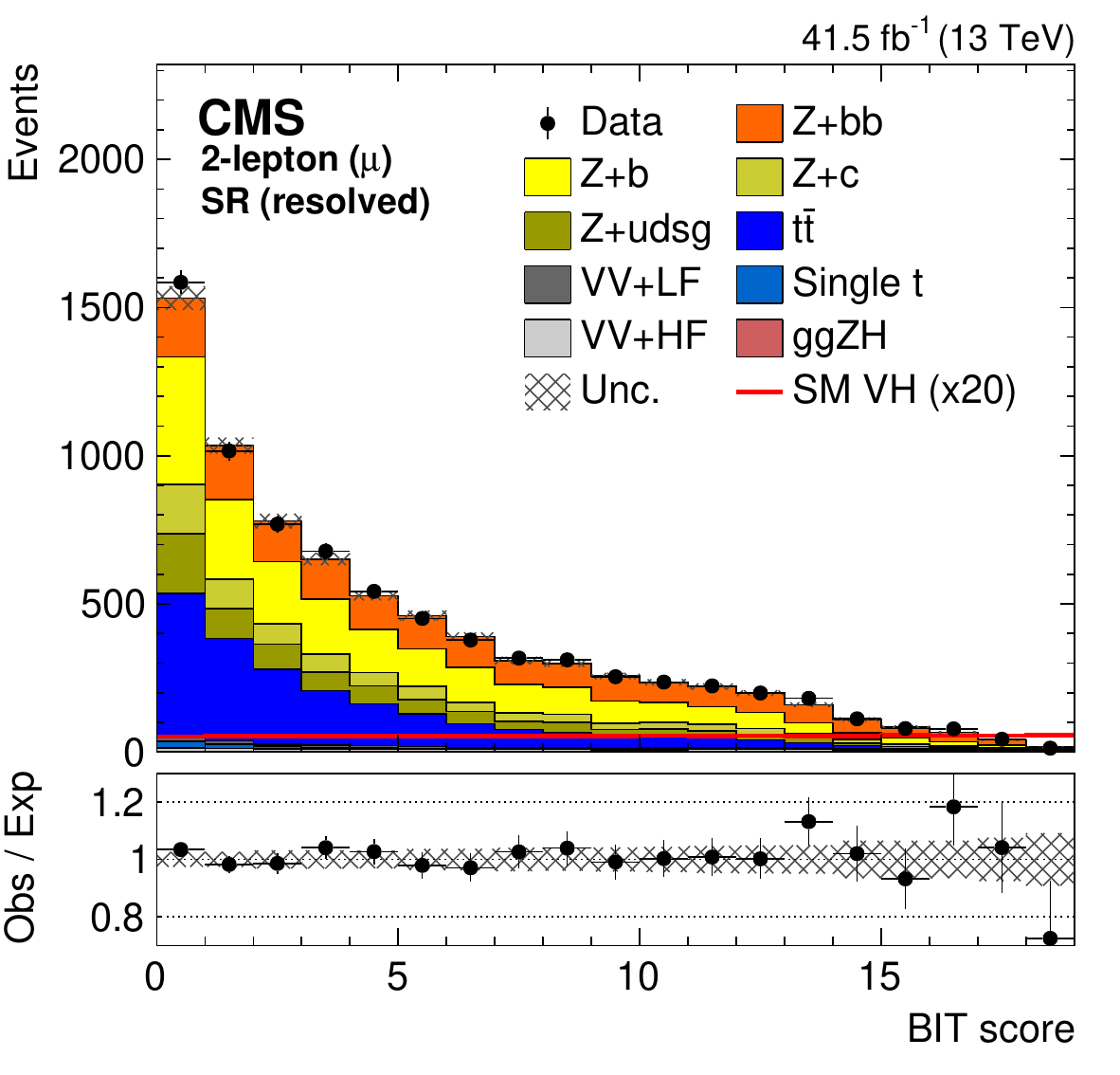}
\includegraphics[width=0.425\textwidth]{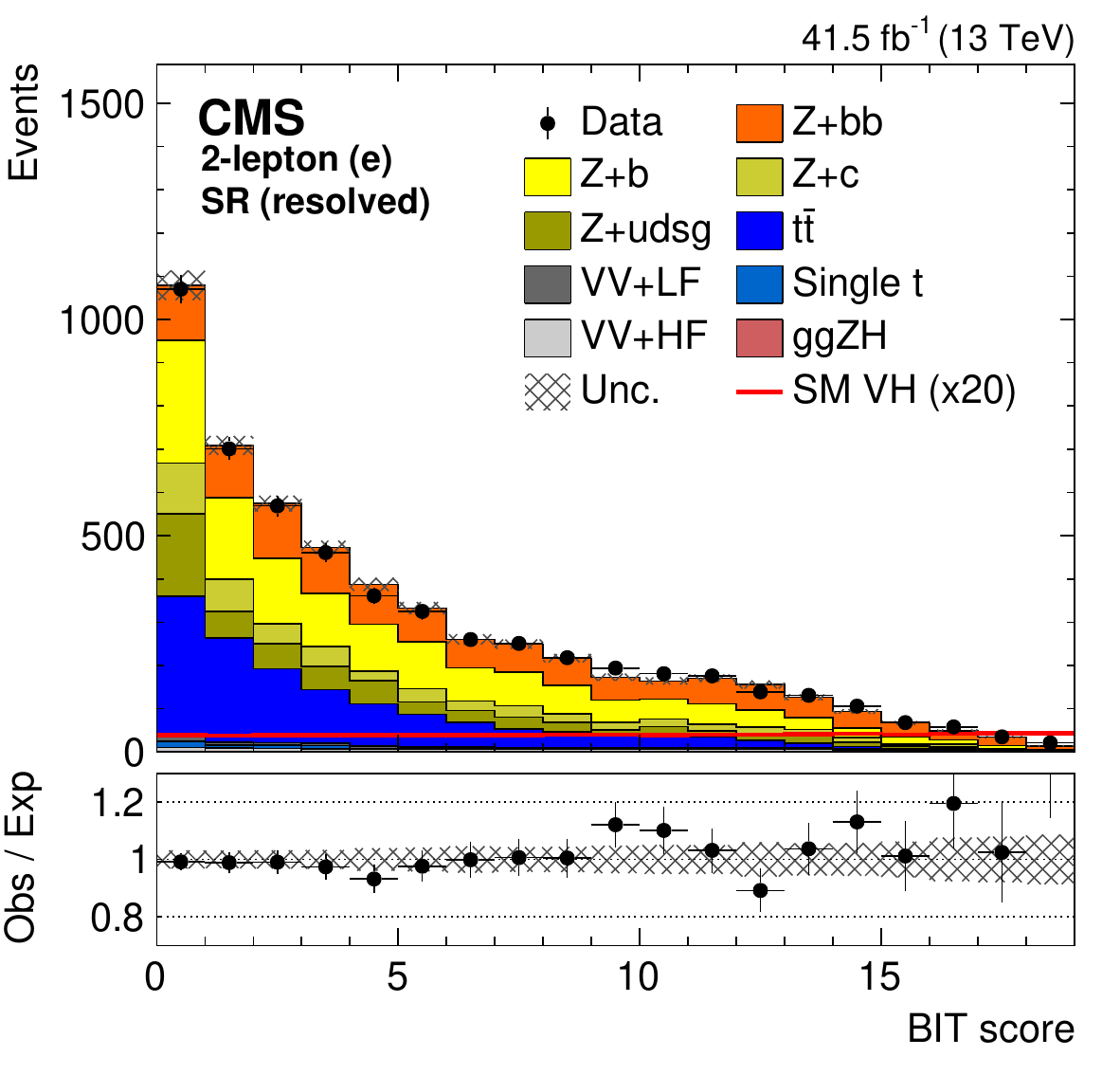}
\includegraphics[width=0.425\textwidth]{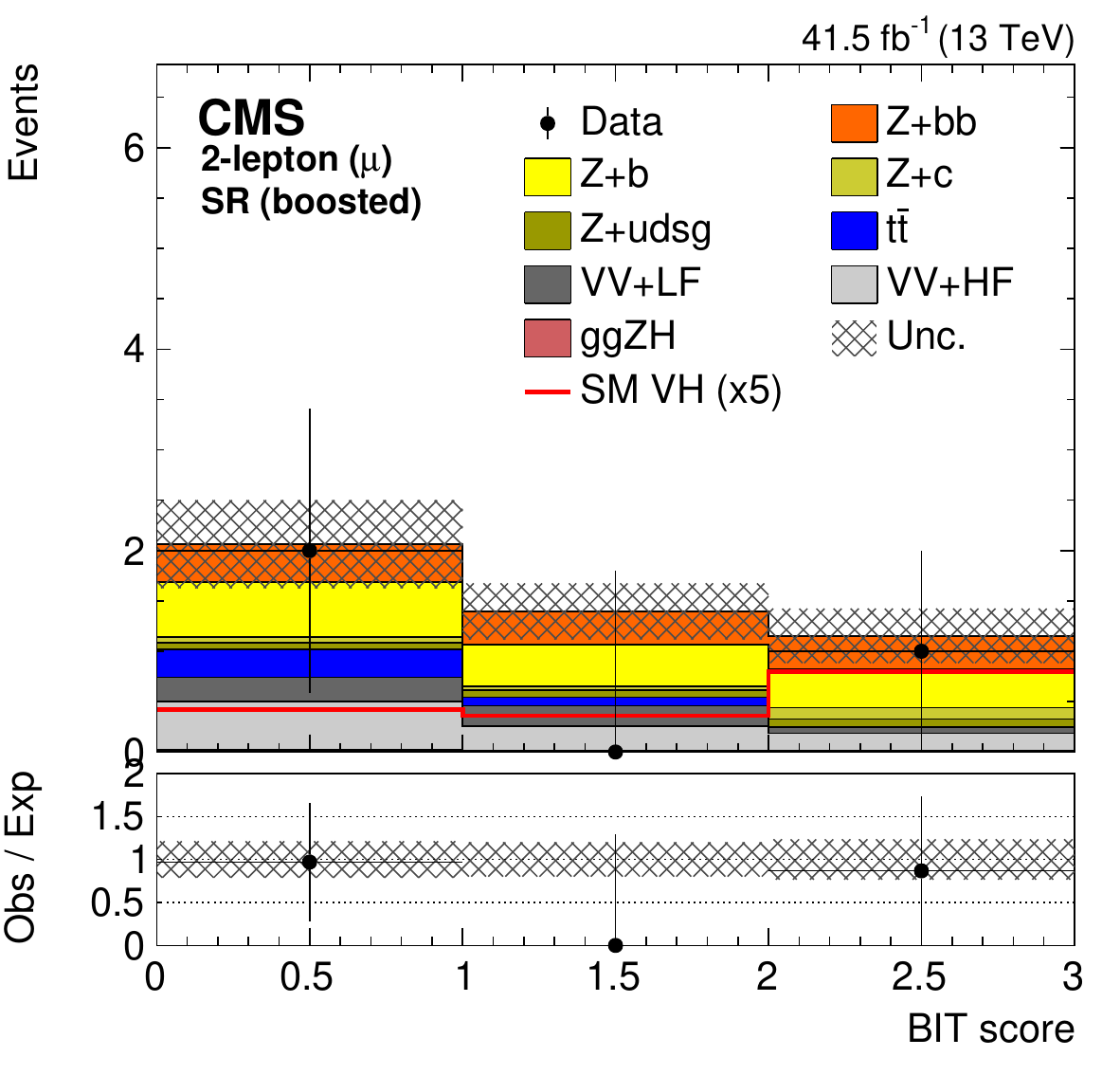}
\includegraphics[width=0.425\textwidth]{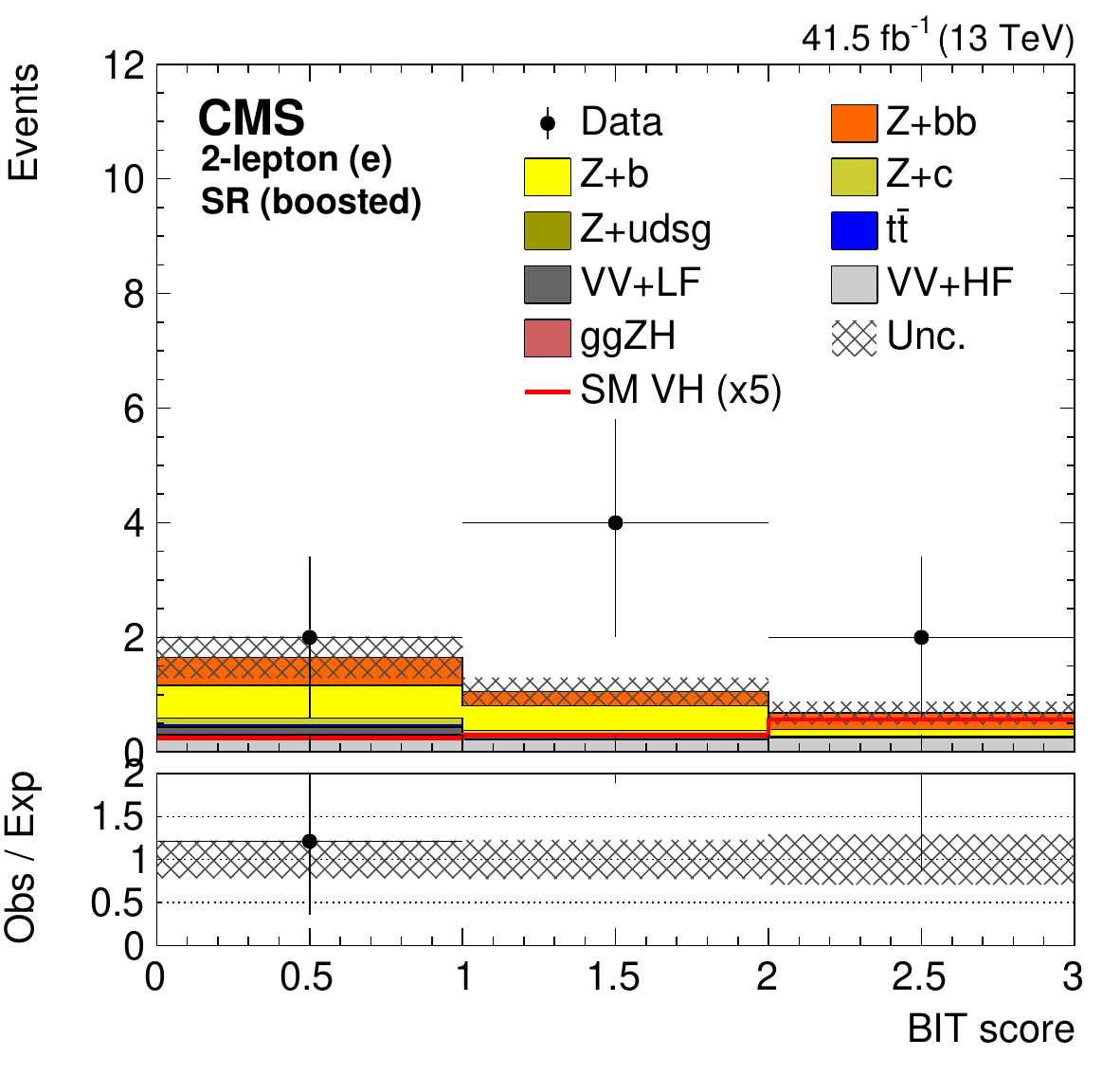}
\caption{The BIT templates obtained using a background-only fit to data in the {\twom} (left) and {\twoe} (right) final states in the SR for resolved (upper row) and boosted (lower row) categories considering the 2017 data set. The SM {\PV}{\PH} signal has been scaled by 20 and 5 for the resolved and boosted BIT templates in the upper and lower row, respectively, for better visualization. The lower panels show the ratio of the data to the background expectation after the background-only fit to the data.
}
\label{fig:Postfit_SR_2lepton}
\end{figure*}

\begin{figure*}[!htb]
\centering
\includegraphics[width=0.425\textwidth]{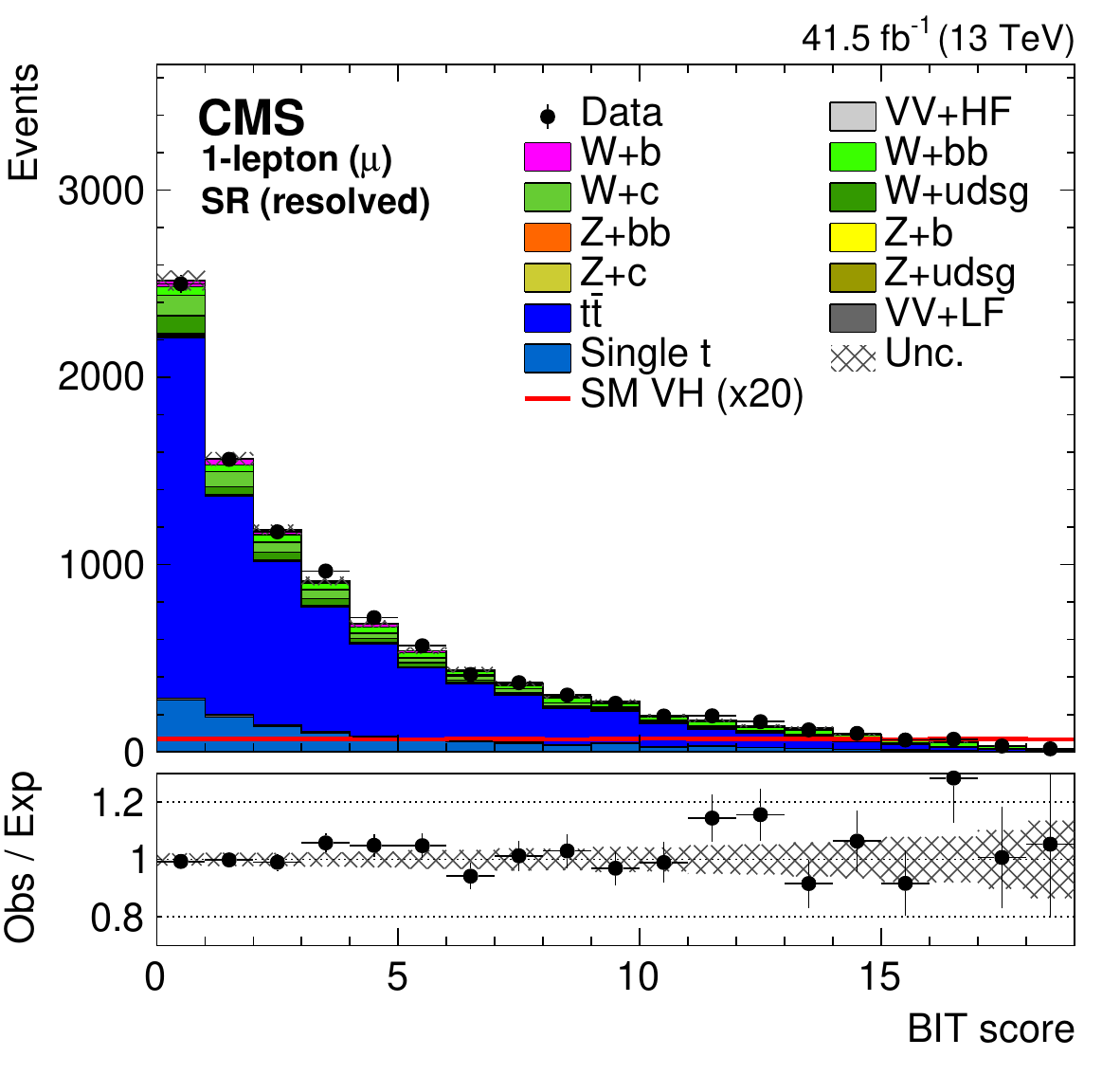}
\includegraphics[width=0.425\textwidth]{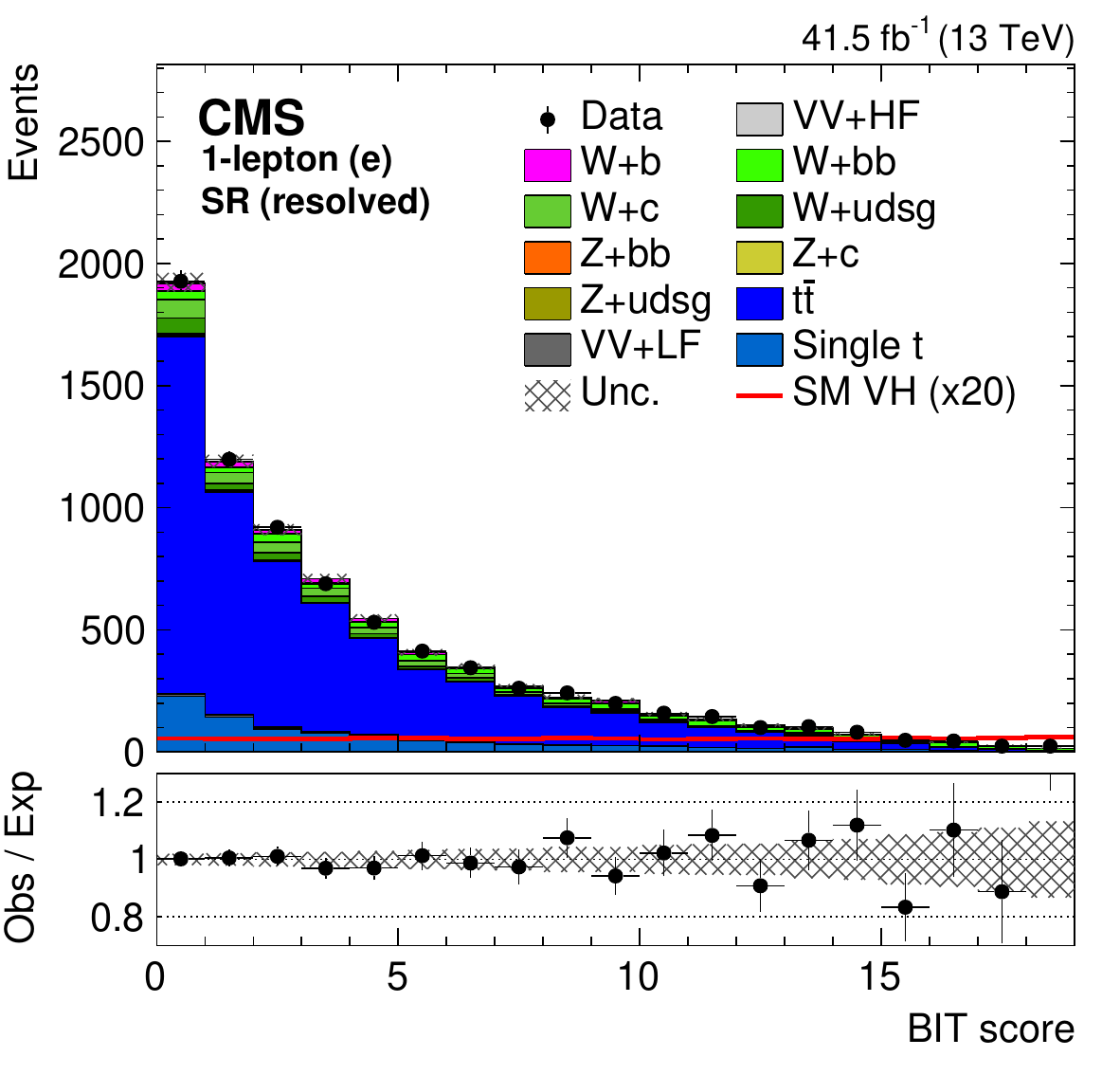}
\includegraphics[width=0.425\textwidth]{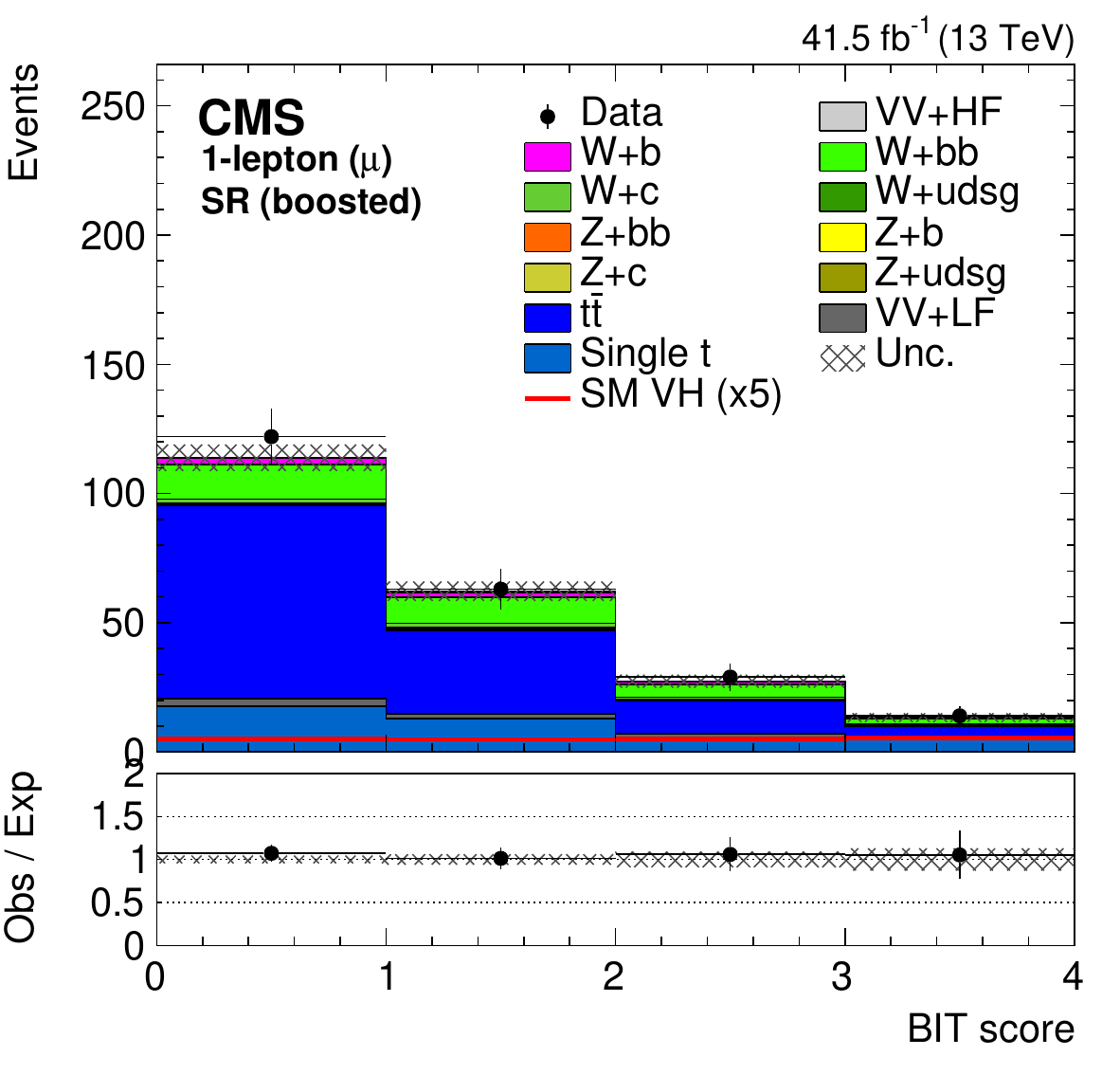}
\includegraphics[width=0.425\textwidth]{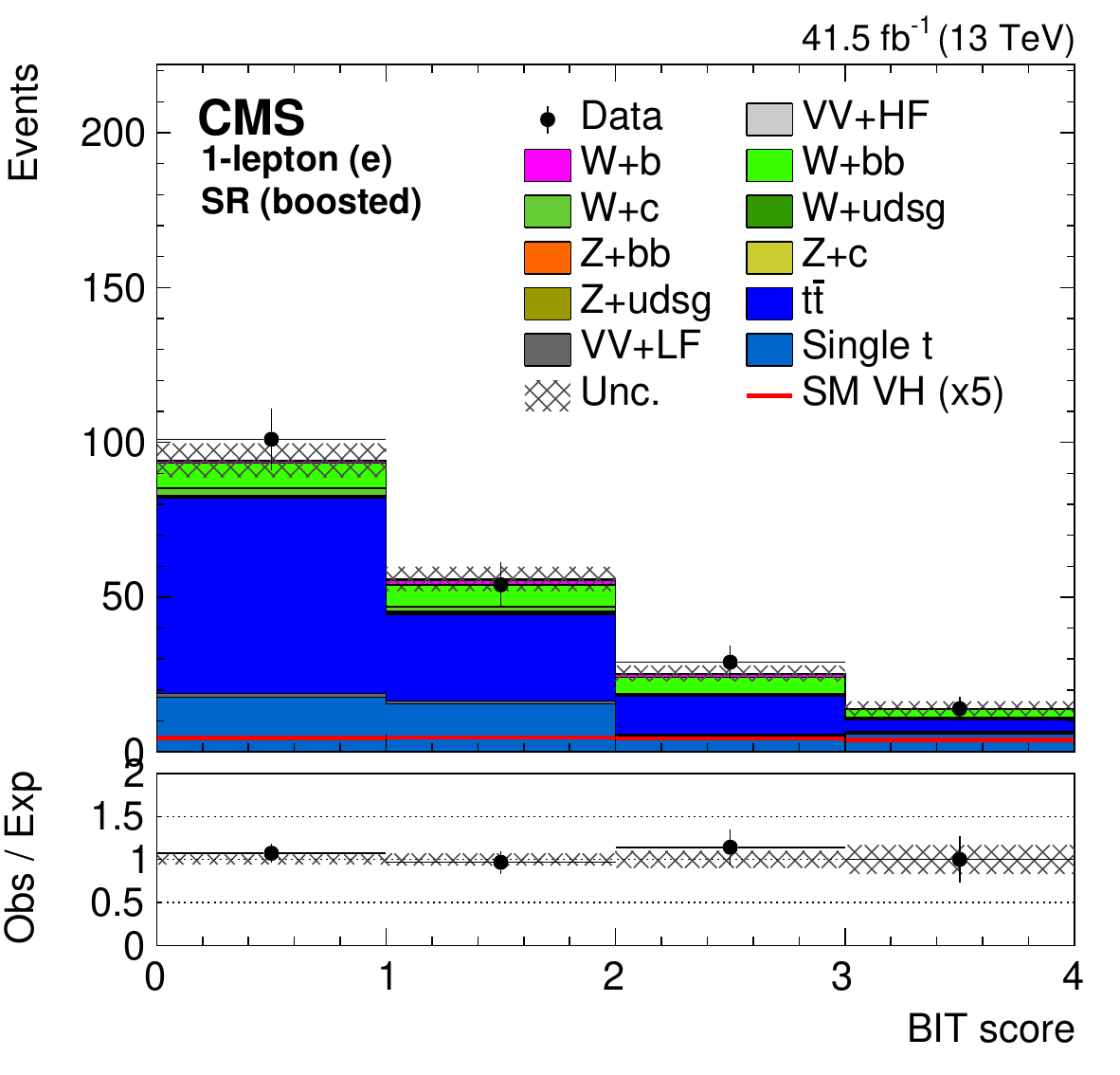}
\caption{The BIT templates obtained using a background-only fit to data in the {\onem} (left) and {\onee} (right) final states in the SR for resolved (upper row) and boosted (lower row) categories considering the 2017 data set. The SM {\PV}{\PH} signal has been scaled by 20 and 5 for the resolved and boosted BIT templates in the upper and lower row, respectively, for better visualization. The lower panels show the ratio of the data to the background expectation after the background-only fit to the data.
}
\label{fig:Postfit_SR_1lepton}
\end{figure*}

\begin{figure*}[!htb]
\centering
\includegraphics[width=0.425\textwidth]{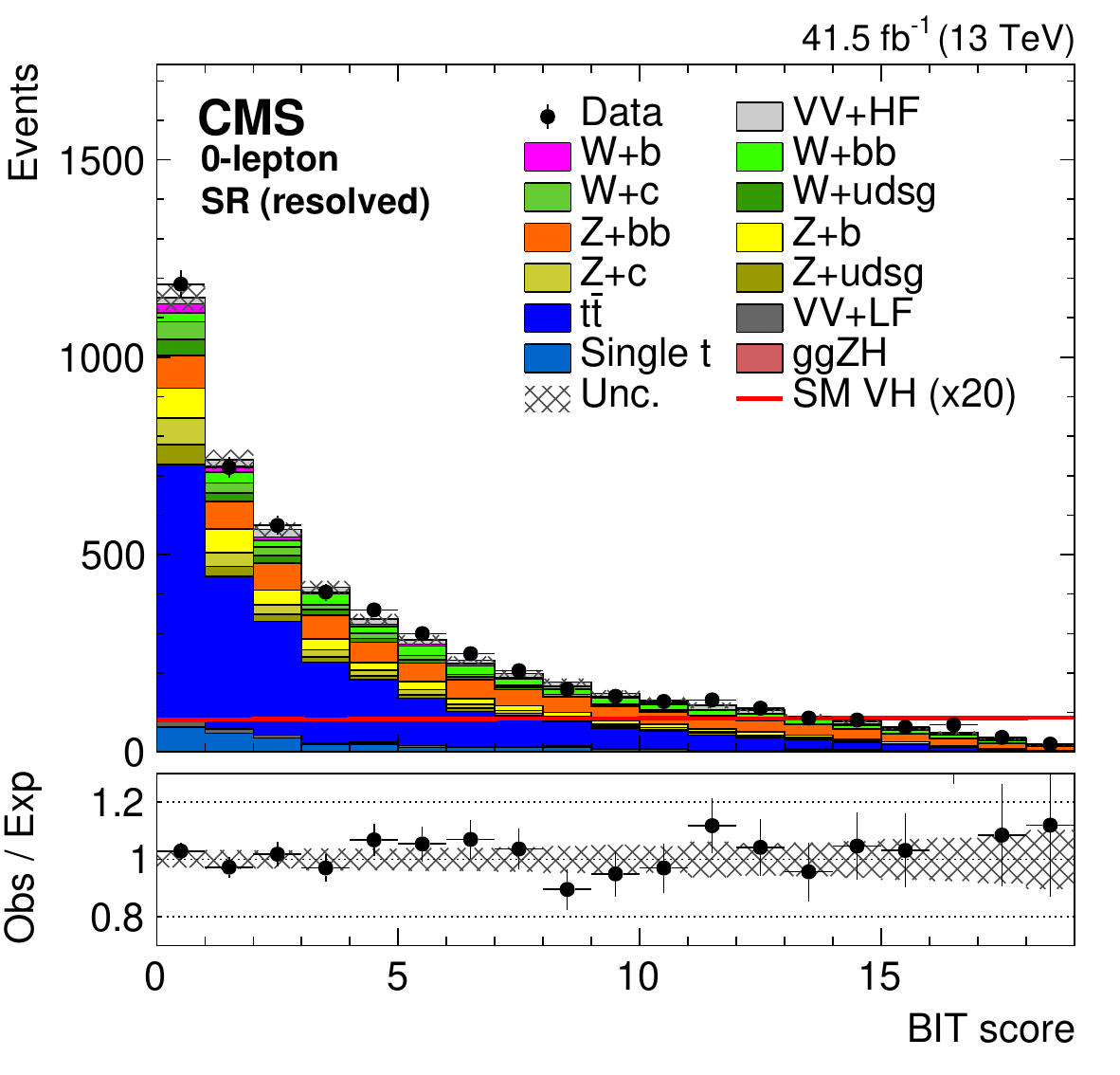}
\includegraphics[width=0.425\textwidth]{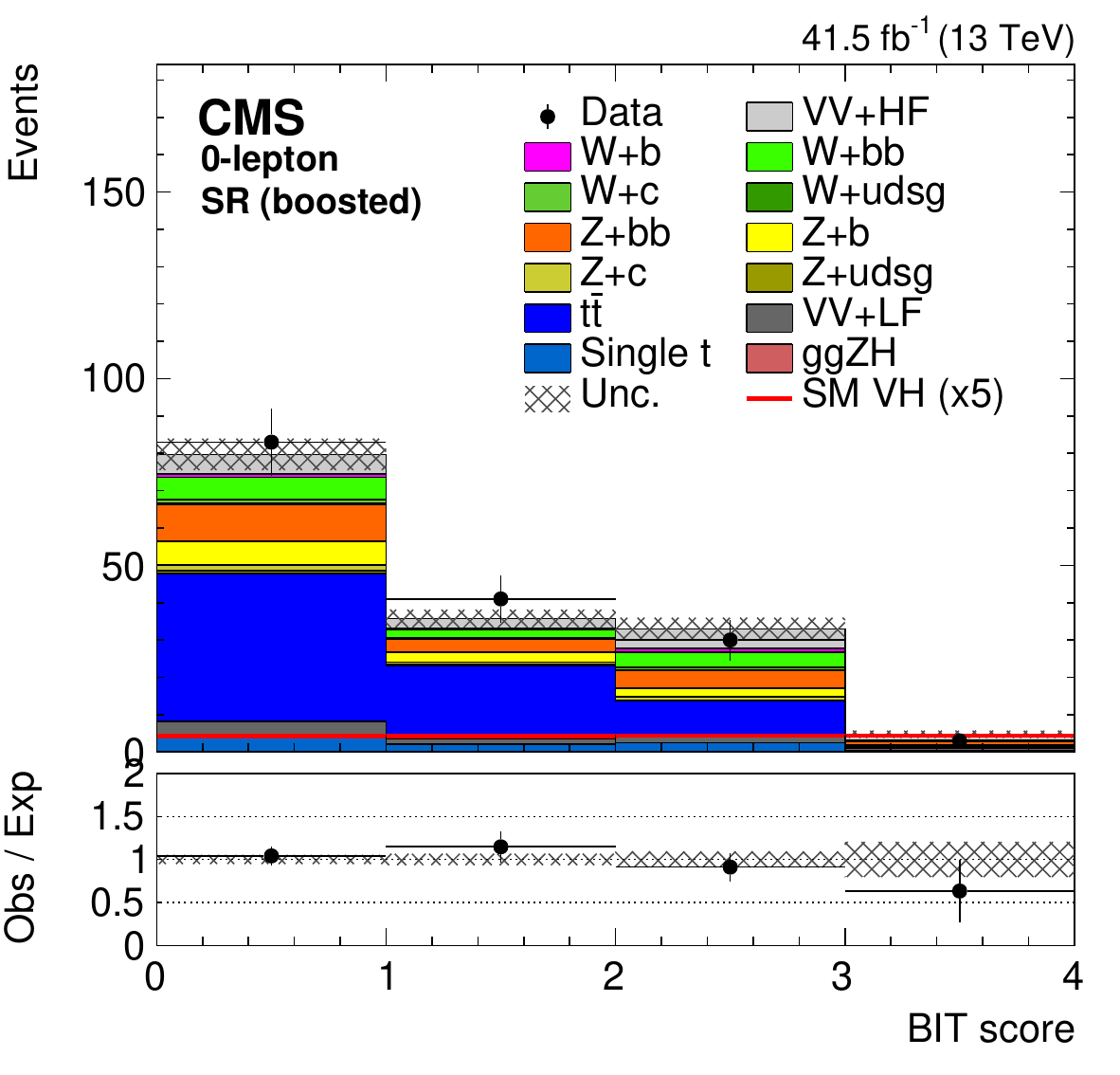}
\caption{The BIT templates obtained using a background-only fit to data in the {\zerol} final state in the SR for resolved (left) and boosted (right) categories considering the 2017 data set. The SM {\PV}{\PH} signal has been scaled by 20 and 5 for the resolved and boosted BIT templates in the upper and lower row, respectively, for better visualization. The lower panels show the ratio of the data to the background expectation after the background-only fit to the data.
}
\label{fig:Postfit_SR_0lepton}
\end{figure*}

Distributions of BIT scores in the SR are shown in Figs.~\ref{fig:Postfit_SR_2lepton},~\ref{fig:Postfit_SR_1lepton}, and~\ref{fig:Postfit_SR_0lepton}  for the resolved and boosted categories of $\twol$, $\onel$, and $\zerol$ final states, respectively, after the background-only fit to data is performed. For the boosted category in Fig.~\ref{fig:Postfit_SR_2lepton}, a smaller number of bins is used compared to the template shown in Fig.~\ref{fig:BIT_templates_2lep} in order to reduce the bin-by-bin uncertainty. 


In the asymptotic regime, the intervals in one dimension, where the test statistic falls below 1 and 4, can be interpreted as 68\% and 95\% confidence interval (CIs), respectively. This regime applies to results obtained under the model with a linear SMEFT parametrization, and the corresponding intervals have the correct coverage. It does not apply to the results obtained with the quadratic model, because the quadratic dependence of the yields in analysis bins on the Wilson coefficients violate the regularity conditions for asymptotic convergence~\cite{Bernlochner:2022oiw} as set out by Wilks theorem~\cite{Wilks}. The likelihood ratio intervals for the quadratic model may thus undercover or overcover. However, for consistency, intervals for both models are reported using the same definition.

\begin{figure*}[!htb]
\centering
\includegraphics[width=0.875\textwidth]{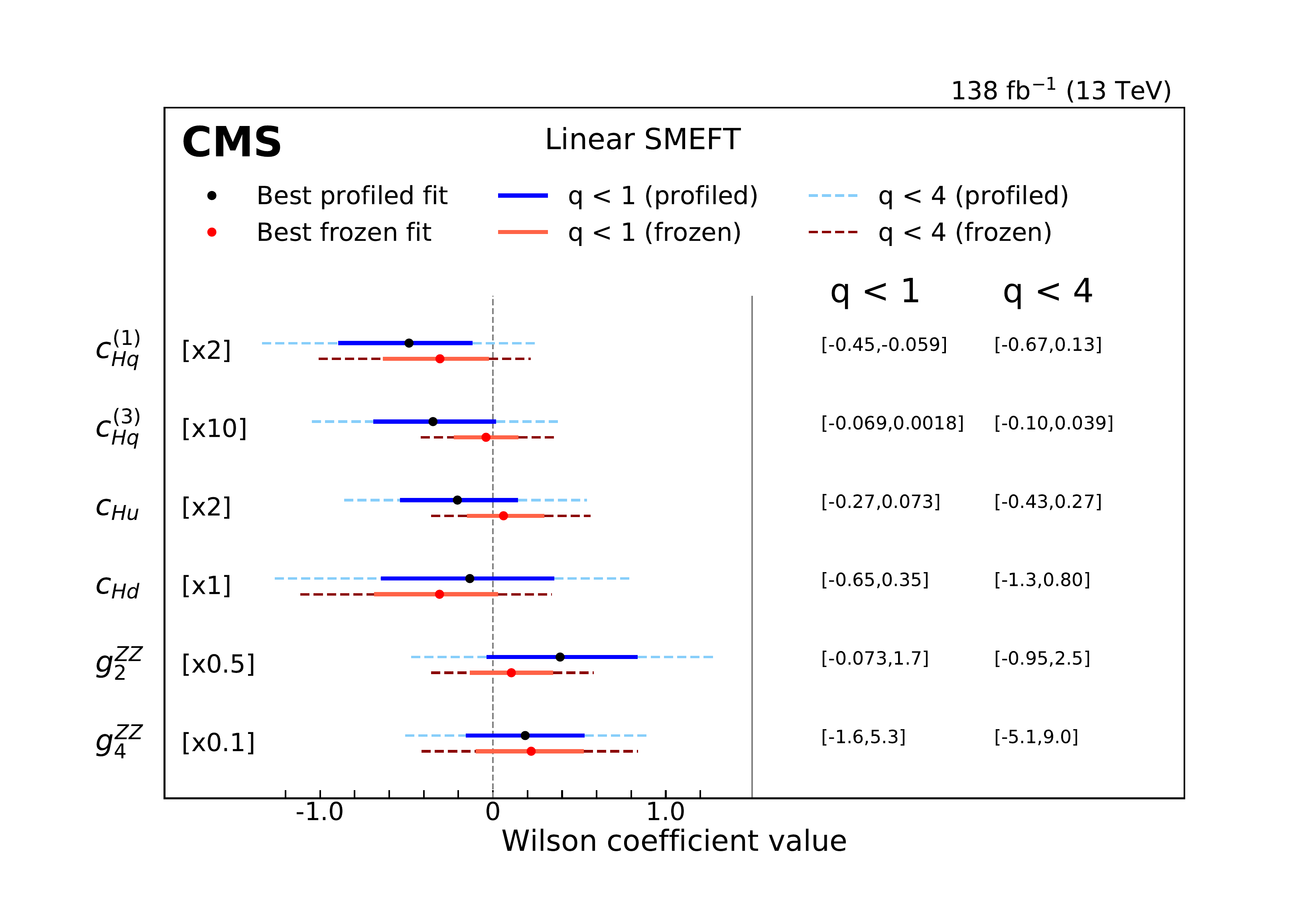}
\includegraphics[width=0.875\textwidth]{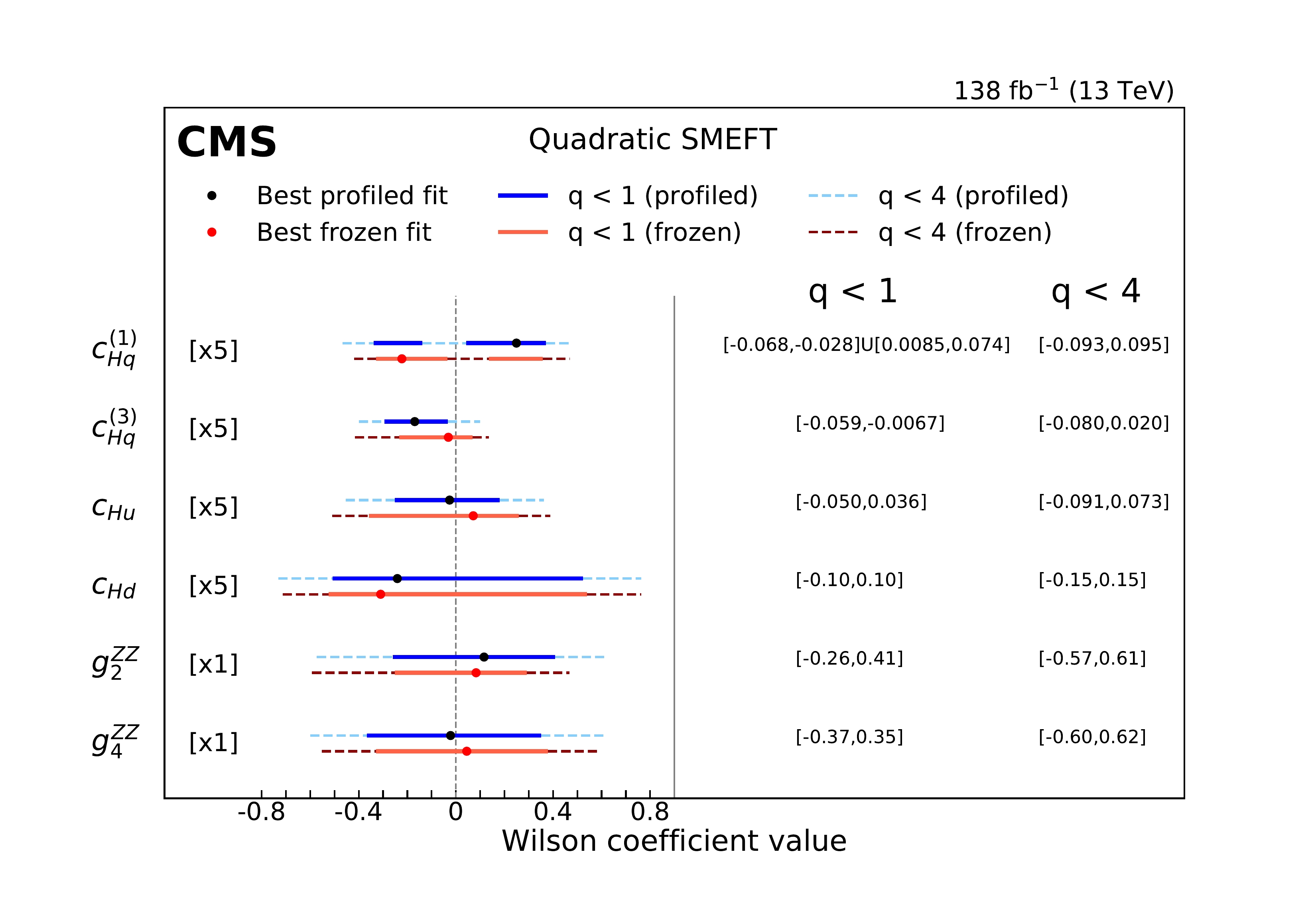}
\caption{Summary of results in terms of best fit value of the Wilson coefficients and the intervals where the test statistic is below 1 and 4, with up to the linear (upper row) and quadratic (lower row) terms in the SMEFT parameterization.  
These results are obtained either by allowing all Wilson coefficients to float freely at every point of the scan (profiled fit), or by keeping all other Wilson coefficients to their SM values, \ie, 0, except for the one that is being considered in the scan (frozen fit). The multiplication factor applies to the sizes of intervals satisfying $q<1$ and $q<4$ but not to the values of the intervals on the right-hand side of the figure, which correspond to the profiled constraints in all cases.}
\label{fig:unbl_1D_summary}
\end{figure*}

One-dimensional likelihood scans for the six Wilson coefficients targeted in the analysis are performed by considering up to linear or quadratic terms due to dimension-six operators in the SMEFT expansion, Eq.~(\ref{eq:factorization}) or (\ref{eq:R_expanded_simple}). For each one-dimensional likelihood scan, all Wilson coefficients are allowed to float freely at every point of the scans to account for correlation across Wilson coefficients, referred to as a profiled scan. 
Results are also provided for the case where all other Wilson coefficients are set to their SM values, \ie, 0, except for the one in the scan, and this scenario is referred to as the frozen fit. In Fig.~\ref{fig:unbl_1D_summary}, the summary of intervals on the Wilson coefficients satisfying $q<1$ and $q<4$ after combining results from all final states and eras with other Wilson coefficients profiled or set to the SM values are presented. Results are shown separately for SMEFT expansions up to linear and quadratic terms. For the linear-only SMEFT expansion, $q=1$ and $q=4$ correspond to 68\% and 95\% CIs on Wilson coefficients, respectively. For all Wilson coefficients, the quadratic components dominate the SMEFT sensitivity, except for $\cHqt$, where the linear and quadratic terms have comparable sensitivity and therefore result in better constraints on the Wilson coefficient observed values.
 
After including quadratic terms in the SMEFT expansion, the constraints for $\cHqo$, corresponding to $q<1$, has two intervals. This is due to the fact that the likelihood function contains two minima after including the quadratic terms due to the interplay of the interference and pure new physics terms, while it has only one minimum when only the linear term in SMEFT expansion is considered. The $p$-value compatibility of the profiled scans with respect to the SM expectation is 73\% while considering only linear SMEFT effects, and 84\% when including SMEFT expansion up to quadratic terms. The $p$-value compatibility of the profiled scans, when including the SMEFT expansion up to quadratic terms, is subject to the same caveat on the coverage discussed earlier regarding the use of the asymptotic approximation for the quadratic model.

\begin{figure*}[!htb]
\centering
\includegraphics[width=0.875\textwidth]{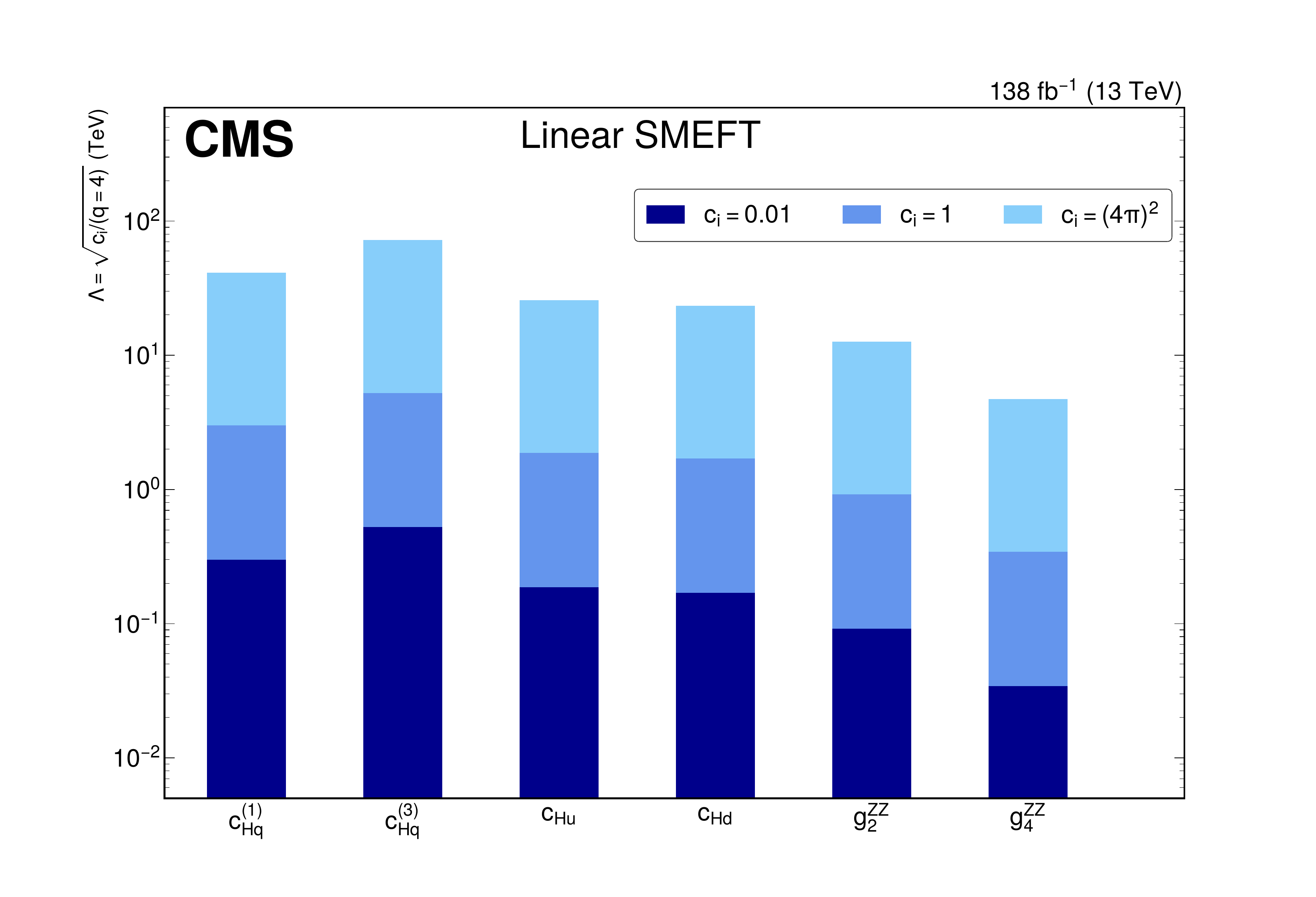}
\includegraphics[width=0.875\textwidth]{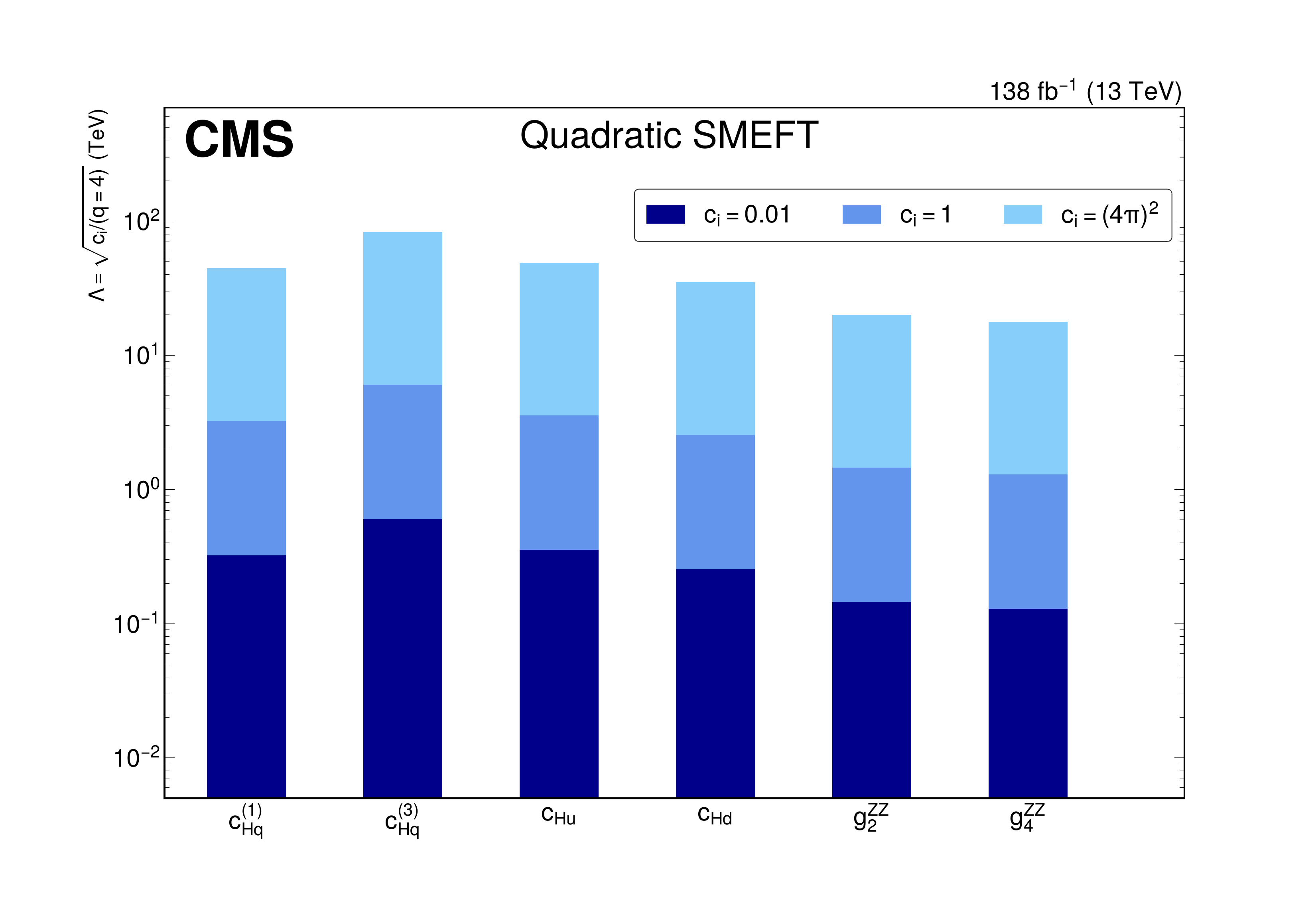}
\caption{Profiled limits on the energy scale $\Lambda$ for three different assumptions for each Wilson coefficient while fixing the other Wilson coefficients to their SM values with up to the linear (upper row) and quadratic (lower row) terms in SMEFT parameterization. The upper limits on the Wilson coefficients corresponding to $q=4$ is used for translating the constraints to $\Lambda$.}
\label{fig:unbl_1D_lambda}
\end{figure*}

Lower limits on the energy scale $\Lambda$ are extracted for different assumptions of the Wilson coefficient values while fixing the other Wilson coefficients to their SM expectations separately for linear and quadratic SMEFT parameterizations and shown in Fig.~\ref{fig:unbl_1D_lambda}. Three assumptions on the values of the Wilson coefficients in Fig.~\ref{fig:unbl_1D_lambda} are chosen to extract lower limits on the energy scale $\Lambda$ under several EFT realizations and those approximately correspond to weakly-coupled models (0.01), no assumptions on the coupling (1), and strong-coupling perturbativity limit (16$\pi^{2}$), respectively. 

The dominant uncertainty in the SMEFT signal extraction is of statistical nature. The main systematic sources impacting the measurement are related to the $\PV$+jets background modeling uncertainties and the limited size of the NLO $\PV$+jets simulated samples in analogy with Ref~\cite{CMS:2023vzh}.

\begin{figure*}[!htb]
\centering
\includegraphics[width=0.425\textwidth]{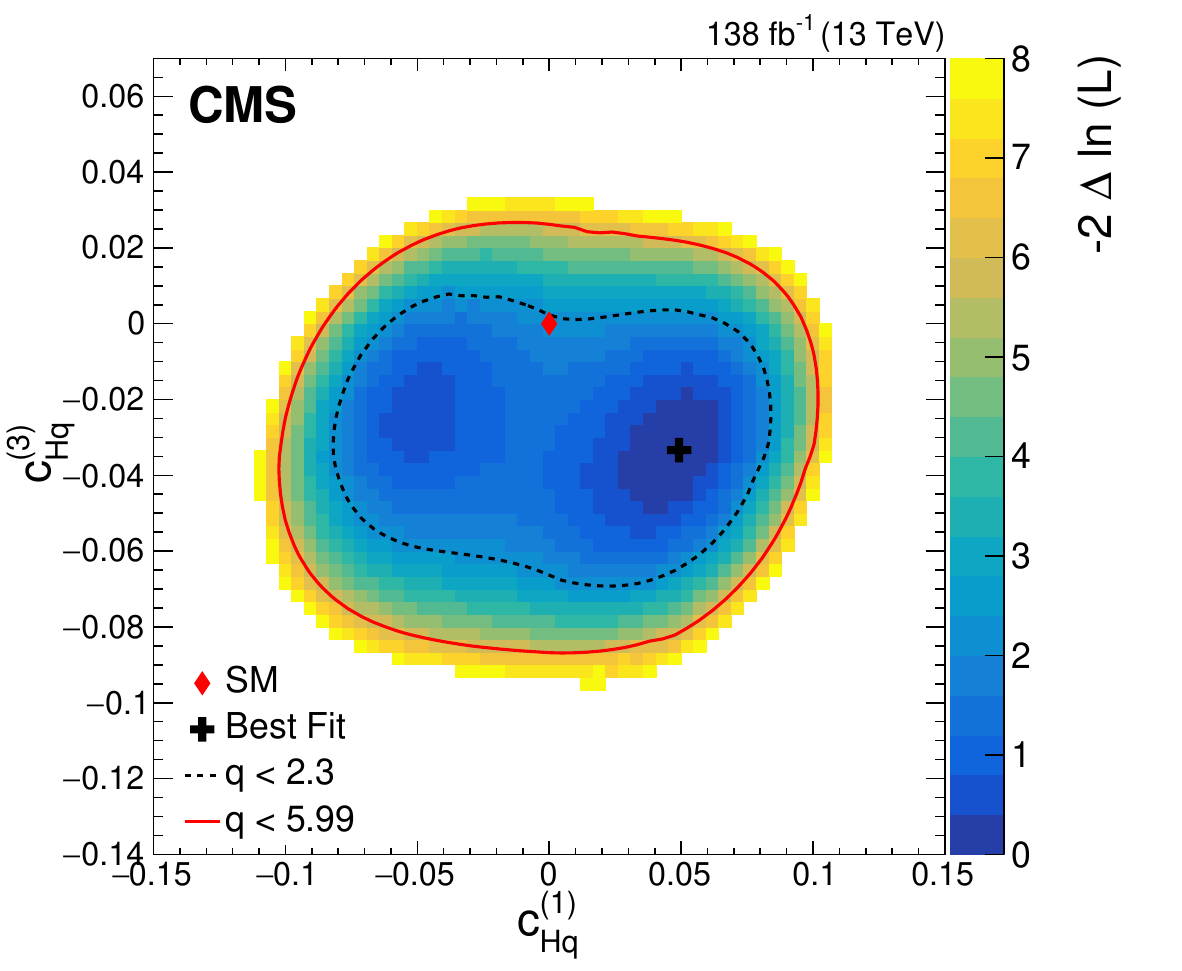}
\includegraphics[width=0.425\textwidth]{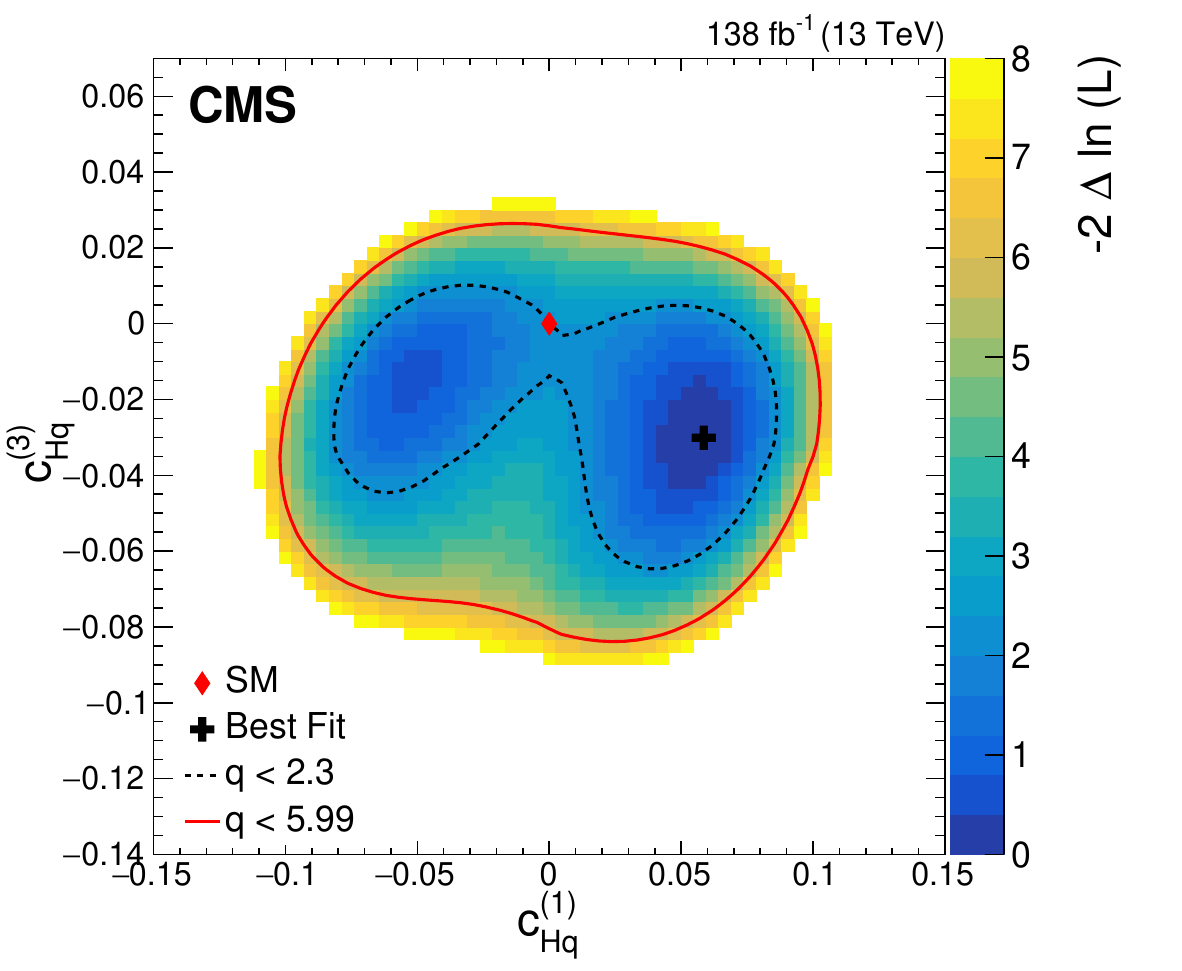}
\includegraphics[width=0.425\textwidth]{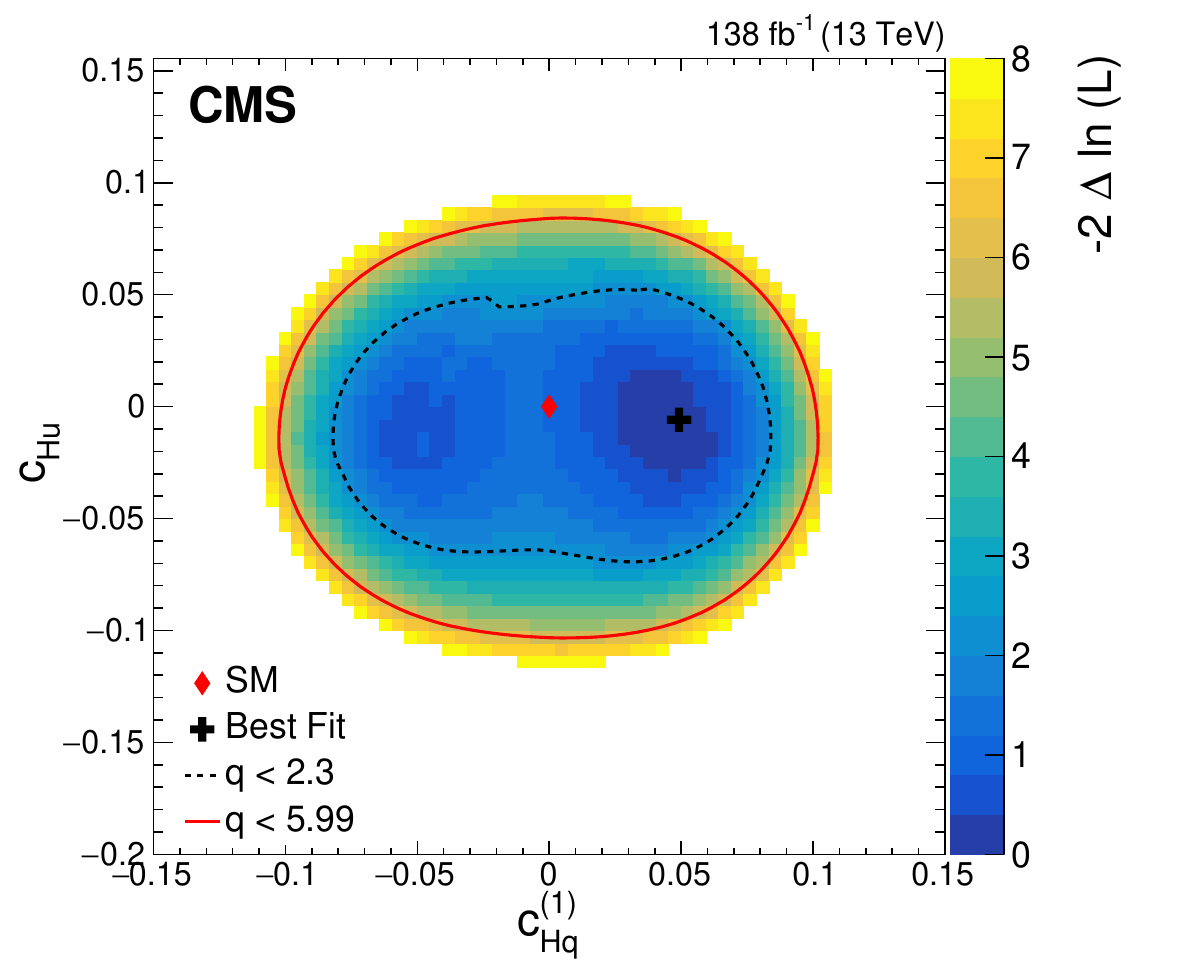}
\includegraphics[width=0.425\textwidth]{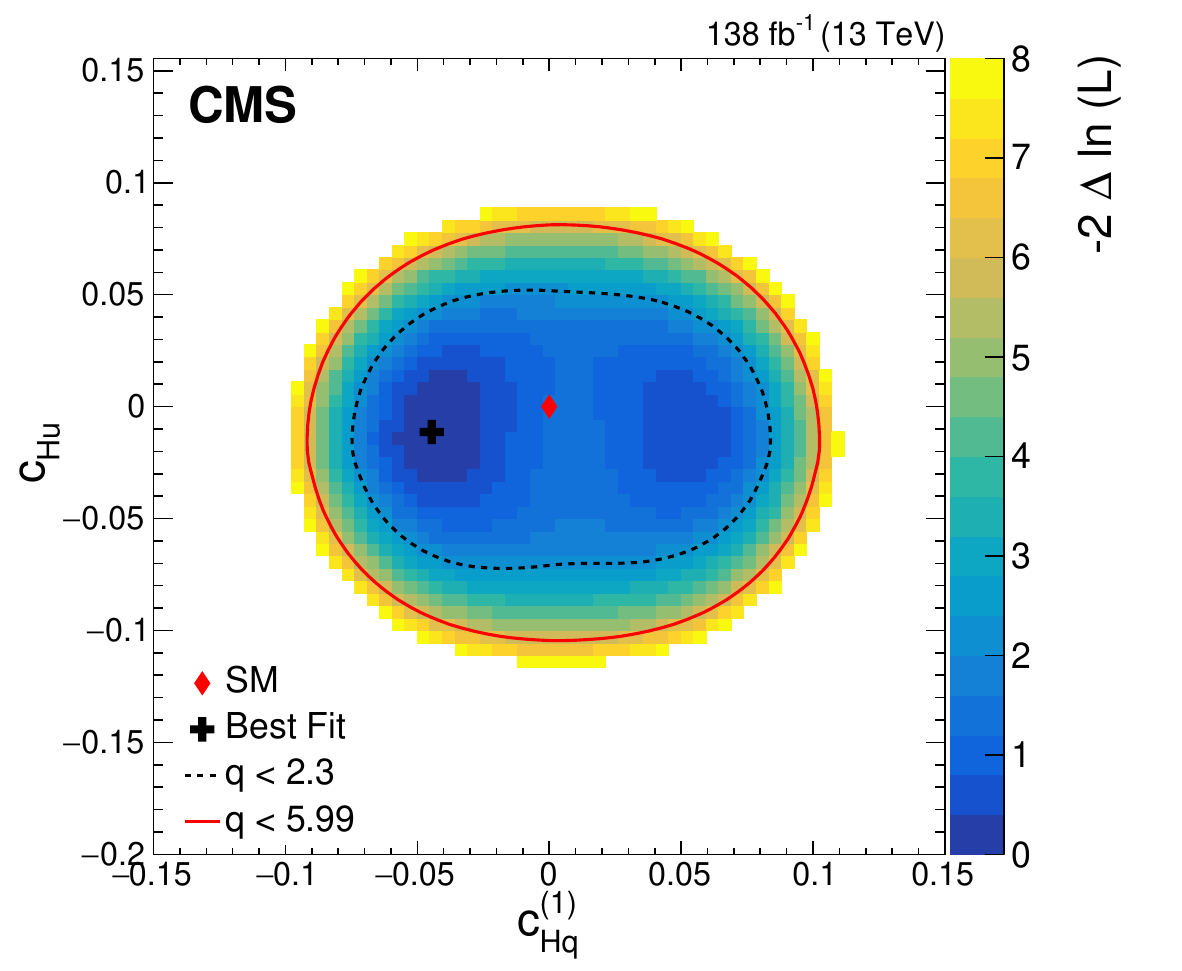}
\includegraphics[width=0.425\textwidth]{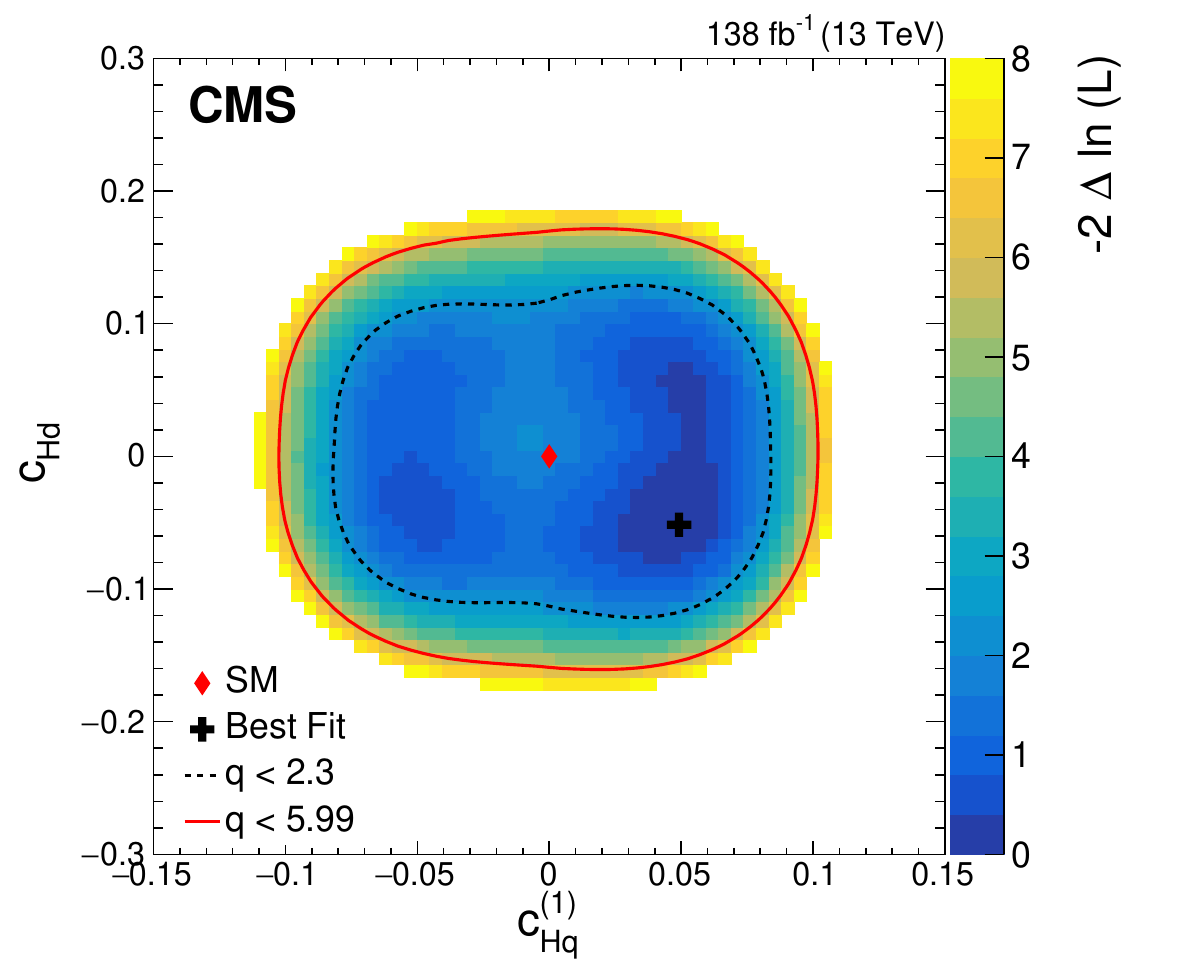}
\includegraphics[width=0.425\textwidth]{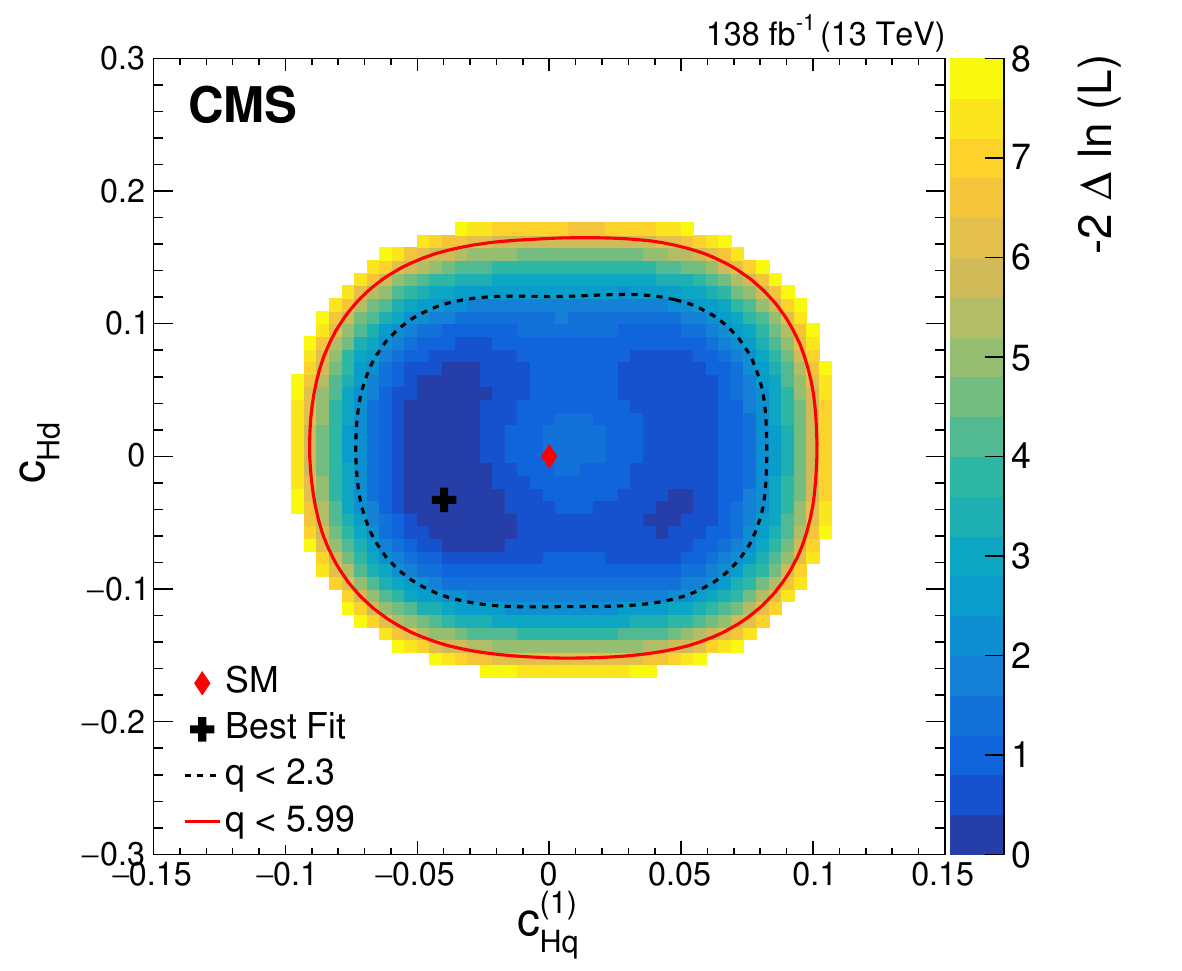}
\caption{
Observed two-dimensional likelihood scans for different pairs of Wilson coefficients: $\cHqo$ vs. $\cHqt$ (upper row), $\cHqo$ vs. $\cHu$ (middle row), $\cHqo$ vs. $\cHd$ (lower row) while allowing the other coefficients to float freely at each point of the scan (left) or fixed at their SM values (right) after combining results from all data-taking years and final states.
}
\label{fig:Scan_2D_a}
\end{figure*}

\begin{figure*}[!htb]
\centering
\includegraphics[width=0.425\textwidth]{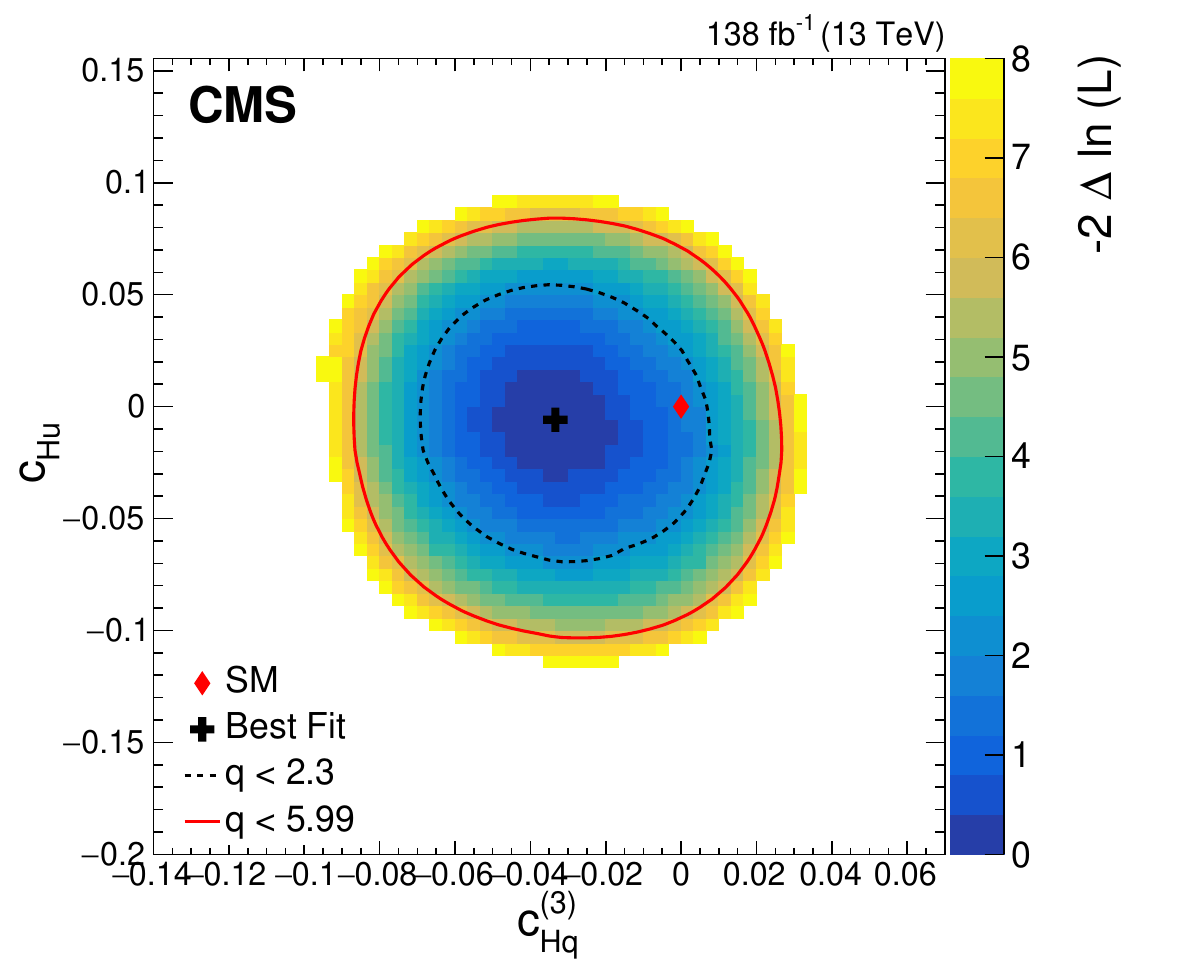}
\includegraphics[width=0.425\textwidth]{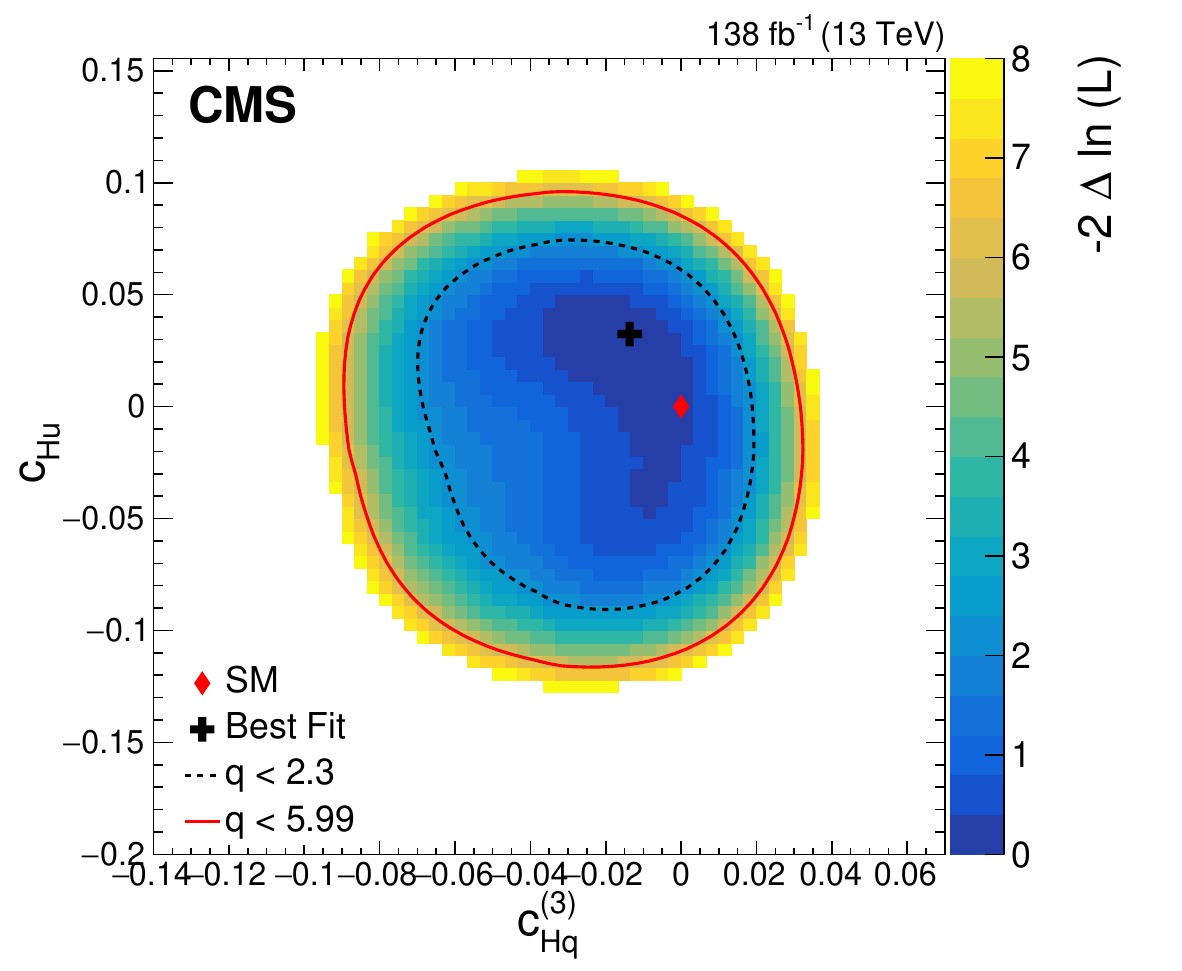}
\includegraphics[width=0.425\textwidth]{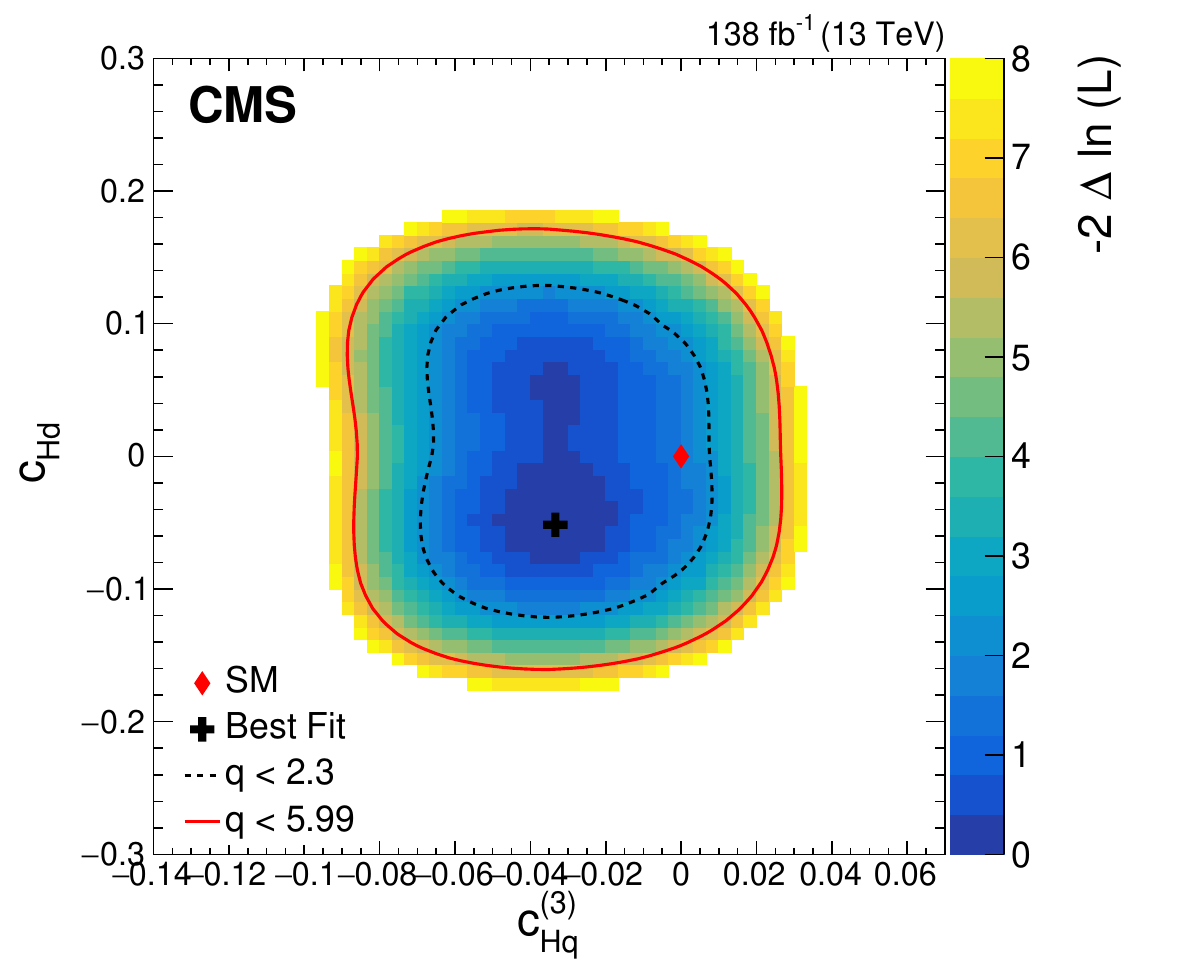}
\includegraphics[width=0.425\textwidth]{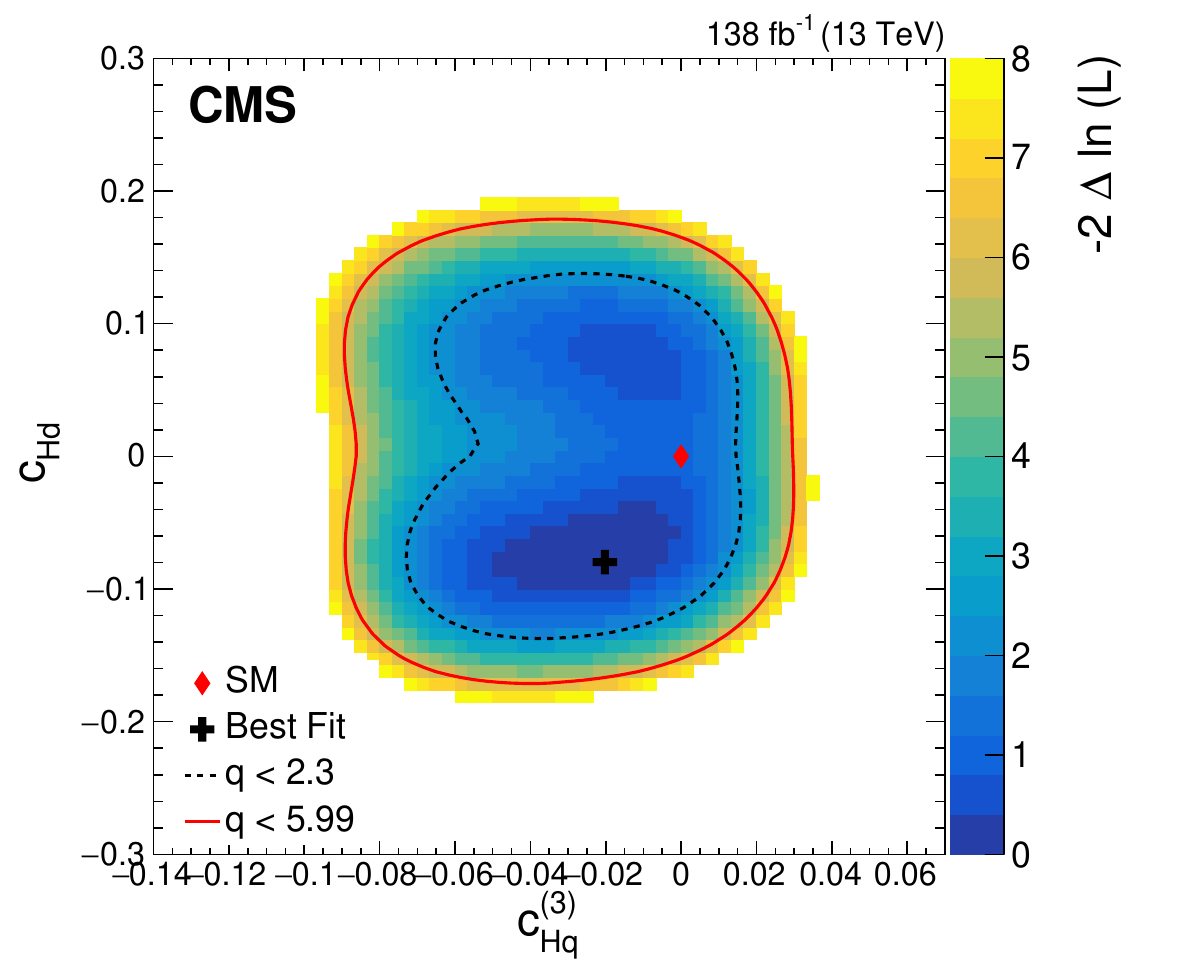}
\includegraphics[width=0.425\textwidth]{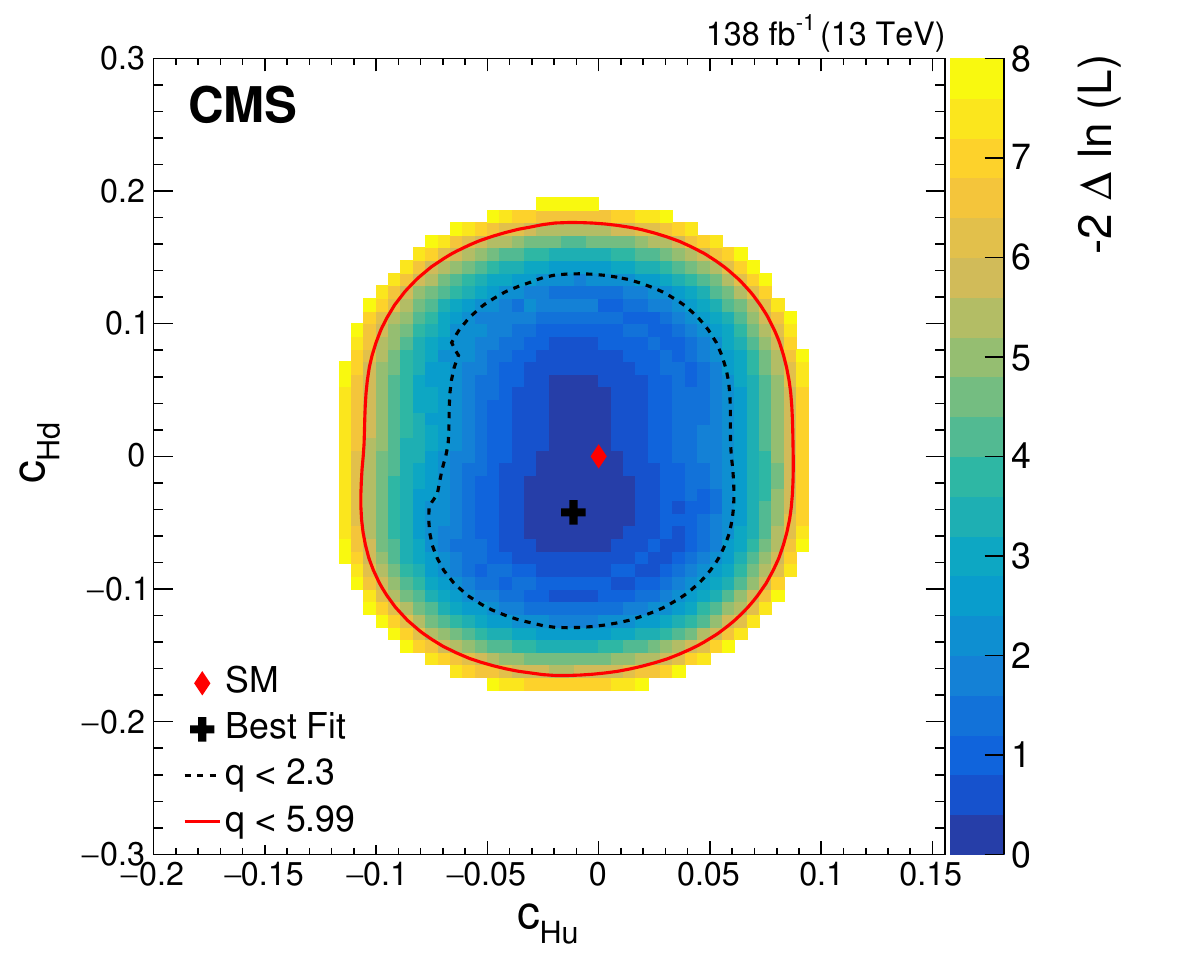}
\includegraphics[width=0.425\textwidth]{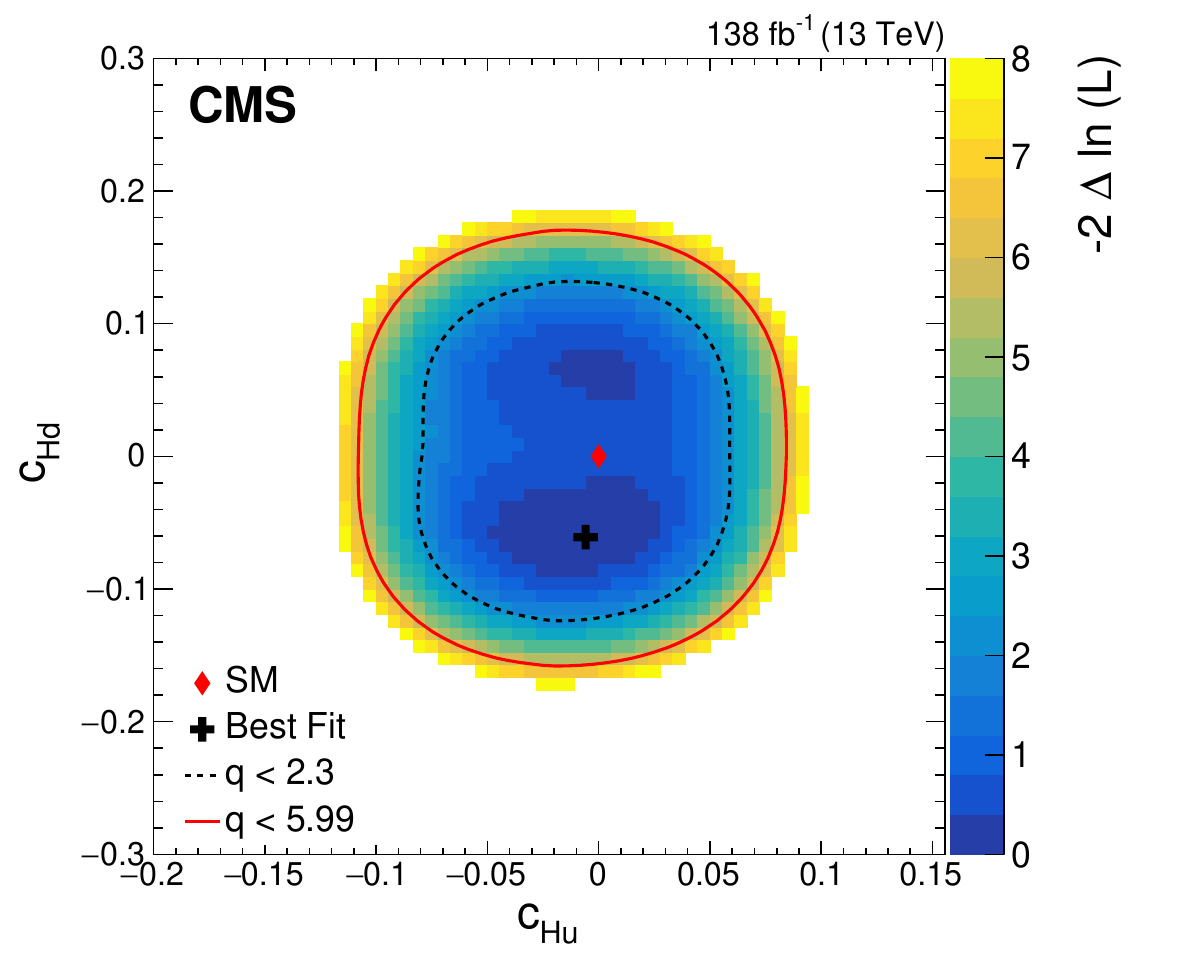}
\caption{
Observed two-dimensional likelihood scans for different pairs of Wilson coefficients: $\cHqt$ vs. $\cHu$ (upper row), $\cHqt$ vs. $\cHd$ (middle row), $\cHu$ vs. $\cHd$ (lower row) while allowing the other coefficients to float freely at each point of the scan (left) or fixed at their SM values (right) after combining results from all data-taking years and final states.
}
\label{fig:Scan_2D_b}
\end{figure*}

\begin{figure*}[!htb]
\centering
\includegraphics[width=0.425\textwidth]{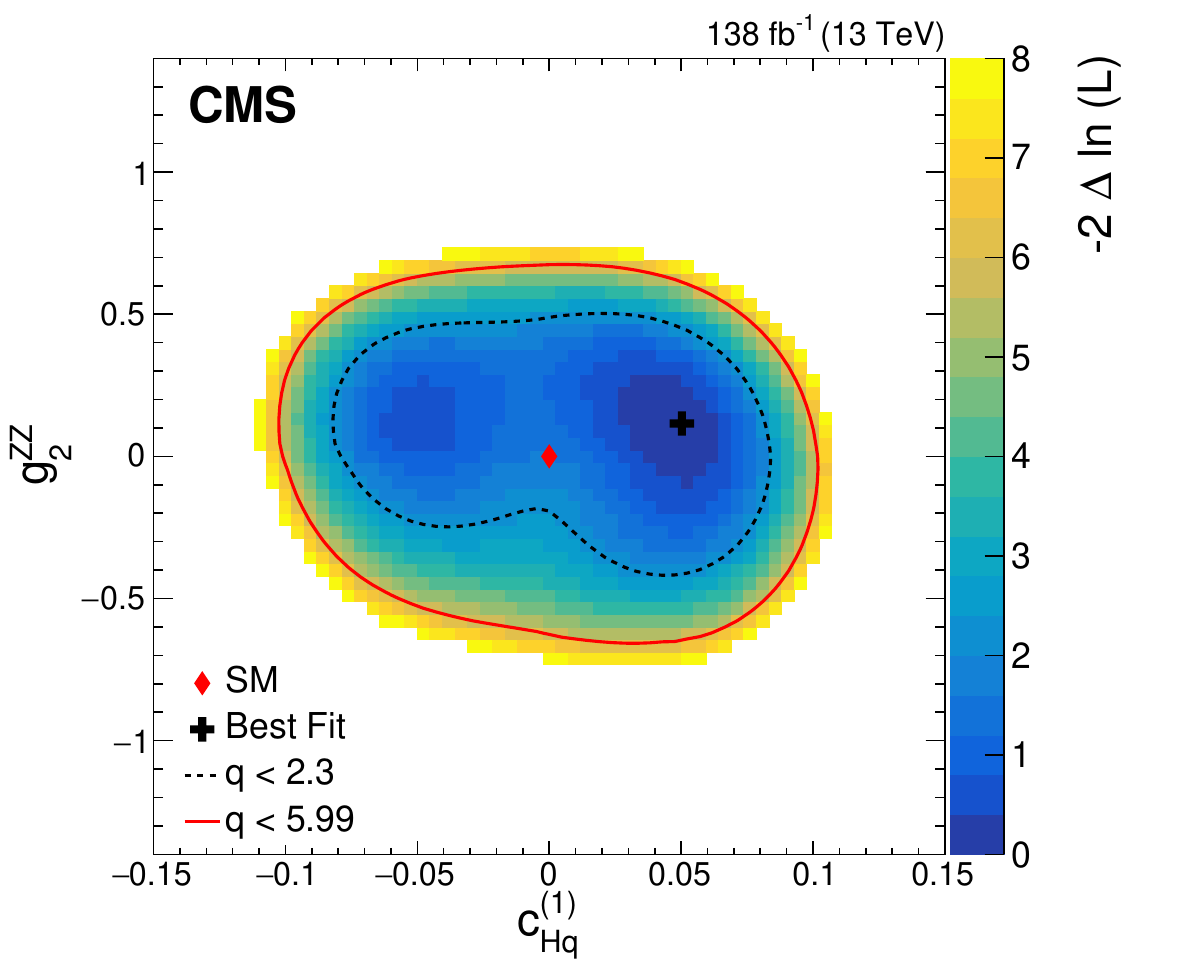}
\includegraphics[width=0.425\textwidth]{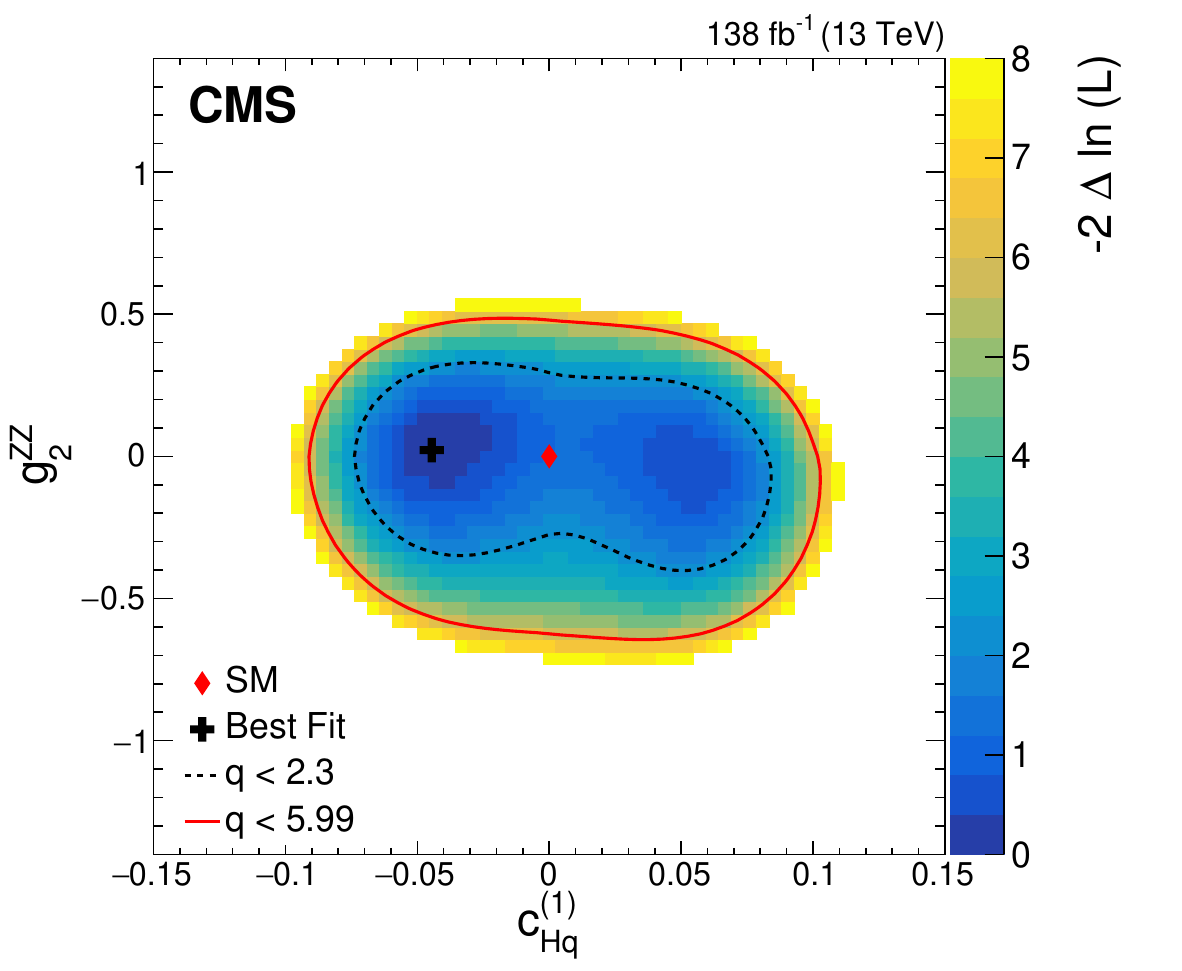}
\includegraphics[width=0.425\textwidth]{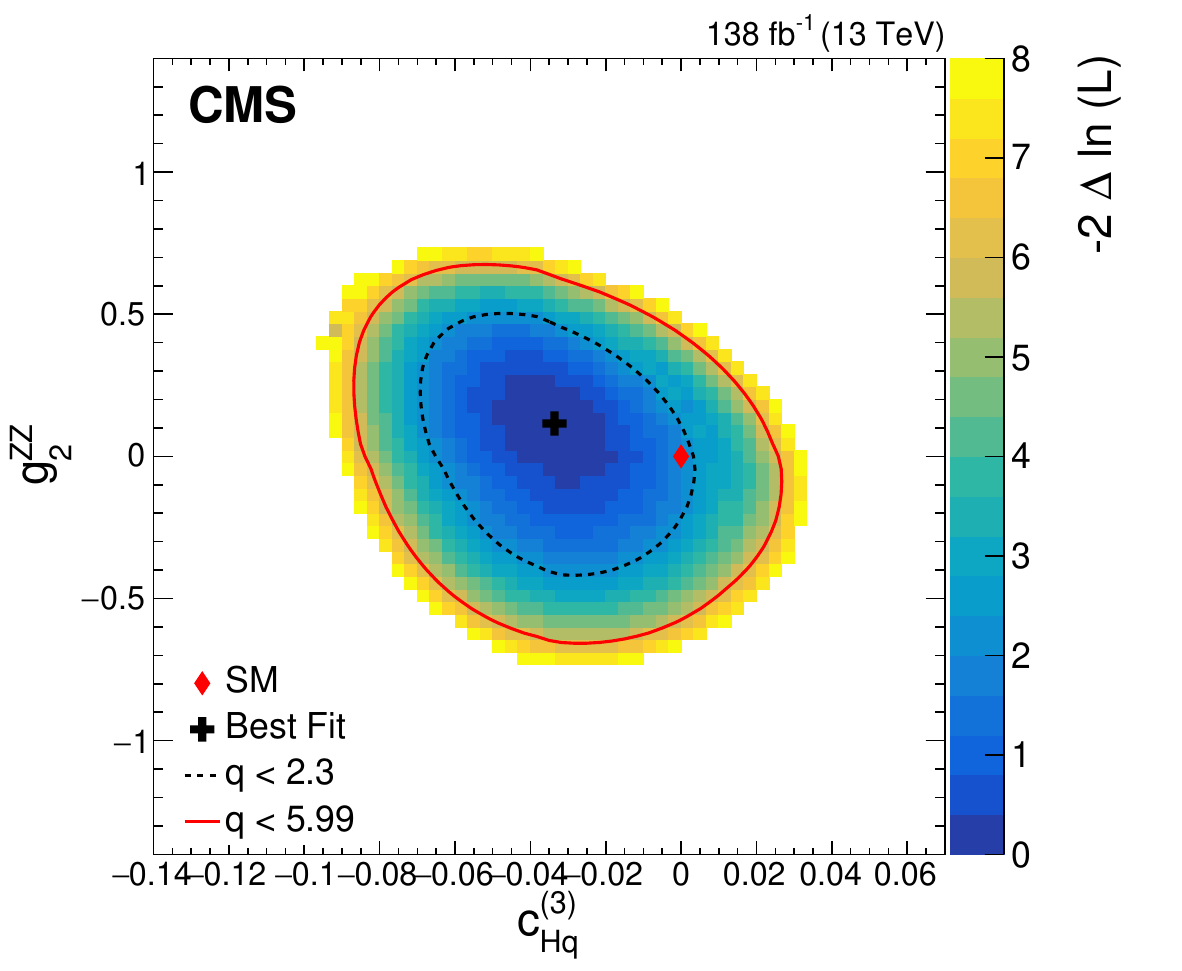}
\includegraphics[width=0.425\textwidth]{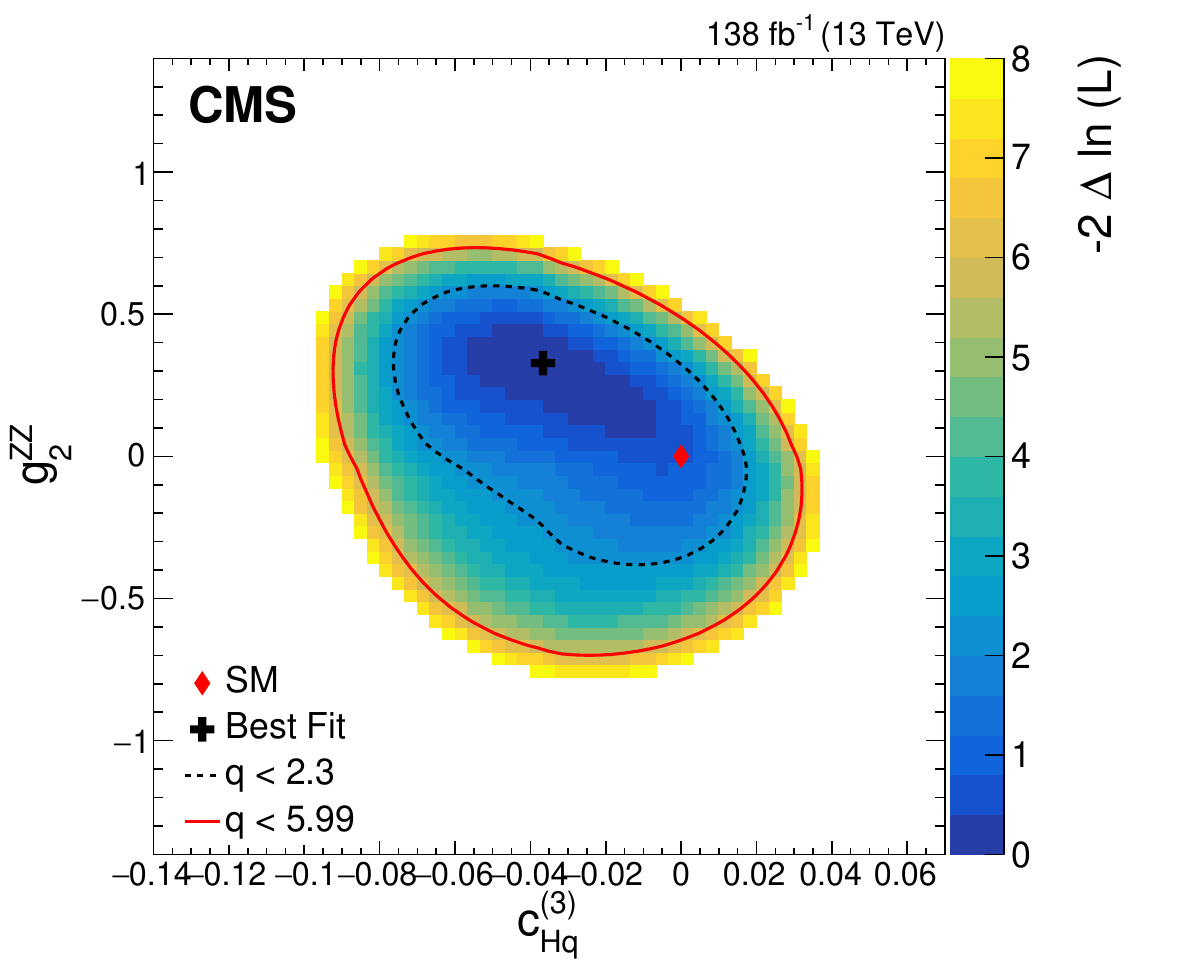}
\includegraphics[width=0.425\textwidth]{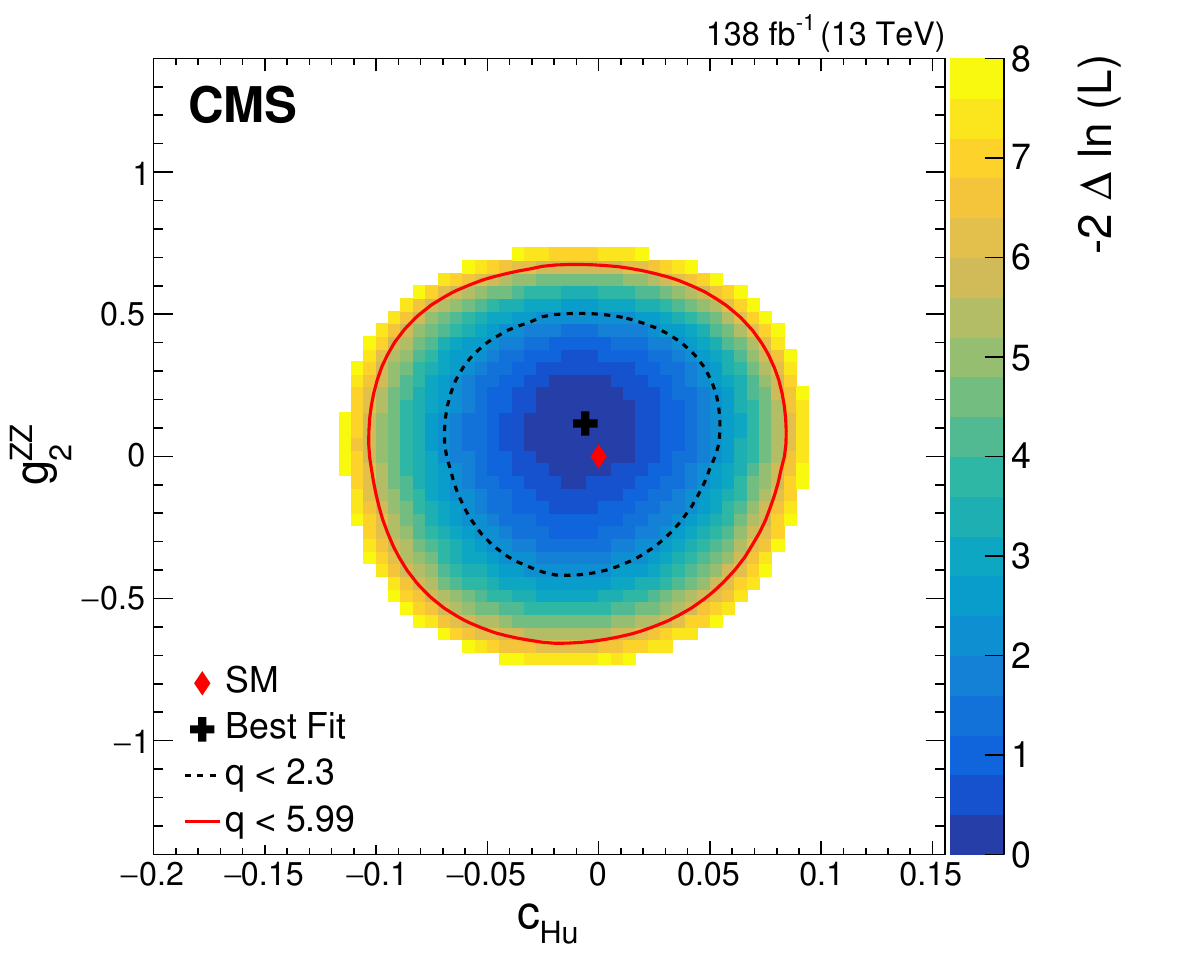}
\includegraphics[width=0.425\textwidth]{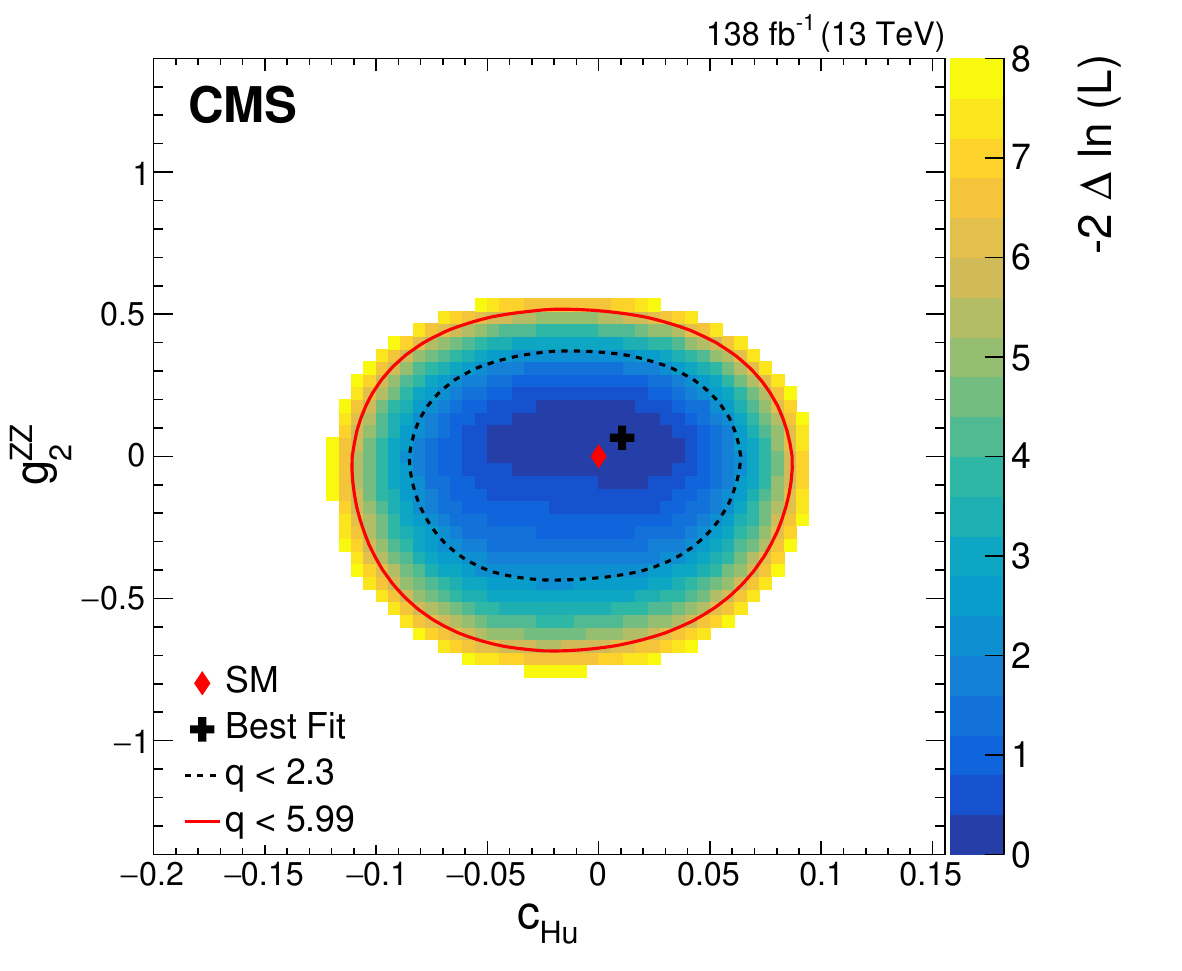}
\caption{
Observed two-dimensional likelihood scans for different pairs of Wilson coefficients: $\cHqo$ vs. $\gtZZ$ (upper row), $\cHqt$ vs. $\gtZZ$ (middle row), $\cHu$ vs. $\gtZZ$ (lower row) while allowing the other coefficients to float freely at each point of the scan (left) or fixed at their SM values (right) after combining results from all data-taking years and final states.
}
\label{fig:Scan_2D_c}
\end{figure*}

\begin{figure*}[!htb]
\centering
\includegraphics[width=0.425\textwidth]{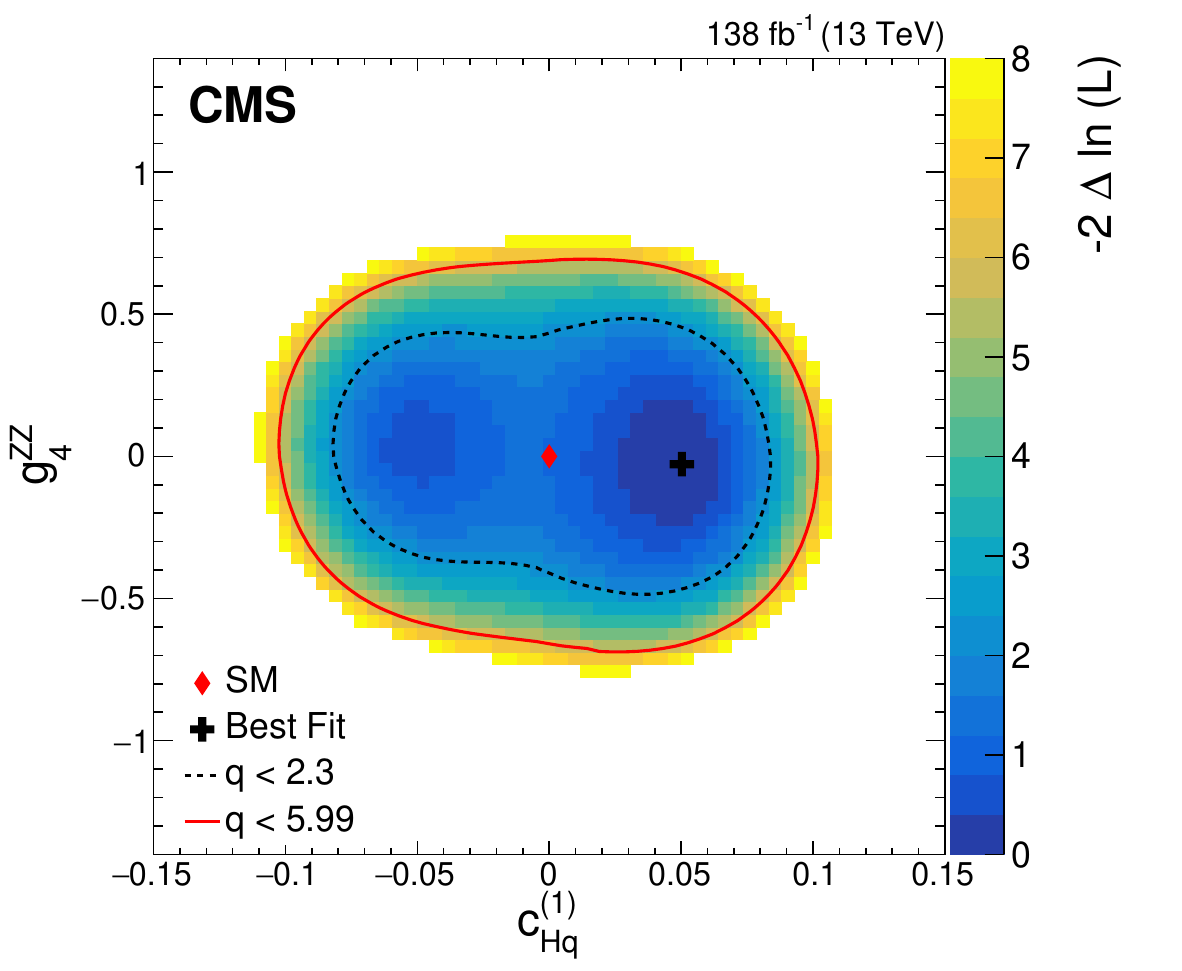}
\includegraphics[width=0.425\textwidth]{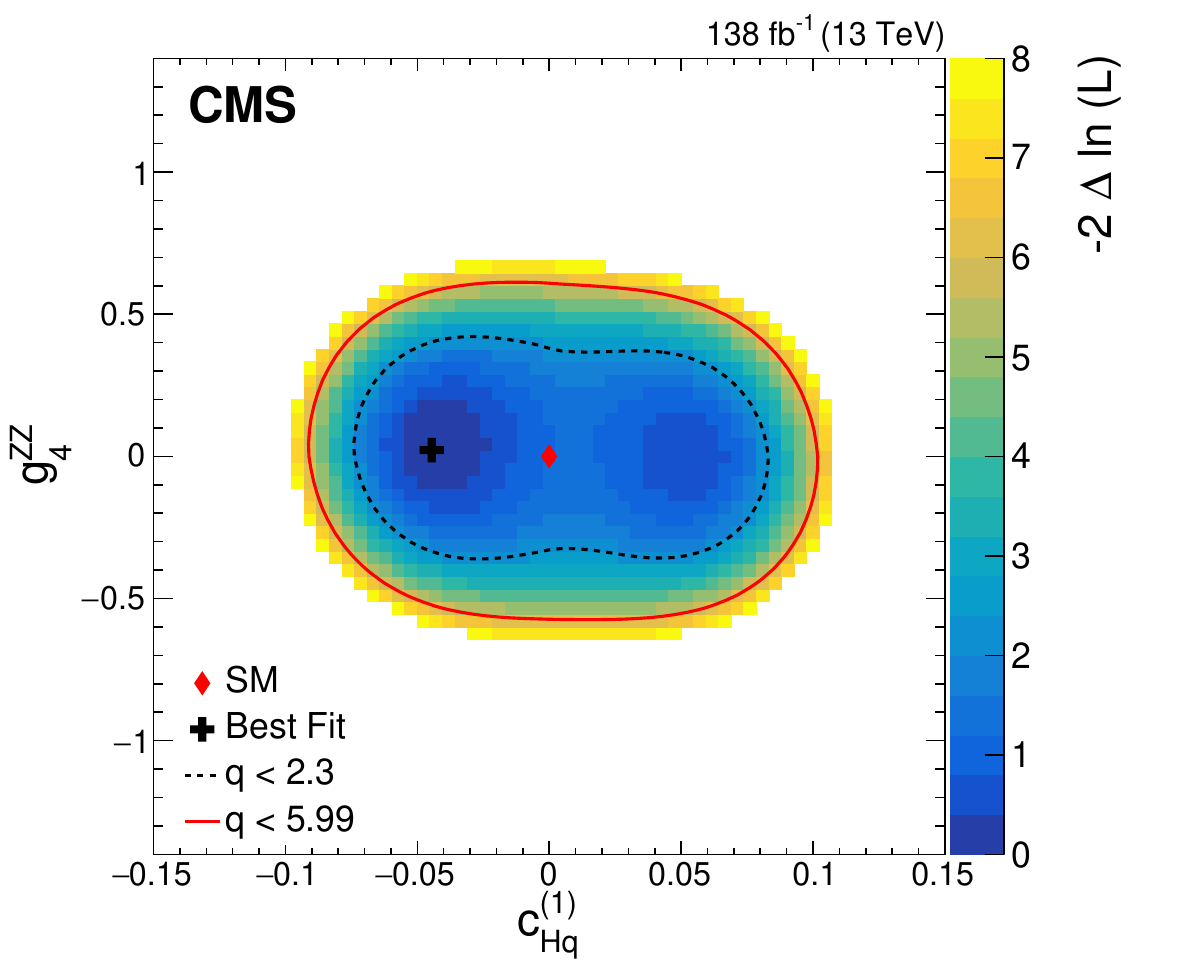}
\includegraphics[width=0.425\textwidth]{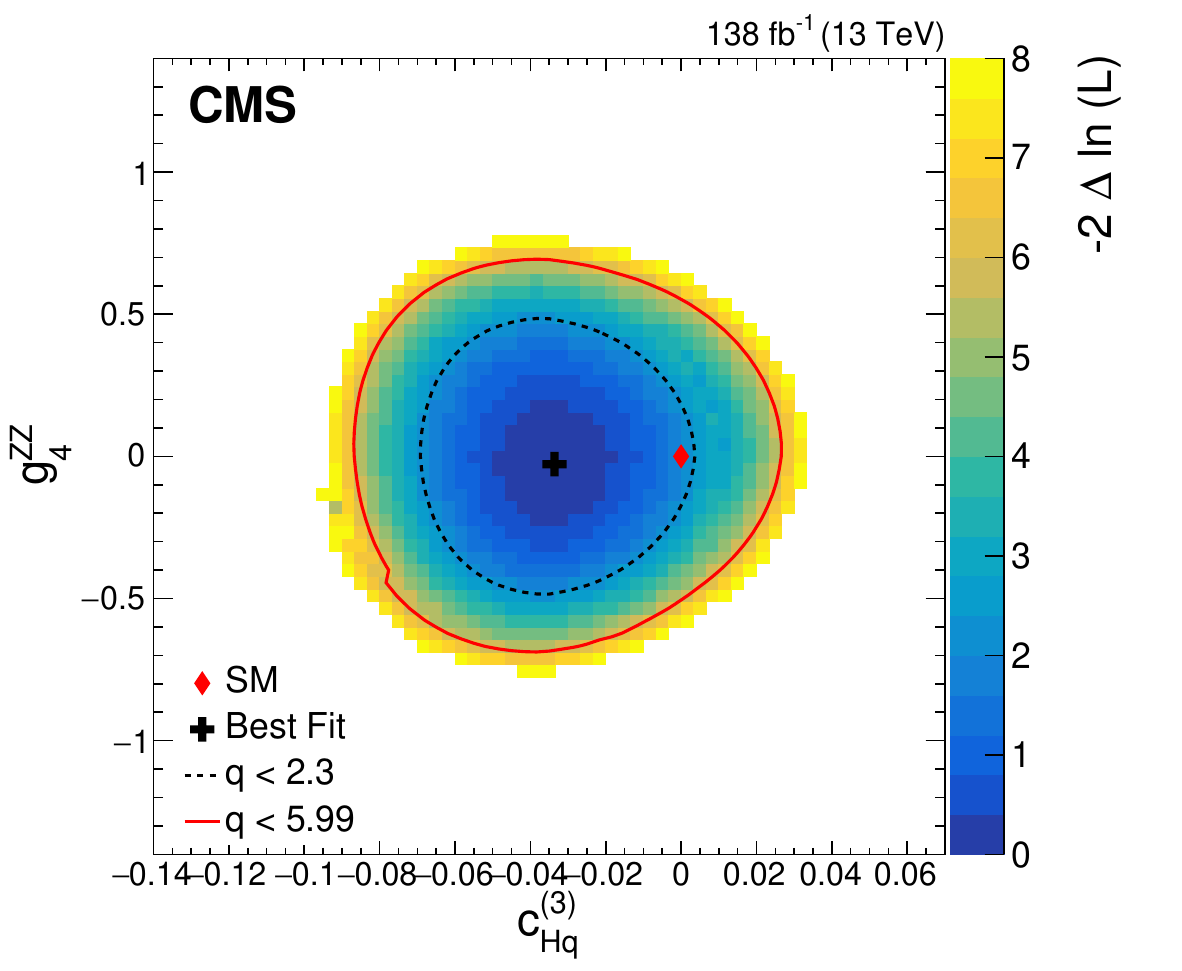}
\includegraphics[width=0.425\textwidth]{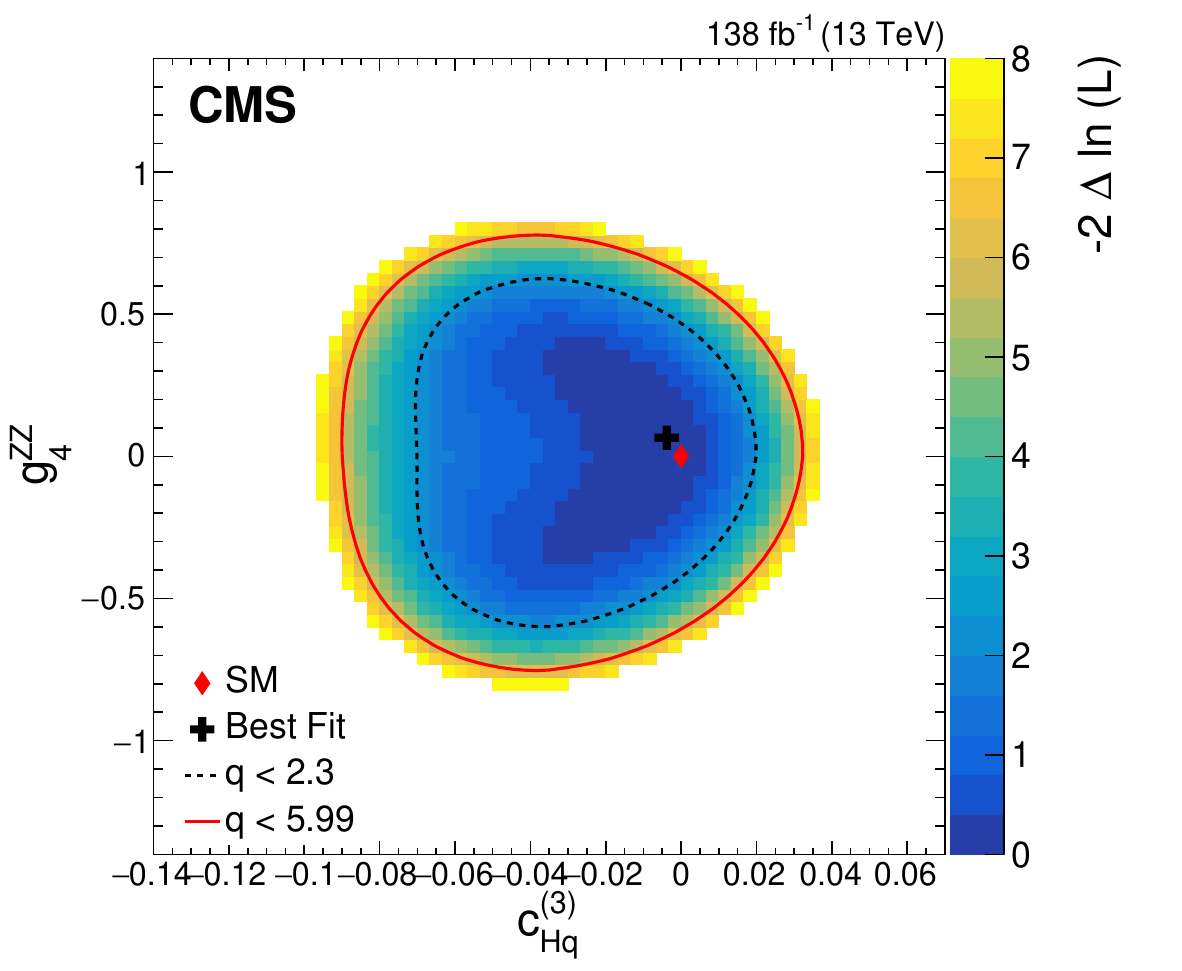}
\includegraphics[width=0.425\textwidth]{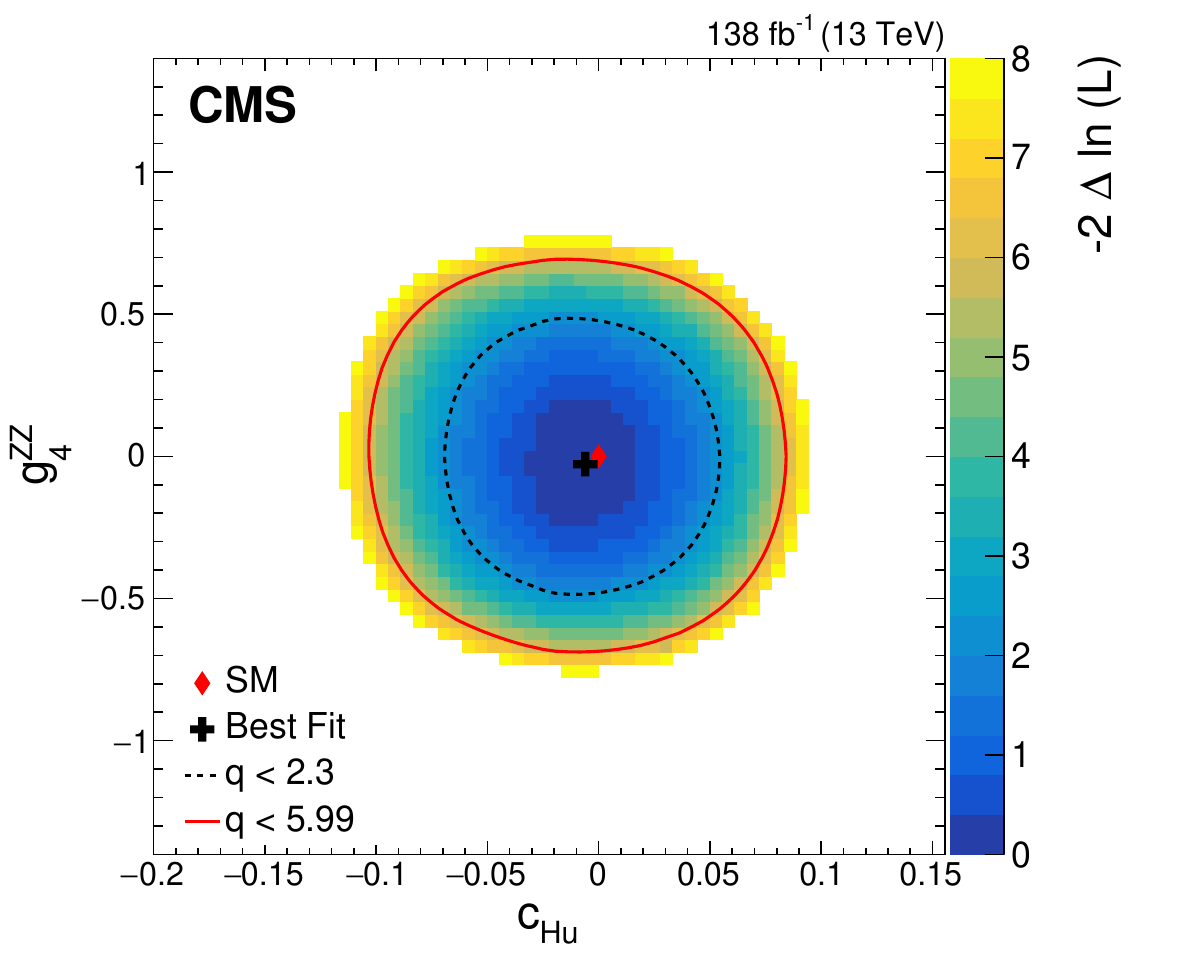}
\includegraphics[width=0.425\textwidth]{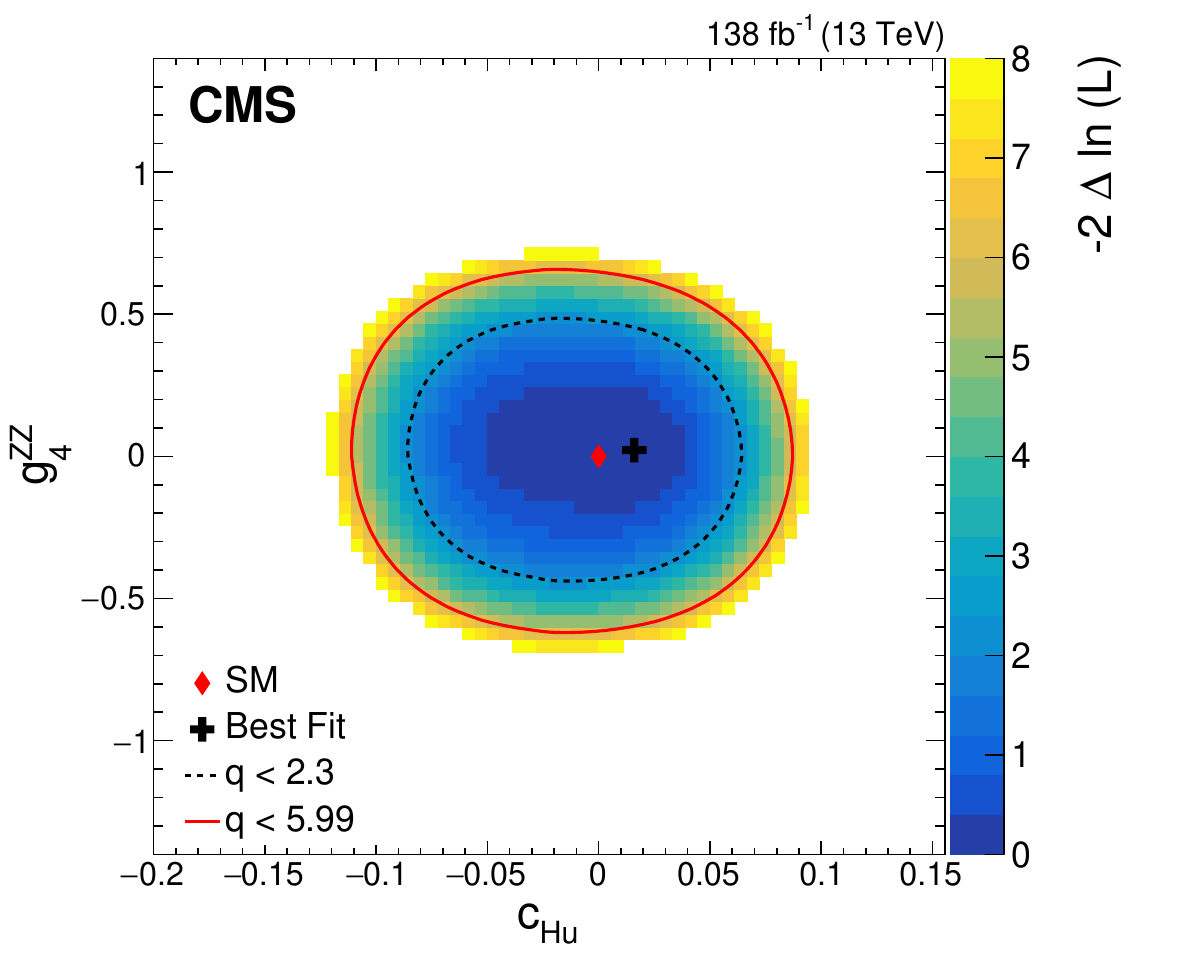}
\caption{
Observed two-dimensional likelihood scans for different pairs of Wilson coefficients: $\cHqo$ vs. $\gfZZ$ (upper row), $\cHqt$ vs. $\gfZZ$ (middle row), $\cHu$ vs. $\gfZZ$ (lower row) while allowing the other coefficients to float freely at each point of the scan (left) or fixed at their SM values (right) after combining results from all data-taking years and final states.
}
\label{fig:Scan_2D_d}
\end{figure*}

\begin{figure*}[!htb]
\centering
\includegraphics[width=0.425\textwidth]{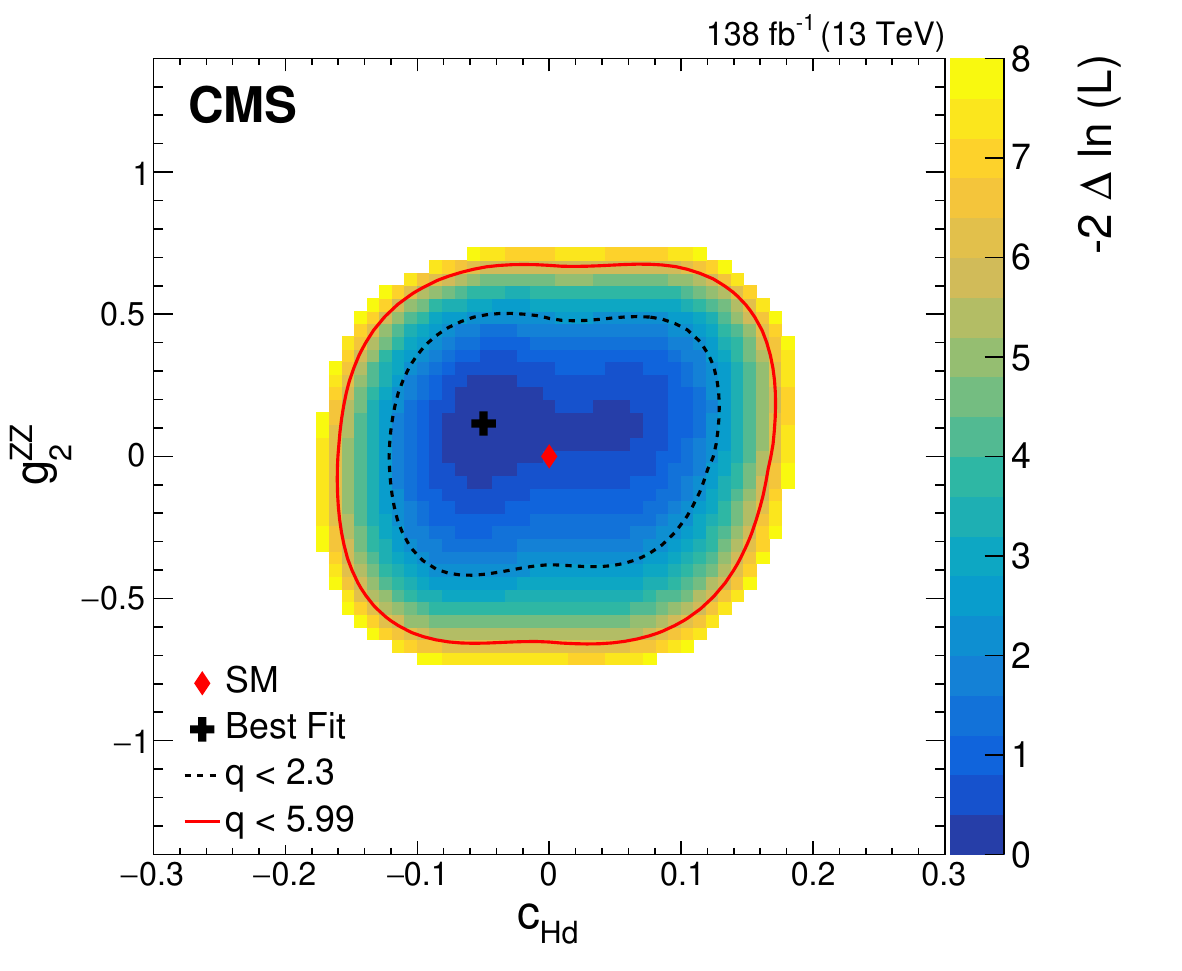}
\includegraphics[width=0.425\textwidth]{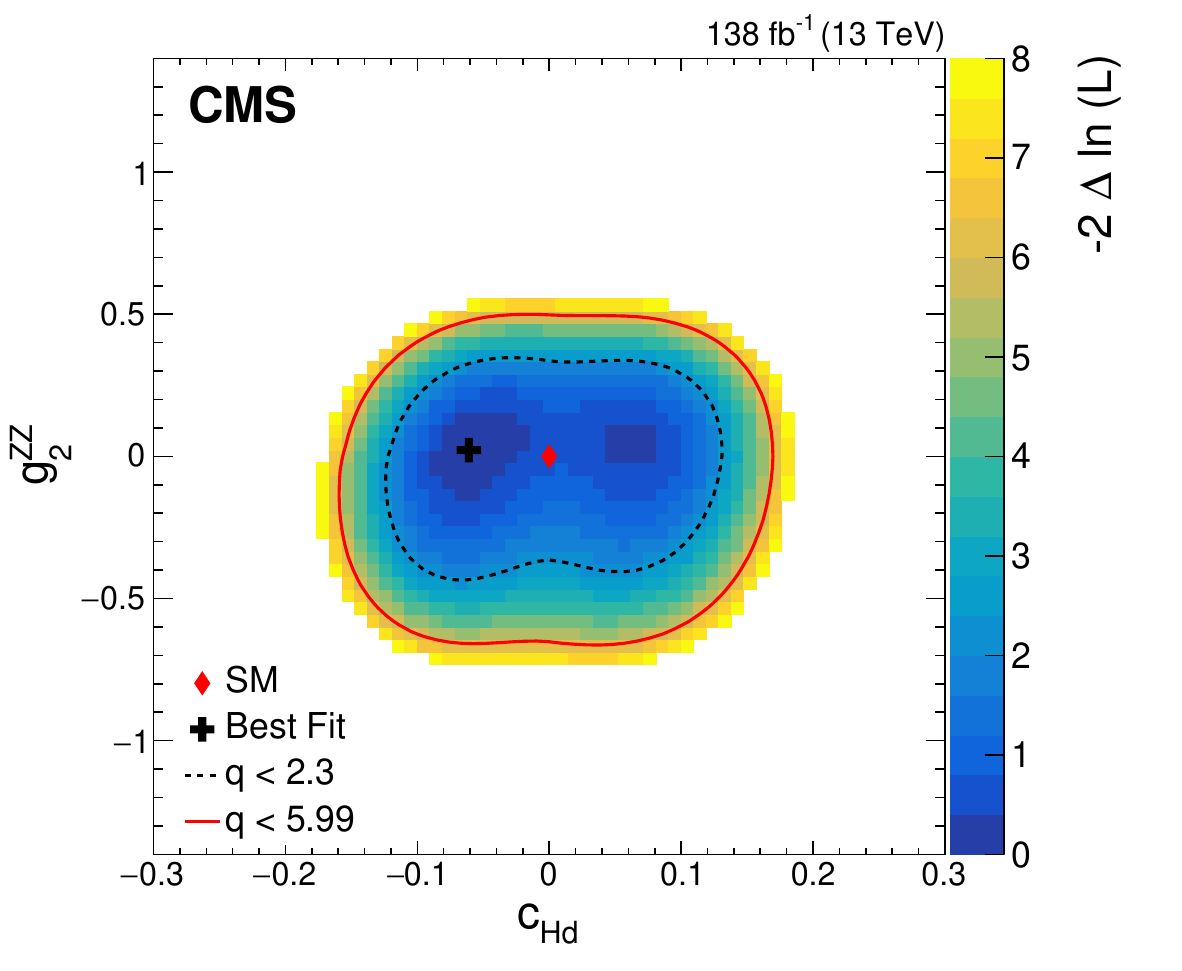}
\includegraphics[width=0.425\textwidth]{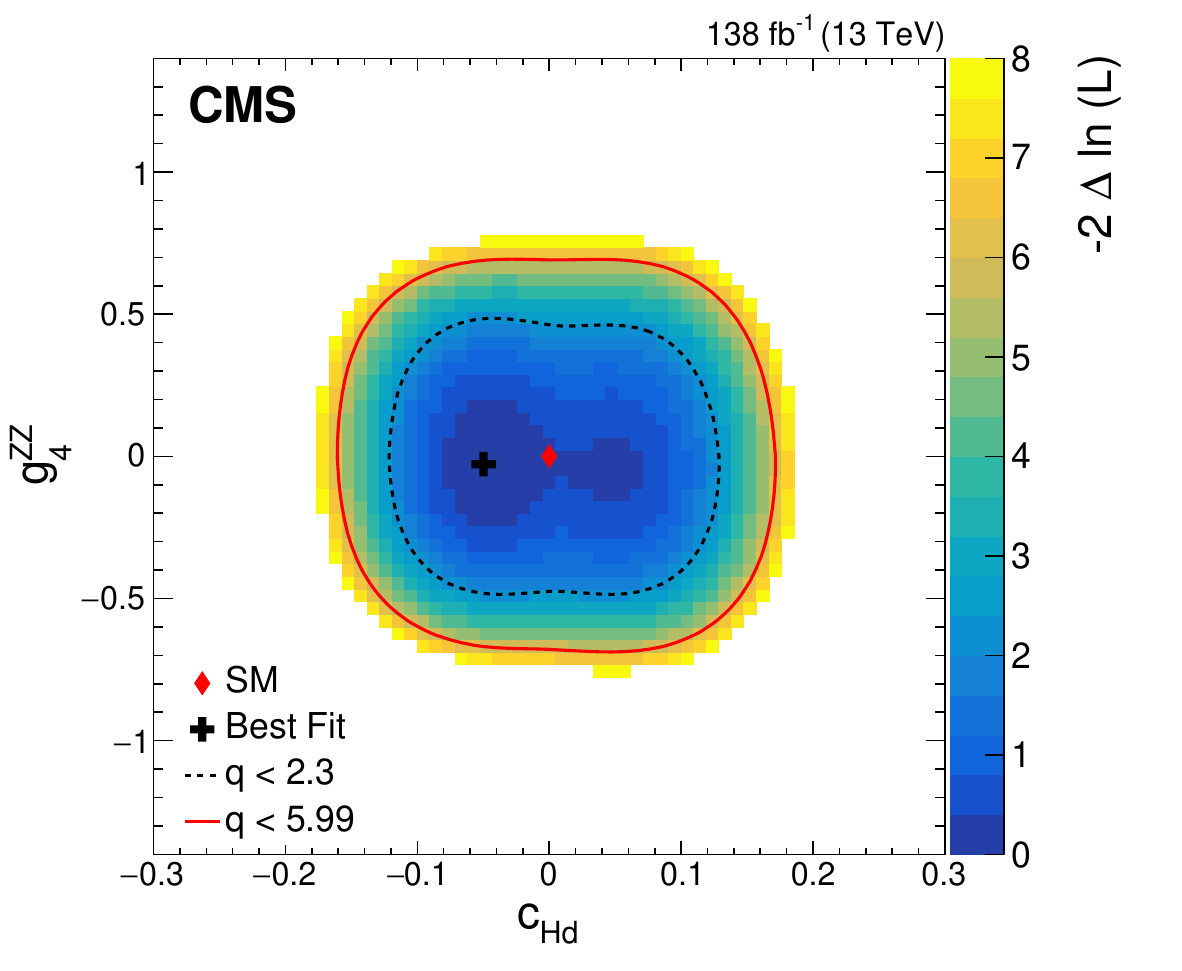}
\includegraphics[width=0.425\textwidth]{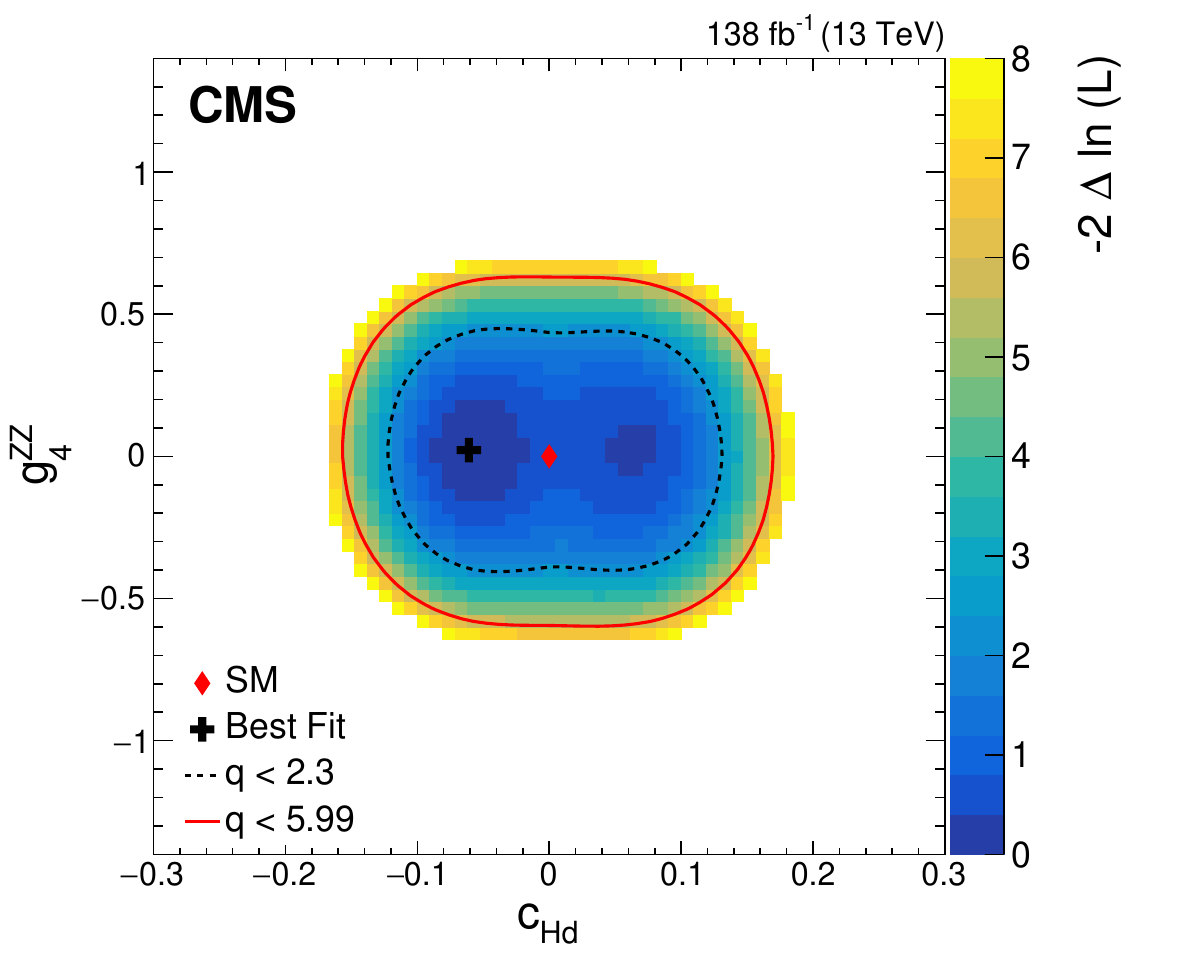}
\includegraphics[width=0.425\textwidth]{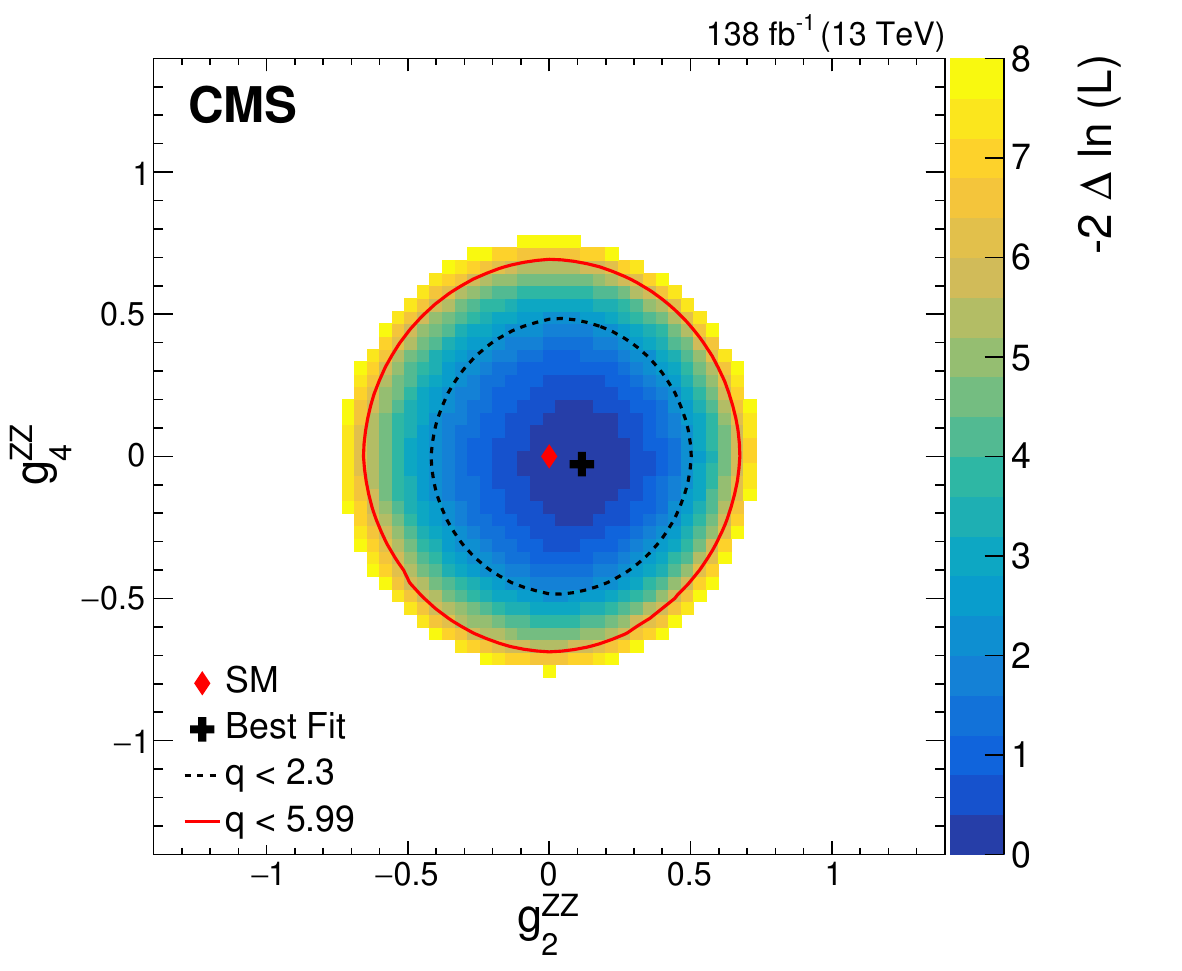}
\includegraphics[width=0.425\textwidth]{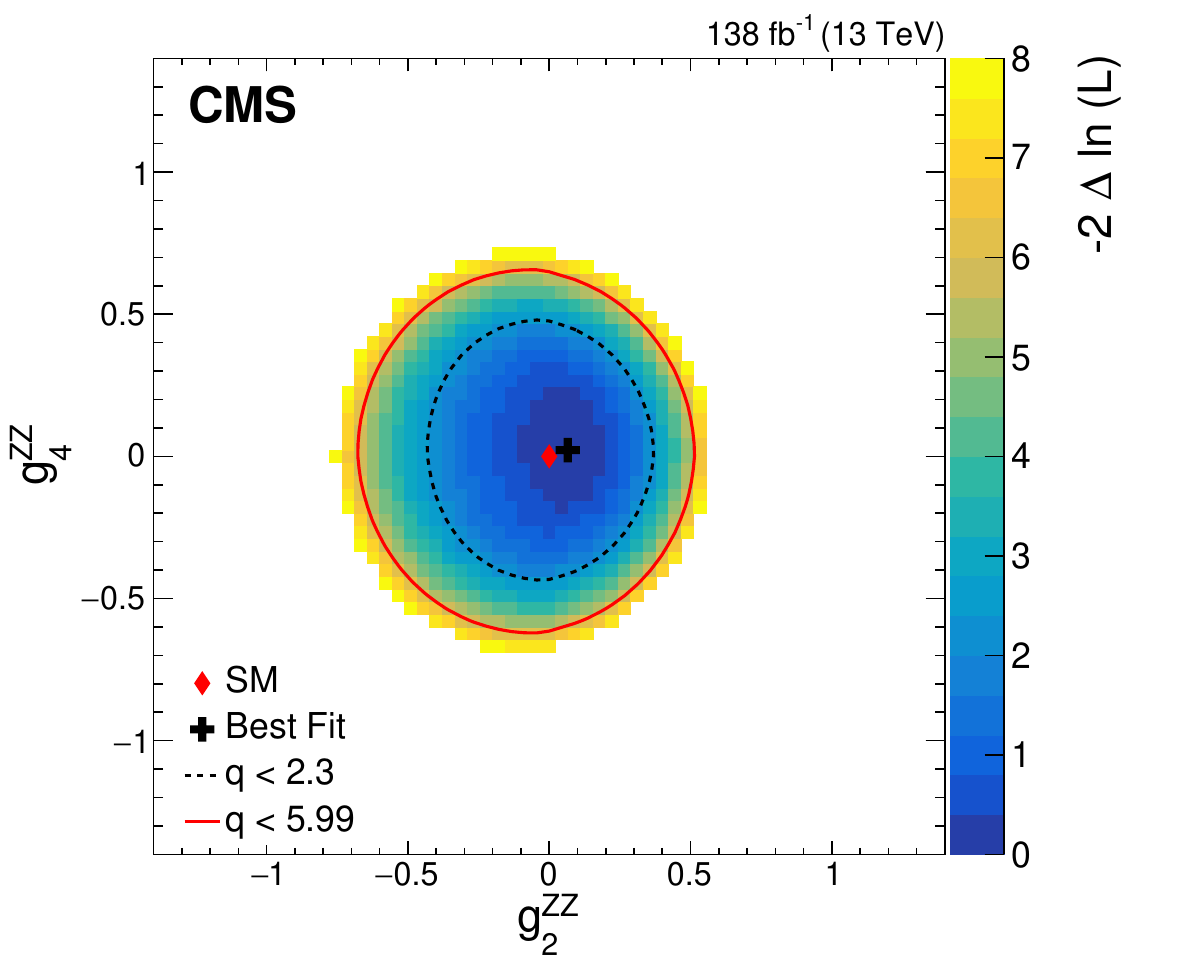}
\caption{
Observed two-dimensional likelihood scans for different pairs of Wilson coefficients: $\cHd$ vs. $\gtZZ$ (upper row), $\cHd$ vs. $\gfZZ$ (middle row), $\gtZZ$ vs. $\gfZZ$ (lower row) while allowing the other coefficients to float freely at each point of the scan (left) or fixed at their SM values (right) after combining results from all data-taking years and final states.
}
\label{fig:Scan_2D_e}
\end{figure*}

To explore correlation between Wilson coefficients, two-dimensional likelihood scans are presented considering different pairs of Wilson coefficients. In Figs.~\ref{fig:Scan_2D_a},~\ref{fig:Scan_2D_b},~\ref{fig:Scan_2D_c},~\ref{fig:Scan_2D_d},~\ref{fig:Scan_2D_e}, two-dimensional likelihood scans for different pairs of Wilson coefficients are shown while setting other coefficients to their SM values or allowing other Wilson coefficients to float freely at every point of the scan. We observe a correlation between different coefficients to various extents depending on the size of cross terms in the SMEFT expansion, which is especially significant for the two-dimensional likelihood scans involving the Wilson coefficients impacting the gauge couplings ($\gtZZ$) and the four-point interaction ($\cHqt$) in Fig.~\ref{fig:Scan_2D_c}.
In Fig.~\ref{fig:Scan_2D_a}, we also observe a double-minimum structure in the likelihood scan for $\cHqo$, which is consistent with the result obtained in one-dimensional likelihood scan including up to quadratic terms in SMEFT expansion as reported in the lower row of Fig.~\ref{fig:unbl_1D_summary}.

\section{Summary}
\label{sec:conclusion}

A standard model effective field theory (SMEFT) analysis is performed in the Higgs-strahlung process, where the Higgs boson is produced in association with a vector boson ($\PV=\PW,\PZ$), probing nonresonant new physics effects. Final states with the Higgs boson decaying to a pair of bottom quarks are targeted. Proton-proton collision data collected by the CMS experiment during 2016--2018 at a center-of-mass energy of 13\TeV are used, corresponding to an integrated luminosity of 138\fbinv. Leptonic decay modes of W and Z bosons ($\WtoLN$, $\ZtoLL$, and $\ZtoNN$) are considered, and both resolved- as well as merged-jet topologies are exploited for the $\HBB$ decay. 
A multivariate analysis strategy based on likelihood-free inference methods is adopted for the first time in the CMS experiment to probe the effects of multiple SMEFT operators including those giving rise to $CP$ violation. The strategy employing boosted decision trees makes use of the angular information which is sensitive to the $CP$ structure of SMEFT operators in this final state. Results are consistent with the standard model expectation. Constraints on the Wilson coefficients of six relevant SMEFT operators ($\cHqo$, $\cHqt$, $\cHu$, $\cHd$, $\gtZZ$, and $\gfZZ$) are obtained by performing a simultaneous fit to the data. Constraints on the vector-coupling operators are slightly more stringent than those on the gauge-coupling operators. Lower limits on the energy scales associated with various SMEFT operators are also presented, offering further constraints on different classes of new physics models. Additionally, constraints on two-dimensional planes of Wilson coefficients for all possible pairs are presented to explore correlations between pairs of Wilson coefficients. This constitutes the most comprehensive SMEFT analysis in this channel to date.

\begin{acknowledgments}
We congratulate our colleagues in the CERN accelerator departments for the excellent performance of the LHC and thank the technical and administrative staffs at CERN and at other CMS institutes for their contributions to the success of the CMS effort. In addition, we gratefully acknowledge the computing centers and personnel of the Worldwide LHC Computing Grid and other centers for delivering so effectively the computing infrastructure essential to our analyses. Finally, we acknowledge the enduring support for the construction and operation of the LHC, the CMS detector, and the supporting computing infrastructure provided by the following funding agencies: SC (Armenia), BMBWF and FWF (Austria); FNRS and FWO (Belgium); CNPq, CAPES, FAPERJ, FAPERGS, and FAPESP (Brazil); MES and BNSF (Bulgaria); CERN; CAS, MoST, and NSFC (China); MINCIENCIAS (Colombia); MSES and CSF (Croatia); RIF (Cyprus); SENESCYT (Ecuador); ERC PRG, RVTT3 and MoER TK202 (Estonia); Academy of Finland, MEC, and HIP (Finland); CEA and CNRS/IN2P3 (France); SRNSF (Georgia); BMBF, DFG, and HGF (Germany); GSRI (Greece); NKFIH (Hungary); DAE and DST (India); IPM (Iran); SFI (Ireland); INFN (Italy); MSIP and NRF (Republic of Korea); MES (Latvia); LMTLT (Lithuania); MOE and UM (Malaysia); BUAP, CINVESTAV, CONACYT, LNS, SEP, and UASLP-FAI (Mexico); MOS (Montenegro); MBIE (New Zealand); PAEC (Pakistan); MES and NSC (Poland); FCT (Portugal);  MESTD (Serbia); MCIN/AEI and PCTI (Spain); MOSTR (Sri Lanka); Swiss Funding Agencies (Switzerland); MST (Taipei); MHESI and NSTDA (Thailand); TUBITAK and TENMAK (Turkey); NASU (Ukraine); STFC (United Kingdom); DOE and NSF (USA).

\hyphenation{Rachada-pisek} Individuals have received support from the Marie-Curie program and the European Research Council and Horizon 2020 Grant, contract Nos.\ 675440, 724704, 752730, 758316, 765710, 824093, 101115353, 101002207, and COST Action CA16108 (European Union); the Leventis Foundation; the Alfred P.\ Sloan Foundation; the Alexander von Humboldt Foundation; the Science Committee, project no. 22rl-037 (Armenia); the Belgian Federal Science Policy Office; the Fonds pour la Formation \`a la Recherche dans l'Industrie et dans l'Agriculture (FRIA-Belgium); the F.R.S.-FNRS and FWO (Belgium) under the ``Excellence of Science -- EOS" -- be.h project n.\ 30820817; the Beijing Municipal Science \& Technology Commission, No. Z191100007219010 and Fundamental Research Funds for the Central Universities (China); the Ministry of Education, Youth and Sports (MEYS) of the Czech Republic; the Shota Rustaveli National Science Foundation, grant FR-22-985 (Georgia); the Deutsche Forschungsgemeinschaft (DFG), among others, under Germany's Excellence Strategy -- EXC 2121 ``Quantum Universe" -- 390833306, and under project number 400140256 - GRK2497; the Hellenic Foundation for Research and Innovation (HFRI), Project Number 2288 (Greece); the Hungarian Academy of Sciences, the New National Excellence Program - \'UNKP, the NKFIH research grants K 131991, K 133046, K 138136, K 143460, K 143477, K 146913, K 146914, K 147048, 2020-2.2.1-ED-2021-00181, TKP2021-NKTA-64, and 2021-4.1.2-NEMZ\_KI (Hungary); the Council of Science and Industrial Research, India; ICSC -- National Research Center for High Performance Computing, Big Data and Quantum Computing and FAIR -- Future Artificial Intelligence Research, funded by the NextGenerationEU program (Italy); the Latvian Council of Science; the Ministry of Education and Science, project no. 2022/WK/14, and the National Science Center, contracts Opus 2021/41/B/ST2/01369 and 2021/43/B/ST2/01552 (Poland); the Funda\c{c}\~ao para a Ci\^encia e a Tecnologia, grant CEECIND/01334/2018 (Portugal); the National Priorities Research Program by Qatar National Research Fund; MCIN/AEI/10.13039/501100011033, ERDF ``a way of making Europe", and the Programa Estatal de Fomento de la Investigaci{\'o}n Cient{\'i}fica y T{\'e}cnica de Excelencia Mar\'{\i}a de Maeztu, grant MDM-2017-0765 and Programa Severo Ochoa del Principado de Asturias (Spain); the Chulalongkorn Academic into Its 2nd Century Project Advancement Project, and the National Science, Research and Innovation Fund via the Program Management Unit for Human Resources \& Institutional Development, Research and Innovation, grant B39G670016 (Thailand); the Kavli Foundation; the Nvidia Corporation; the SuperMicro Corporation; the Welch Foundation, contract C-1845; and the Weston Havens Foundation (USA).
\end{acknowledgments}

\bibliography{auto_generated}
\cleardoublepage \appendix\section{The CMS Collaboration \label{app:collab}}\begin{sloppypar}\hyphenpenalty=5000\widowpenalty=500\clubpenalty=5000\input{HIG-23-016-public-authorlist.tex}\end{sloppypar}
%%% END EDITABLE REGION %%%
% skeleton_end
\end{document}

%% file: HIG-23-016-public-authorlist.tex
\cmsinstitute{Yerevan Physics Institute, Yerevan, Armenia}
{\tolerance=6000
V.~Chekhovsky, A.~Hayrapetyan, V.~Makarenko\cmsorcid{0000-0002-8406-8605}, A.~Tumasyan\cmsAuthorMark{1}\cmsorcid{0009-0000-0684-6742}
\par}
\cmsinstitute{Institut f\"{u}r Hochenergiephysik, Vienna, Austria}
{\tolerance=6000
W.~Adam\cmsorcid{0000-0001-9099-4341}, J.W.~Andrejkovic, L.~Benato\cmsorcid{0000-0001-5135-7489}, T.~Bergauer\cmsorcid{0000-0002-5786-0293}, S.~Chatterjee\cmsorcid{0000-0003-2660-0349}, K.~Damanakis\cmsorcid{0000-0001-5389-2872}, M.~Dragicevic\cmsorcid{0000-0003-1967-6783}, P.S.~Hussain\cmsorcid{0000-0002-4825-5278}, M.~Jeitler\cmsAuthorMark{2}\cmsorcid{0000-0002-5141-9560}, N.~Krammer\cmsorcid{0000-0002-0548-0985}, A.~Li\cmsorcid{0000-0002-4547-116X}, D.~Liko\cmsorcid{0000-0002-3380-473X}, I.~Mikulec\cmsorcid{0000-0003-0385-2746}, J.~Schieck\cmsAuthorMark{2}\cmsorcid{0000-0002-1058-8093}, R.~Sch\"{o}fbeck\cmsAuthorMark{2}\cmsorcid{0000-0002-2332-8784}, D.~Schwarz\cmsorcid{0000-0002-3821-7331}, M.~Sonawane\cmsorcid{0000-0003-0510-7010}, W.~Waltenberger\cmsorcid{0000-0002-6215-7228}, C.-E.~Wulz\cmsAuthorMark{2}\cmsorcid{0000-0001-9226-5812}
% ...
\par}
\vskip\cmsinstskip
\dag:~Deceased\\
$^{1}$Also at Yerevan State University, Yerevan, Armenia\\
$^{2}$Also at TU Wien, Vienna, Austria\\
% ...


%% file: HIG-23-016_temp.bbl
\providecommand{\href}[2]{#2}\begingroup\raggedright\begin{thebibliography}{100}%
\makeatletter
\providecommand{\hrefCMSnoop }[0]{\@secondoftwo}%
\makeatother
\providecommand{\doi}{\texttt{doi:}\begingroup \urlstyle{tt}\Url}

\bibitem{Englert:1964et}
\hrefCMSnoop {}{F.~Englert and R.~Brout, ``{Broken symmetry and the mass of
  gauge vector mesons}'',} \textit{ Phys. Rev. Lett.} \textbf{ 13} (1964) 321,
  \href{http://dx.doi.org/10.1103/PhysRevLett.13.321}{\doi{10.1103/PhysRevLett.13.321}}.

\bibitem{Higgs:1964ia}
\hrefCMSnoop {}{P.~W. Higgs, ``{Broken symmetries, massless particles and gauge
  fields}'',} \textit{ Phys. Lett.} \textbf{ 12} (1964) 132,
  \href{http://dx.doi.org/10.1016/0031-9163(64)91136-9}{\doi{10.1016/0031-9163(64)91136-9}}.

\bibitem{Higgs:1964pj}
\hrefCMSnoop {}{P.~W. Higgs, ``{Broken symmetries and the masses of gauge
  bosons}'',} \textit{ Phys. Rev. Lett.} \textbf{ 13} (1964) 508,
  \href{http://dx.doi.org/10.1103/PhysRevLett.13.508}{\doi{10.1103/PhysRevLett.13.508}}.

\bibitem{Guralnik:1964eu}
\hrefCMSnoop {}{G.~S. Guralnik, C.~R. Hagen, and T.~W.~B. Kibble, ``{Global
  conservation laws and massless particles}'',} \textit{ Phys. Rev. Lett.}
  \textbf{ 13} (1964) 585,
  \href{http://dx.doi.org/10.1103/PhysRevLett.13.585}{\doi{10.1103/PhysRevLett.13.585}}.

\bibitem{ATLAS:2012yve}
\hrefCMSnoop {}{{ATLAS Collaboration}, ``{Observation of a new particle in the
  search for the Standard Model Higgs boson with the ATLAS detector at the
  LHC}'',} \textit{ Phys. Lett. B} \textbf{ 716} (2012) 1,
  \href{http://dx.doi.org/10.1016/j.physletb.2012.08.020}{\doi{10.1016/j.physletb.2012.08.020}},
  \href{http://www.arXiv.org/abs/1207.7214}{\texttt{arXiv:1207.7214}}.

\bibitem{CMS:2012qbp}
\hrefCMSnoop {}{{CMS Collaboration}, ``{Observation of a new boson at a mass of
  125 GeV with the CMS experiment at the LHC}'',} \textit{ Phys. Lett. B}
  \textbf{ 716} (2012) 30,
  \href{http://dx.doi.org/10.1016/j.physletb.2012.08.021}{\doi{10.1016/j.physletb.2012.08.021}},
  \href{http://www.arXiv.org/abs/1207.7235}{\texttt{arXiv:1207.7235}}.

\bibitem{CMS:2013btf}
\hrefCMSnoop {}{{CMS Collaboration}, ``{Observation of a new boson with mass
  near 125 GeV in {pp} collisions at $\sqrt{s}$ = 7 and 8 TeV}'',} \textit{
  JHEP} \textbf{ 06} (2013) 081,
  \href{http://dx.doi.org/10.1007/JHEP06(2013)081}{\doi{10.1007/JHEP06(2013)081}},
  \href{http://www.arXiv.org/abs/1303.4571}{\texttt{arXiv:1303.4571}}.

\bibitem{ATLAS:2018kot}
\hrefCMSnoop {}{{ATLAS Collaboration}, ``{Observation of {$\rm{H} \rightarrow
  \PQb\PAQb$} decays and {VH} production with the ATLAS detector}'',} \textit{
  Phys. Lett. B} \textbf{ 786} (2018) 59,
  \href{http://dx.doi.org/10.1016/j.physletb.2018.09.013}{\doi{10.1016/j.physletb.2018.09.013}},
  \href{http://www.arXiv.org/abs/1808.08238}{\texttt{arXiv:1808.08238}}.

\bibitem{CMS:2018nsn}
\hrefCMSnoop {}{{CMS Collaboration}, ``{Observation of Higgs boson decay to
  bottom quarks}'',} \textit{ Phys. Rev. Lett.} \textbf{ 121} (2018) 121801,
  \href{http://dx.doi.org/10.1103/PhysRevLett.121.121801}{\doi{10.1103/PhysRevLett.121.121801}},
  \href{http://www.arXiv.org/abs/1808.08242}{\texttt{arXiv:1808.08242}}.

\bibitem{ATLAS:2024yzu}
\hrefCMSnoop {}{{ATLAS Collaboration}, ``{Measurements of WH and ZH production
  with Higgs boson decays into bottom quarks and direct constraints on the
  charm Yukawa coupling in $13\,\mathrm{TeV}$ pp collisions with the ATLAS
  detector}'',} 2024.
  \href{http://www.arXiv.org/abs/2410.19611}{\texttt{arXiv:2410.19611}}.
  submitted to {\it JHEP\/}.

\bibitem{CMS:2023vzh}
\hrefCMSnoop {}{{CMS Collaboration}, ``{Measurement of simplified template
  cross sections of the Higgs boson produced in association with W or Z bosons
  in the H $\to$$\mathrm{b\bar{b}}$ decay channel in proton-proton collisions
  at $\sqrt{s}$ =13 TeV}'',} \textit{ Phys. Rev. D} \textbf{ 109} (2024)
  092011,
  \href{http://dx.doi.org/0.1103/PhysRevD.109.092011}{\doi{0.1103/PhysRevD.109.092011}},
  \href{http://www.arXiv.org/abs/2312.07562}{\texttt{arXiv:2312.07562}}.

\bibitem{Buchmuller:1985jz}
\hrefCMSnoop {}{W.~Buchmuller and D.~Wyler, ``{Effective Lagrangian analysis of
  new interactions and flavor conservation}'',} \textit{ Nucl. Phys. B}
  \textbf{ 268} (1986) 621,
  \href{http://dx.doi.org/10.1016/0550-3213(86)90262-2}{\doi{10.1016/0550-3213(86)90262-2}}.

\bibitem{Grinstein:1991cd}
\hrefCMSnoop {}{B.~Grinstein and M.~B. Wise, ``{Operator analysis for precision
  electroweak physics}'',} \textit{ Phys. Lett. B} \textbf{ 265} (1991) 326,
  \href{http://dx.doi.org/10.1016/0370-2693(91)90061-T}{\doi{10.1016/0370-2693(91)90061-T}}.

\bibitem{Chiu:2007dg}
\hrefCMSnoop {}{J.-y. Chiu, F.~Golf, R.~Kelley, and A.~V. Manohar,
  ``{Electroweak corrections in high energy processes using effective field
  theory}'',} \textit{ Phys. Rev. D} \textbf{ 77} (2008) 053004,
  \href{http://dx.doi.org/10.1103/PhysRevD.77.053004}{\doi{10.1103/PhysRevD.77.053004}},
  \href{http://www.arXiv.org/abs/0712.0396}{\texttt{arXiv:0712.0396}}.

\bibitem{Degrande:2012wf}
C.~Degrande\hrefCMSnoop {}{ { et~al.}, ``{Effective field theory: A modern
  approach to anomalous couplings}'',} \textit{ Annals Phys.} \textbf{ 335}
  (2013) 21,
  \href{http://dx.doi.org/10.1016/j.aop.2013.04.016}{\doi{10.1016/j.aop.2013.04.016}},
  \href{http://www.arXiv.org/abs/1205.4231}{\texttt{arXiv:1205.4231}}.

\bibitem{Jenkins:2013zja}
\hrefCMSnoop {}{E.~E. Jenkins, A.~V. Manohar, and M.~Trott, ``{Renormalization
  group evolution of the standard model dimension six operators I: formalism
  and lambda dependence}'',} \textit{ JHEP} \textbf{ 10} (2013) 087,
  \href{http://dx.doi.org/10.1007/JHEP10(2013)087}{\doi{10.1007/JHEP10(2013)087}},
  \href{http://www.arXiv.org/abs/1308.2627}{\texttt{arXiv:1308.2627}}.

\bibitem{Alonso:2013hga}
\hrefCMSnoop {}{R.~Alonso, E.~E. Jenkins, A.~V. Manohar, and M.~Trott,
  ``{Renormalization group evolution of the standard model dimension six
  operators III: gauge coupling dependence and phenomenology}'',} \textit{
  JHEP} \textbf{ 04} (2014) 159,
  \href{http://dx.doi.org/10.1007/JHEP04(2014)159}{\doi{10.1007/JHEP04(2014)159}},
  \href{http://www.arXiv.org/abs/1312.2014}{\texttt{arXiv:1312.2014}}.

\bibitem{Jenkins:2013wua}
\hrefCMSnoop {}{E.~E. Jenkins, A.~V. Manohar, and M.~Trott, ``{Renormalization
  group evolution of the standard model dimension six operators II: Yukawa
  dependence}'',} \textit{ JHEP} \textbf{ 01} (2014) 035,
  \href{http://dx.doi.org/10.1007/JHEP01(2014)035}{\doi{10.1007/JHEP01(2014)035}},
  \href{http://www.arXiv.org/abs/1310.4838}{\texttt{arXiv:1310.4838}}.

\bibitem{Englert:2014cva}
\hrefCMSnoop {}{C.~Englert and M.~Spannowsky, ``{Effective theories and
  measurements at colliders}'',} \textit{ Phys. Lett. B} \textbf{ 740} (2015)
  8,
  \href{http://dx.doi.org/10.1016/j.physletb.2014.11.035}{\doi{10.1016/j.physletb.2014.11.035}},
  \href{http://www.arXiv.org/abs/1408.5147}{\texttt{arXiv:1408.5147}}.

\bibitem{Brivio:2017vri}
\hrefCMSnoop {}{I.~Brivio and M.~Trott, ``{The standard model as an effective
  field theory}'',} \textit{ Phys. Rept.} \textbf{ 793} (2019) 1,
  \href{http://dx.doi.org/10.1016/j.physrep.2018.11.002}{\doi{10.1016/j.physrep.2018.11.002}},
  \href{http://www.arXiv.org/abs/1706.08945}{\texttt{arXiv:1706.08945}}.

\bibitem{Isidori:2023pyp}
\hrefCMSnoop {}{G.~Isidori, F.~Wilsch, and D.~Wyler, ``{The standard model
  effective field theory at work}'',} \textit{ Rev. Mod. Phys.} \textbf{ 96}
  (2024) 015006,
  \href{http://dx.doi.org/10.1103/RevModPhys.96.015006}{\doi{10.1103/RevModPhys.96.015006}},
  \href{http://www.arXiv.org/abs/2303.16922}{\texttt{arXiv:2303.16922}}.

\bibitem{CMS:2021nnc}
\hrefCMSnoop {}{{CMS Collaboration}, ``{Constraints on anomalous Higgs boson
  couplings to vector bosons and fermions in its production and decay using the
  four-lepton final state}'',} \textit{ Phys. Rev. D} \textbf{ 104} (2021)
  052004,
  \href{http://dx.doi.org/10.1103/PhysRevD.104.052004}{\doi{10.1103/PhysRevD.104.052004}},
  \href{http://www.arXiv.org/abs/2104.12152}{\texttt{arXiv:2104.12152}}.

\bibitem{CMS:2024bua}
\hrefCMSnoop {}{{CMS Collaboration}, ``{Constraints on anomalous Higgs boson
  couplings from its production and decay using the WW channel in
  proton\textendash{}proton collisions at $\sqrt{s} = 13~\text {TeV}$}'',}
  \textit{ Eur. Phys. J. C} \textbf{ 84} (2024) 779,
  \href{http://dx.doi.org/10.1140/epjc/s10052-024-12925-0}{\doi{10.1140/epjc/s10052-024-12925-0}},
  \href{http://www.arXiv.org/abs/2403.00657}{\texttt{arXiv:2403.00657}}.

\bibitem{MELAcmsHTT}
\hrefCMSnoop {}{{CMS Collaboration}, ``{Constraints on anomalous Higgs boson
  couplings to vector bosons and fermions from the production of Higgs bosons
  using the $\tau\tau$ final state}'',} \textit{ Phys. Rev. D} \textbf{ 108}
  (2023) 032013,
  \href{http://dx.doi.org/10.1103/PhysRevD.108.032013}{\doi{10.1103/PhysRevD.108.032013}},
  \href{http://www.arXiv.org/abs/2205.05120}{\texttt{arXiv:2205.05120}}.

\bibitem{Gao:2010qx}
Y.~Gao\hrefCMSnoop {}{ { et~al.}, ``{Spin determination of single-produced
  resonances at hadron colliders}'',} \textit{ Phys. Rev. D} \textbf{ 81}
  (2010) 075022,
  \href{http://dx.doi.org/10.1103/PhysRevD.81.075022}{\doi{10.1103/PhysRevD.81.075022}},
  \href{http://www.arXiv.org/abs/1001.3396}{\texttt{arXiv:1001.3396}}.

\bibitem{Bolognesi:2012mm}
S.~Bolognesi\hrefCMSnoop {}{ { et~al.}, ``{On the spin and parity of a
  single-produced resonance at the LHC}'',} \textit{ Phys. Rev. D} \textbf{ 86}
  (2012) 095031,
  \href{http://dx.doi.org/10.1103/PhysRevD.86.095031}{\doi{10.1103/PhysRevD.86.095031}},
  \href{http://www.arXiv.org/abs/1208.4018}{\texttt{arXiv:1208.4018}}.

\bibitem{Anderson:2013afp}
\hrefCMSnoop {}{I.~Anderson { et~al.}, ``{Constraining anomalous HVV
  interactions at proton and lepton colliders}'',} \textit{ Phys. Rev. D}
  \textbf{ 89} (2014) 035007,
  \href{http://dx.doi.org/10.1103/PhysRevD.89.035007}{\doi{10.1103/PhysRevD.89.035007}},
  \href{http://www.arXiv.org/abs/1309.4819}{\texttt{arXiv:1309.4819}}.

\bibitem{ATLAS:2019yhn}
\hrefCMSnoop {}{{ATLAS Collaboration}, ``{Measurement of VH, $ \mathrm{H}\to
  \mathrm{b}\overline{\mathrm{b}} $ production as a function of the
  vector-boson transverse momentum in 13 TeV pp collisions with the ATLAS
  detector}'',} \textit{ JHEP} \textbf{ 05} (2019) 141,
  \href{http://dx.doi.org/10.1007/JHEP05(2019)141}{\doi{10.1007/JHEP05(2019)141}},
  \href{http://www.arXiv.org/abs/1903.04618}{\texttt{arXiv:1903.04618}}.

\bibitem{ATLAS:2020fcp}
\hrefCMSnoop {}{{ATLAS Collaboration}, ``{Measurements of {WH} and {ZH}
  production in the {$\rm{H} \rightarrow \PQb\PAQb$} decay channel in {pp}
  collisions at 13 TeV with the ATLAS detector}'',} \textit{ Eur. Phys. J. C}
  \textbf{ 81} (2021) 178,
  \href{http://dx.doi.org/10.1140/epjc/s10052-020-08677-2}{\doi{10.1140/epjc/s10052-020-08677-2}},
  \href{http://www.arXiv.org/abs/2007.02873}{\texttt{arXiv:2007.02873}}.

\bibitem{CMS-PAS-HIG-19-005}
\href {http://cds.cern.ch/record/2706103}{{CMS Collaboration}, ``{Combined
  Higgs boson production and decay measurements with up to 137 fb$^{-1}$ of
  proton-proton collision data at $\sqrt{s} = 13$ TeV}'',} CMS Physics Analysis
  Summary CMS-PAS-HIG-19-005, 2020.

\bibitem{Ellis:2020unq}
J.~Ellis\hrefCMSnoop {}{ { et~al.}, ``{Top, higgs, diboson and electroweak fit
  to the standard model effective field theory}'',} \textit{ JHEP} \textbf{ 04}
  (2021) 279,
  \href{http://dx.doi.org/10.1007/JHEP04(2021)279}{\doi{10.1007/JHEP04(2021)279}},
  \href{http://www.arXiv.org/abs/2012.02779}{\texttt{arXiv:2012.02779}}.

\bibitem{Ethier:2021bye}
\hrefCMSnoop {}{{SMEFiT} Collaboration, ``{Combined SMEFT interpretation of
  Higgs, diboson, and top quark data from the LHC}'',} \textit{ JHEP} \textbf{
  11} (2021) 089,
  \href{http://dx.doi.org/10.1007/JHEP11(2021)089}{\doi{10.1007/JHEP11(2021)089}},
  \href{http://www.arXiv.org/abs/2105.00006}{\texttt{arXiv:2105.00006}}.

\bibitem{hepdata}
\hrefCMSnoop {}{``{HEPD}ata record for this analysis'',} 2024.
\newblock
  \href{http://dx.doi.org/10.17182/hepdata.155497}{\doi{10.17182/hepdata.155497}}.

\bibitem{PhysRevLett.43.1566}
\hrefCMSnoop {}{S.~Weinberg, ``Baryon- and lepton-nonconserving processes'',}
  \textit{ Phys. Rev. Lett.} \textbf{ 43} (1979) 1566,
  \href{http://dx.doi.org/10.1103/PhysRevLett.43.1566}{\doi{10.1103/PhysRevLett.43.1566}}.

\bibitem{Grzadkowski:2010es}
\hrefCMSnoop {}{B.~Grzadkowski, M.~Iskrzynski, M.~Misiak, and J.~Rosiek,
  ``{Dimension-six terms in the standard model Lagrangian}'',} \textit{ JHEP}
  \textbf{ 10} (2010) 085,
  \href{http://dx.doi.org/10.1007/JHEP10(2010)085}{\doi{10.1007/JHEP10(2010)085}},
  \href{http://www.arXiv.org/abs/1008.4884}{\texttt{arXiv:1008.4884}}.

\bibitem{Falkowski:2014tna}
\hrefCMSnoop {}{A.~Falkowski and F.~Riva, ``{Model-independent precision
  constraints on dimension-6 operators}'',} \textit{ JHEP} \textbf{ 02} (2015)
  039,
  \href{http://dx.doi.org/10.1007/JHEP02(2015)039}{\doi{10.1007/JHEP02(2015)039}},
  \href{http://www.arXiv.org/abs/1411.0669}{\texttt{arXiv:1411.0669}}.

\bibitem{Banerjee:2018bio}
\hrefCMSnoop {}{S.~Banerjee, C.~Englert, R.~S. Gupta, and M.~Spannowsky,
  ``{Probing electroweak precision physics via boosted Higgs-strahlung at the
  LHC}'',} \textit{ Phys. Rev. D} \textbf{ 98} (2018) 095012,
  \href{http://dx.doi.org/10.1103/PhysRevD.98.095012}{\doi{10.1103/PhysRevD.98.095012}},
  \href{http://www.arXiv.org/abs/1807.01796}{\texttt{arXiv:1807.01796}}.

\bibitem{Banerjee:2019twi}
S.~Banerjee\hrefCMSnoop {}{ { et~al.}, ``{Towards the ultimate differential
  SMEFT analysis}'',} \textit{ JHEP} \textbf{ 09} (2020) 170,
  \href{http://dx.doi.org/10.1007/JHEP09(2020)170}{\doi{10.1007/JHEP09(2020)170}},
  \href{http://www.arXiv.org/abs/1912.07628}{\texttt{arXiv:1912.07628}}.

\bibitem{Brehmer:2018kdj}
\hrefCMSnoop {}{J.~Brehmer, K.~Cranmer, G.~Louppe, and J.~Pavez,
  ``{Constraining effective field theories with machine learning}'',} \textit{
  Phys. Rev. Lett.} \textbf{ 121} (2018) 111801,
  \href{http://dx.doi.org/10.1103/PhysRevLett.121.111801}{\doi{10.1103/PhysRevLett.121.111801}},
  \href{http://www.arXiv.org/abs/1805.00013}{\texttt{arXiv:1805.00013}}.

\bibitem{Brehmer:2018hga}
\hrefCMSnoop {}{J.~Brehmer, G.~Louppe, J.~Pavez, and K.~Cranmer, ``{Mining gold
  from implicit models to improve likelihood-free inference}'',} \textit{ Proc.
  Nat. Acad. Sci.} \textbf{ 117} (2020) 5242,
  \href{http://dx.doi.org/10.1073/pnas.1915980117}{\doi{10.1073/pnas.1915980117}},
  \href{http://www.arXiv.org/abs/1805.12244}{\texttt{arXiv:1805.12244}}.

\bibitem{Chen:2020mev}
\hrefCMSnoop {}{S.~Chen, A.~Glioti, G.~Panico, and A.~Wulzer, ``{Parametrized
  classifiers for optimal EFT sensitivity}'',} \textit{ JHEP} \textbf{ 05}
  (2021) 247,
  \href{http://dx.doi.org/10.1007/JHEP05(2021)247}{\doi{10.1007/JHEP05(2021)247}},
  \href{http://www.arXiv.org/abs/2007.10356}{\texttt{arXiv:2007.10356}}.

\bibitem{Chatterjee:2021nms}
S.~Chatterjee\hrefCMSnoop {}{ { et~al.}, ``{Tree boosting for learning EFT
  parameters}'',} \textit{ Comput. Phys. Commun.} \textbf{ 277} (2022) 108385,
  \href{http://dx.doi.org/10.1016/j.cpc.2022.108385}{\doi{10.1016/j.cpc.2022.108385}},
  \href{http://www.arXiv.org/abs/2107.10859}{\texttt{arXiv:2107.10859}}.

\bibitem{Chatterjee:2022oco}
\hrefCMSnoop {}{S.~Chatterjee, S.~Rohshap, R.~Sch{\"o}fbeck, and D.~Schwarz,
  ``{Learning the EFT likelihood with tree boosting}'',} 2022.
  \href{http://www.arXiv.org/abs/2205.12976}{\texttt{arXiv:2205.12976}}.

\bibitem{GomezAmbrosio:2022mpm}
R.~Gomez~Ambrosio\hrefCMSnoop {}{ { et~al.}, ``{Unbinned multivariate
  observables for global SMEFT analyses from machine learning}'',} \textit{
  JHEP} \textbf{ 03} (2023) 033,
  \href{http://dx.doi.org/10.1007/JHEP03(2023)033}{\doi{10.1007/JHEP03(2023)033}},
  \href{http://www.arXiv.org/abs/2211.02058}{\texttt{arXiv:2211.02058}}.

\bibitem{deFlorian:2016spz}
\hrefCMSnoop {}{{LHC Higgs Cross Section Working Group}, ``Handbook of {LHC}
  {H}iggs cross sections: 4. {D}eciphering the nature of the {H}iggs sector'',}
  CERN Report CERN-2017-002-M, 2016.
\newblock
  \href{http://dx.doi.org/10.23731/CYRM-2017-002}{\doi{10.23731/CYRM-2017-002}},
  \href{http://www.arXiv.org/abs/1610.07922}{\texttt{arXiv:1610.07922}}.

\bibitem{Davis:2021tiv}
J.~Davis\hrefCMSnoop {}{ { et~al.}, ``{Constraining anomalous Higgs boson
  couplings to virtual photons}'',} \textit{ Phys. Rev. D} \textbf{ 105} (2022)
  096027,
  \href{http://dx.doi.org/10.1103/PhysRevD.105.096027}{\doi{10.1103/PhysRevD.105.096027}},
  \href{http://www.arXiv.org/abs/2109.13363}{\texttt{arXiv:2109.13363}}.

\bibitem{Rossia:2023hen}
\hrefCMSnoop {}{A.~Rossia, M.~Thomas, and E.~Vryonidou, ``{Diboson production
  in the SMEFT from gluon fusion}'',} \textit{ JHEP} \textbf{ 11} (2023) 132,
  \href{http://dx.doi.org/10.1007/JHEP11(2023)132}{\doi{10.1007/JHEP11(2023)132}},
  \href{http://www.arXiv.org/abs/2306.09963}{\texttt{arXiv:2306.09963}}.

\bibitem{CMS:2008xjf}
\hrefCMSnoop {}{{CMS Collaboration}, ``{The CMS experiment at the CERN LHC}'',}
  \textit{ JINST} \textbf{ 3} (2008) S08004,
  \href{http://dx.doi.org/10.1088/1748-0221/3/08/S08004}{\doi{10.1088/1748-0221/3/08/S08004}}.

\bibitem{CMS:2023gfb}
\hrefCMSnoop {}{{CMS Collaboration}, ``{Development of the CMS detector for the
  CERN LHC Run 3}'',} \textit{ JINST} \textbf{ 19} (2024) P05064,
  \href{http://dx.doi.org/10.1088/1748-0221/19/05/P05064}{\doi{10.1088/1748-0221/19/05/P05064}},
  \href{http://www.arXiv.org/abs/2309.05466}{\texttt{arXiv:2309.05466}}.

\bibitem{CMS:2014pgm}
\hrefCMSnoop {}{{CMS Collaboration}, ``{Description and performance of track
  and primary-vertex reconstruction with the CMS tracker}'',} \textit{ JINST}
  \textbf{ 9} (2014) P10009,
  \href{http://dx.doi.org/10.1088/1748-0221/9/10/P10009}{\doi{10.1088/1748-0221/9/10/P10009}},
  \href{http://www.arXiv.org/abs/1405.6569}{\texttt{arXiv:1405.6569}}.

\bibitem{Phase1Pixel}
\hrefCMSnoop {}{{CMS Tracker Group}, ``The {CMS} phase-1 pixel detector
  upgrade'',} \textit{ JINST} \textbf{ 16} (2021) P02027,
  \href{http://dx.doi.org/10.1088/1748-0221/16/02/P02027}{\doi{10.1088/1748-0221/16/02/P02027}},
  \href{http://www.arXiv.org/abs/2012.14304}{\texttt{arXiv:2012.14304}}.

\bibitem{DP-2020-049}
\href {https://cds.cern.ch/record/2743740}{{CMS Collaboration}, ``Track impact
  parameter resolution for the full pseudo rapidity coverage in the 2017
  dataset with the {CMS} phase-1 pixel detector'',} CMS Detector Performance
  Summary CMS-DP-2020-049, 2020.

\bibitem{CMS:2017yfk}
\hrefCMSnoop {}{{CMS Collaboration}, ``{Particle-flow reconstruction and global
  event description with the CMS detector}'',} \textit{ JINST} \textbf{ 12}
  (2017) P10003,
  \href{http://dx.doi.org/10.1088/1748-0221/12/10/P10003}{\doi{10.1088/1748-0221/12/10/P10003}},
  \href{http://www.arXiv.org/abs/1706.04965}{\texttt{arXiv:1706.04965}}.

\bibitem{CMS:2020cmk}
\hrefCMSnoop {}{{CMS Collaboration}, ``{Performance of the CMS Level-1 trigger
  in proton-proton collisions at $\sqrt{s} =$ 13 TeV}'',} \textit{ JINST}
  \textbf{ 15} (2020) P10017,
  \href{http://dx.doi.org/10.1088/1748-0221/15/10/P10017}{\doi{10.1088/1748-0221/15/10/P10017}},
  \href{http://www.arXiv.org/abs/2006.10165}{\texttt{arXiv:2006.10165}}.

\bibitem{CMS:2016ngn}
\hrefCMSnoop {}{{CMS Collaboration}, ``{The CMS trigger system}'',} \textit{
  JINST} \textbf{ 12} (2017) P01020,
  \href{http://dx.doi.org/10.1088/1748-0221/12/01/P01020}{\doi{10.1088/1748-0221/12/01/P01020}},
  \href{http://www.arXiv.org/abs/1609.02366}{\texttt{arXiv:1609.02366}}.

\bibitem{muontrigger}
\hrefCMSnoop {}{{CMS Collaboration}, ``{Performance of the CMS muon trigger
  system in proton-proton collisions at 13 {TeV}}'',} \textit{ JINST} \textbf{
  16} (2021) P07001,
  \href{http://dx.doi.org/10.1088/1748-0221/16/07/P07001}{\doi{10.1088/1748-0221/16/07/P07001}},
  \href{http://www.arXiv.org/abs/2102.04790}{\texttt{arXiv:2102.04790}}.

\bibitem{Nason:2004rx}
\hrefCMSnoop {}{P.~Nason, ``{A New method for combining NLO QCD with shower
  Monte Carlo algorithms}'',} \textit{ JHEP} \textbf{ 11} (2004) 040,
  \href{http://dx.doi.org/10.1088/1126-6708/2004/11/040}{\doi{10.1088/1126-6708/2004/11/040}},
\href{http://www.arXiv.org/abs/hep-ph/0409146}{\texttt{arXiv:hep-ph/0409146}}.
%%CITATION = HEP-PH/0409146;%%.

\bibitem{Frixione:2007vw}
\hrefCMSnoop {}{S.~Frixione, P.~Nason, and C.~Oleari, ``{Matching NLO QCD
  computations with parton shower simulations: the POWHEG method}'',} \textit{
  JHEP} \textbf{ 11} (2007) 070,
  \href{http://dx.doi.org/10.1088/1126-6708/2007/11/070}{\doi{10.1088/1126-6708/2007/11/070}},
\href{http://www.arXiv.org/abs/0709.2092}{\texttt{arXiv:0709.2092}}.
%%CITATION = ARXIV:0709.2092;%%.

\bibitem{Alioli:2010xd}
\hrefCMSnoop {}{S.~Alioli, P.~Nason, C.~Oleari, and E.~Re, ``{A general
  framework for implementing NLO calculations in shower Monte Carlo programs:
  the POWHEG BOX}'',} \textit{ JHEP} \textbf{ 06} (2010) 043,
  \href{http://dx.doi.org/10.1007/JHEP06(2010)043}{\doi{10.1007/JHEP06(2010)043}},
\href{http://www.arXiv.org/abs/1002.2581}{\texttt{arXiv:1002.2581}}.
%%CITATION = ARXIV:1002.2581;%%.

\bibitem{Frixione:2007nw}
\hrefCMSnoop {}{S.~Frixione, P.~Nason, and G.~Ridolfi, ``{A positive-weight
  next-to-leading-order Monte Carlo for heavy flavour hadroproduction}'',}
  \textit{ JHEP} \textbf{ 09} (2007) 126,
  \href{http://dx.doi.org/10.1088/1126-6708/2007/09/126}{\doi{10.1088/1126-6708/2007/09/126}},
\href{http://www.arXiv.org/abs/0707.3088}{\texttt{arXiv:0707.3088}}.
%%CITATION = ARXIV:0707.3088;%%.

\bibitem{Czakon:2011xx}
\hrefCMSnoop {}{M.~Czakon and A.~Mitov, ``{Top++: a program for the calculation
  of the top-pair cross-section at hadron colliders}'',} \textit{ Comput. Phys.
  Commun.} \textbf{ 185} (2014) 2930,
  \href{http://dx.doi.org/10.1016/j.cpc.2014.06.021}{\doi{10.1016/j.cpc.2014.06.021}},
\href{http://www.arXiv.org/abs/1112.5675}{\texttt{arXiv:1112.5675}}.
%%CITATION = ARXIV:1112.5675;%%.

\bibitem{Frederix:2012dh}
\hrefCMSnoop {}{R.~Frederix, E.~Re, and P.~Torrielli, ``{Single-top $t$-channel
  hadroproduction in the four-flavour scheme with POWHEG and aMC@NLO}'',}
  \textit{ JHEP} \textbf{ 09} (2012) 130,
  \href{http://dx.doi.org/10.1007/JHEP09(2012)130}{\doi{10.1007/JHEP09(2012)130}},
  \href{http://www.arXiv.org/abs/1207.5391}{\texttt{arXiv:1207.5391}}.

\bibitem{Re:2010bp}
\hrefCMSnoop {}{E.~Re, ``{Single-top $\rm Wt$-channel production matched with
  parton showers using the POWHEG method}'',} \textit{ Eur. Phys. J. C}
  \textbf{ 71} (2011) 1547,
  \href{http://dx.doi.org/10.1140/epjc/s10052-011-1547-z}{\doi{10.1140/epjc/s10052-011-1547-z}},
\href{http://www.arXiv.org/abs/1009.2450}{\texttt{arXiv:1009.2450}}.
%%CITATION = 1009.2450;%%.

\bibitem{Alwall:2014hca}
J.~Alwall\hrefCMSnoop {}{ { et~al.}, ``{The automated computation of tree-level
  and next-to-leading order differential cross sections, and their matching to
  parton shower simulations}'',} \textit{ JHEP} \textbf{ 07} (2014) 079,
  \href{http://dx.doi.org/10.1007/JHEP07(2014)079}{\doi{10.1007/JHEP07(2014)079}},
\href{http://www.arXiv.org/abs/1405.0301}{\texttt{arXiv:1405.0301}}.
%%CITATION = ARXIV:1405.0301;%%.

\bibitem{Artoisenet:2012st}
\hrefCMSnoop {}{P.~Artoisenet, R.~Frederix, O.~Mattelaer, and R.~Rietkerk,
  ``Automatic spin-entangled decays of heavy resonances in {Monte Carlo}
  simulations'',} \textit{ JHEP} \textbf{ 03} (2013) 015,
  \href{http://dx.doi.org/10.1007/JHEP03(2013)015}{\doi{10.1007/JHEP03(2013)015}},
  \href{http://www.arXiv.org/abs/1212.3460}{\texttt{arXiv:1212.3460}}.

\bibitem{Alwall:2007fs}
\hrefCMSnoop {}{J.~Alwall { et~al.}, ``{Comparative study of various algorithms
  for the merging of parton showers and matrix elements in hadronic
  collisions}'',} \textit{ Eur. Phys. J. C} \textbf{ 53} (2008) 473,
  \href{http://dx.doi.org/10.1140/epjc/s10052-007-0490-5}{\doi{10.1140/epjc/s10052-007-0490-5}},
  \href{http://www.arXiv.org/abs/0706.2569}{\texttt{arXiv:0706.2569}}.

\bibitem{Frederix:2012ps}
\hrefCMSnoop {}{R.~Frederix and S.~Frixione, ``{Merging meets matching in
  MC@NLO}'',} \textit{ JHEP} \textbf{ 12} (2012) 061,
  \href{http://dx.doi.org/10.1007/JHEP12(2012)061}{\doi{10.1007/JHEP12(2012)061}},
  \href{http://www.arXiv.org/abs/1209.6215}{\texttt{arXiv:1209.6215}}.

\bibitem{Belvedere:2024nzh}
\href {https://arxiv.org/abs/2406.14620}{A.~Belvedere { et~al.}, ``{LHC EFT WG
  Note: SMEFT predictions, event reweighting, and simulation}'',} CERN Report
  CERN-LHCEFTWG-2024-001, 2024.
\newblock
  \href{http://www.arXiv.org/abs/2406.14620}{\texttt{arXiv:2406.14620}}.

\bibitem{Hamilton:2012np}
\hrefCMSnoop {}{K.~Hamilton, P.~Nason, and G.~Zanderighi, ``{MINLO}:
  multi-scale improved {NLO}'',} \textit{ JHEP} \textbf{ 10} (2012) 155,
  \href{http://dx.doi.org/10.1007/JHEP10(2012)155}{\doi{10.1007/JHEP10(2012)155}},
  \href{http://www.arXiv.org/abs/1206.3572}{\texttt{arXiv:1206.3572}}.

\bibitem{Luisoni:2013cuh}
\hrefCMSnoop {}{G.~Luisoni, P.~Nason, C.~Oleari, and F.~Tramontano, ``{HW/HZ} +
  0 and 1 jet at {NLO} with the {POWHEG BOX} interfaced to {GoSam} and their
  merging within {MiNLO}'',} \textit{ JHEP} \textbf{ 10} (2013) 083,
  \href{http://dx.doi.org/10.1007/JHEP10(2013)083}{\doi{10.1007/JHEP10(2013)083}},
  \href{http://www.arXiv.org/abs/1306.2542}{\texttt{arXiv:1306.2542}}.

\bibitem{Brivio:2017btx}
\hrefCMSnoop {}{I.~Brivio, Y.~Jiang, and M.~Trott, ``{The SMEFTsim package,
  theory and tools}'',} \textit{ JHEP} \textbf{ 12} (2017) 070,
  \href{http://dx.doi.org/10.1007/JHEP12(2017)070}{\doi{10.1007/JHEP12(2017)070}},
  \href{http://www.arXiv.org/abs/1709.06492}{\texttt{arXiv:1709.06492}}.

\bibitem{Brivio:2020onw}
\hrefCMSnoop {}{I.~Brivio, ``{SMEFTsim 3.0 \textemdash{} a practical guide}'',}
  \textit{ JHEP} \textbf{ 04} (2021) 073,
  \href{http://dx.doi.org/10.1007/JHEP04(2021)073}{\doi{10.1007/JHEP04(2021)073}},
  \href{http://www.arXiv.org/abs/2012.11343}{\texttt{arXiv:2012.11343}}.

\bibitem{Brivio:2019myy}
\hrefCMSnoop {}{I.~Brivio, T.~Corbett, and M.~Trott, ``{The Higgs width in the
  SMEFT}'',} \textit{ JHEP} \textbf{ 10} (2019) 056,
  \href{http://dx.doi.org/10.1007/JHEP10(2019)056}{\doi{10.1007/JHEP10(2019)056}},
  \href{http://www.arXiv.org/abs/1906.06949}{\texttt{arXiv:1906.06949}}.

\bibitem{Artoisenet:2008zz}
\hrefCMSnoop {}{P.~Artoisenet and O.~Mattelaer, ``{MadWeight: automatic event
  reweighting with matrix elements}'',} in \textit{ Proc. 2nd International
  Workshop on Prospects for Charged Higgs Discovery at Colliders (CHARGED
  2008)}, T.~Ekelof and J.~Rathsman, eds., p.~025.
\newblock 2008.
\newblock
  \href{http://dx.doi.org/10.22323/1.073.0025}{\doi{10.22323/1.073.0025}}.

\bibitem{Ball:2017nwa}
\hrefCMSnoop {}{{NNPDF} Collaboration, ``{Parton distributions from
  high-precision collider data}'',} \textit{ Eur. Phys. J. C} \textbf{ 77}
  (2017) 663,
  \href{http://dx.doi.org/10.1140/epjc/s10052-017-5199-5}{\doi{10.1140/epjc/s10052-017-5199-5}},
  \href{http://www.arXiv.org/abs/1706.00428}{\texttt{arXiv:1706.00428}}.

\bibitem{CMS:2019csb}
\hrefCMSnoop {}{{CMS Collaboration}, ``{Extraction and validation of a new set
  of CMS PYTHIA8 tunes from underlying-event measurements}'',} \textit{ Eur.
  Phys. J. C} \textbf{ 80} (2020) 4,
  \href{http://dx.doi.org/10.1140/epjc/s10052-019-7499-4}{\doi{10.1140/epjc/s10052-019-7499-4}},
  \href{http://www.arXiv.org/abs/1903.12179}{\texttt{arXiv:1903.12179}}.

\bibitem{CMS:2018mlc}
\hrefCMSnoop {}{{CMS Collaboration}, ``{Measurement of the inelastic
  proton-proton cross section at $ \sqrt{s}=13 $ TeV}'',} \textit{ JHEP}
  \textbf{ 07} (2018) 161,
  \href{http://dx.doi.org/10.1007/JHEP07(2018)161}{\doi{10.1007/JHEP07(2018)161}},
  \href{http://www.arXiv.org/abs/1802.02613}{\texttt{arXiv:1802.02613}}.

\bibitem{Agostinelli:2002hh}
\hrefCMSnoop {}{{GEANT4} Collaboration, ``{\GEANTfour}---a simulation
  toolkit'',} \textit{ Nucl. Instrum. Meth. A} \textbf{ 506} (2003) 250,
\href{http://dx.doi.org/10.1016/S0168-9002(03)01368-8}{\doi{10.1016/S0168-9002(03)01368-8}}.
%%CITATION = NUIMA,A506,250;%%.

\bibitem{CMS-TDR-15-02}
\href {http://cds.cern.ch/record/2020886}{{CMS Collaboration}, ``Technical
  proposal for the {Phase-II} upgrade of the {Compact Muon Solenoid}'',} CMS
  Technical Proposal CERN-LHCC-2015-010, CMS-TDR-15-02, 2015.

\bibitem{CMS:2020uim}
\hrefCMSnoop {}{{CMS Collaboration}, ``{Electron and photon reconstruction and
  identification with the CMS experiment at the CERN LHC}'',} \textit{ JINST}
  \textbf{ 16} (2021) P05014,
  \href{http://dx.doi.org/10.1088/1748-0221/16/05/P05014}{\doi{10.1088/1748-0221/16/05/P05014}},
  \href{http://www.arXiv.org/abs/2012.06888}{\texttt{arXiv:2012.06888}}.

\bibitem{CMS-DP-2020-021}
\href {https://cds.cern.ch/record/2717925}{{CMS Collaboration}, ``{ECAL} 2016
  refined calibration and {Run2} summary plots'',} CMS Detector Performance
  Summary CMS-DP-2020-021, 2020.

\bibitem{CMS:2018rym}
\hrefCMSnoop {}{{CMS Collaboration}, ``{Performance of the CMS muon detector
  and muon reconstruction with proton-proton collisions at $\sqrt{s}=$ 13
  TeV}'',} \textit{ JINST} \textbf{ 13} (2018) P06015,
  \href{http://dx.doi.org/10.1088/1748-0221/13/06/P06015}{\doi{10.1088/1748-0221/13/06/P06015}},
  \href{http://www.arXiv.org/abs/1804.04528}{\texttt{arXiv:1804.04528}}.

\bibitem{Cacciari:2008gp}
\hrefCMSnoop {}{M.~Cacciari, G.~P. Salam, and G.~Soyez, ``{The anti-\kt jet
  clustering algorithm}'',} \textit{ JHEP} \textbf{ 04} (2008) 063,
  \href{http://dx.doi.org/10.1088/1126-6708/2008/04/063}{\doi{10.1088/1126-6708/2008/04/063}},
\href{http://www.arXiv.org/abs/0802.1189}{\texttt{arXiv:0802.1189}}.
%%CITATION = ARXIV:0802.1189;%%.

\bibitem{Cacciari:2011ma}
\hrefCMSnoop {}{M.~Cacciari, G.~P. Salam, and G.~Soyez, ``{FastJet user
  manual}'',} \textit{ Eur. Phys. J. C} \textbf{ 72} (2012) 1896,
  \href{http://dx.doi.org/10.1140/epjc/s10052-012-1896-2}{\doi{10.1140/epjc/s10052-012-1896-2}},
\href{http://www.arXiv.org/abs/1111.6097}{\texttt{arXiv:1111.6097}}.
%%CITATION = ARXIV:1111.6097;%%.

\bibitem{CMS-DP-2021-033}
\href {http://cds.cern.ch/record/2792322}{{CMS Collaboration}, ``Jet energy
  scale and resolution measurement with {Run~2} legacy data collected by {CMS}
  at {13~TeV}'',} CMS Detector Performance Summary CMS-DP-2021-033, 2021.

\bibitem{Cacciari:2007fd}
\hrefCMSnoop {}{M.~Cacciari and G.~P. Salam, ``{Pileup subtraction using jet
  areas}'',} \textit{ Phys. Lett. B} \textbf{ 659} (2008) 119,
  \href{http://dx.doi.org/10.1016/j.physletb.2007.09.077}{\doi{10.1016/j.physletb.2007.09.077}},
\href{http://www.arXiv.org/abs/0707.1378}{\texttt{arXiv:0707.1378}}.
%%CITATION = ARXIV:0707.1378;%%.

\bibitem{CMS:2016lmd}
\hrefCMSnoop {}{{CMS Collaboration}, ``{Jet energy scale and resolution in the
  CMS experiment in pp collisions at 8 TeV}'',} \textit{ JINST} \textbf{ 12}
  (2017) P02014,
  \href{http://dx.doi.org/10.1088/1748-0221/12/02/P02014}{\doi{10.1088/1748-0221/12/02/P02014}},
  \href{http://www.arXiv.org/abs/1607.03663}{\texttt{arXiv:1607.03663}}.

\bibitem{Bertolini:2014bba}
\hrefCMSnoop {}{D.~Bertolini, P.~Harris, M.~Low, and N.~Tran, ``{Pileup per
  particle identification}'',} \textit{ JHEP} \textbf{ 10} (2014) 059,
  \href{http://dx.doi.org/10.1007/JHEP10(2014)059}{\doi{10.1007/JHEP10(2014)059}},
\href{http://www.arXiv.org/abs/1407.6013}{\texttt{arXiv:1407.6013}}.
%%CITATION = ARXIV:1407.6013;%%.

\bibitem{CMS:2020ebo}
\hrefCMSnoop {}{{CMS Collaboration}, ``{Pileup mitigation at CMS in 13 TeV
  data}'',} \textit{ JINST} \textbf{ 15} (2020) P09018,
  \href{http://dx.doi.org/10.1088/1748-0221/15/09/P09018}{\doi{10.1088/1748-0221/15/09/P09018}},
  \href{http://www.arXiv.org/abs/2003.00503}{\texttt{arXiv:2003.00503}}.

\bibitem{CMS-PAS-JME-16-003}
\href {https://cds.cern.ch/record/2256875}{{CMS Collaboration}, ``{Jet
  algorithms performance in 13 TeV data}'',} {CMS Physics Analysis Summary}
  CMS-PAS-JME-16-003, 2017.

\bibitem{Bols:2020bkb}
E.~Bols\hrefCMSnoop {}{ { et~al.}, ``{Jet flavour classification using
  DeepJet}'',} \textit{ JINST} \textbf{ 15} (2020) P12012,
  \href{http://dx.doi.org/10.1088/1748-0221/15/12/P12012}{\doi{10.1088/1748-0221/15/12/P12012}},
  \href{http://www.arXiv.org/abs/2008.10519}{\texttt{arXiv:2008.10519}}.

\bibitem{CMS-DP-2023-005}
\href {http://cds.cern.ch/record/2854609}{{CMS Collaboration}, ``{Performance
  summary of AK4 jet b tagging with data from proton-proton collisions at 13
  TeV}'',} {CMS Detector Performance Report} CMS-DP-2023-005, 2023.

\bibitem{CMS:2019uxx}
\hrefCMSnoop {}{{CMS Collaboration}, ``{A deep neural network for simultaneous
  estimation of b jet energy and resolution}'',} \textit{ Comput. Softw. Big
  Sci.} \textbf{ 4} (2020) 10,
  \href{http://dx.doi.org/10.1007/s41781-020-00041-z}{\doi{10.1007/s41781-020-00041-z}},
  \href{http://www.arXiv.org/abs/1912.06046}{\texttt{arXiv:1912.06046}}.

\bibitem{Qu:2019gqs}
\hrefCMSnoop {}{H.~Qu and L.~Gouskos, ``{Jet tagging via particle clouds}'',}
  \textit{ Phys. Rev. D} \textbf{ 101} (2020) 056019,
  \href{http://dx.doi.org/10.1103/PhysRevD.101.056019}{\doi{10.1103/PhysRevD.101.056019}},
  \href{http://www.arXiv.org/abs/1902.08570}{\texttt{arXiv:1902.08570}}.

\bibitem{Butterworth:2008iy}
\hrefCMSnoop {}{J.~M. Butterworth, A.~R. Davison, M.~Rubin, and G.~P. Salam,
  ``Jet substructure as a new {Higgs} search channel at the {LHC}'',} \textit{
  Phys. Rev. Lett.} \textbf{ 100} (2008) 242001,
  \href{http://dx.doi.org/10.1103/PhysRevLett.100.242001}{\doi{10.1103/PhysRevLett.100.242001}},
\href{http://www.arXiv.org/abs/0802.2470}{\texttt{arXiv:0802.2470}}.
%%CITATION = ARXIV:0802.2470;%%.

\bibitem{Dasgupta:2013ihk}
\hrefCMSnoop {}{M.~Dasgupta, A.~Fregoso, S.~Marzani, and G.~P. Salam, ``Towards
  an understanding of jet substructure'',} \textit{ JHEP} \textbf{ 09} (2013)
  029,
  \href{http://dx.doi.org/10.1007/JHEP09(2013)029}{\doi{10.1007/JHEP09(2013)029}},
\href{http://www.arXiv.org/abs/1307.0007}{\texttt{arXiv:1307.0007}}.
%%CITATION = ARXIV:1307.0007;%%.

\bibitem{Larkoski:2014wba}
\hrefCMSnoop {}{A.~J. Larkoski, S.~Marzani, G.~Soyez, and J.~Thaler, ``{Soft
  Drop}'',} \textit{ JHEP} \textbf{ 05} (2014) 146,
  \href{http://dx.doi.org/10.1007/JHEP05(2014)146}{\doi{10.1007/JHEP05(2014)146}},
\href{http://www.arXiv.org/abs/1402.2657}{\texttt{arXiv:1402.2657}}.
%%CITATION = ARXIV:1402.2657;%%.

\bibitem{Dokshitzer:1997in}
\hrefCMSnoop {}{Y.~L. Dokshitzer, G.~D. Leder, S.~Moretti, and B.~R. Webber,
  ``Better jet clustering algorithms'',} \textit{ JHEP} \textbf{ 08} (1997)
  001,
  \href{http://dx.doi.org/10.1088/1126-6708/1997/08/001}{\doi{10.1088/1126-6708/1997/08/001}},
\href{http://www.arXiv.org/abs/hep-ph/9707323}{\texttt{arXiv:hep-ph/9707323}}.
%%CITATION = HEP-PH/9707323;%%.

\bibitem{Wobisch:1998wt}
\href {https://inspirehep.net/record/484872}{M.~Wobisch and T.~Wengler,
  ``Hadronization corrections to jet cross-sections in deep inelastic
  scattering'',} in \textit{ {Workshop on Monte Carlo Generators for HERA
  Physics, Hamburg, Germany}}, p.~270.
\newblock 1998.
\newblock
\href{http://www.arXiv.org/abs/hep-ph/9907280}{\texttt{arXiv:hep-ph/9907280}}.
\newblock
%%CITATION = HEP-PH/9907280;%%.

\bibitem{CMS-DP-2020-002}
\href {https://cds.cern.ch/record/2707946}{{CMS Collaboration},
  ``{Identification of highly Lorentz-boosted heavy particles using graph
  neural networks and new mass decorrelation techniques}'',} {CMS Detector
  Performance Report} CMS-DP-2020-002, 2020.

\bibitem{CMS-DP-2022-005}
\href {https://cds.cern.ch/record/2805611}{{CMS Collaboration}, ``{Calibration
  of the mass-decorrelated ParticleNet tagger for boosted
  $\mathrm{b}\bar{\mathrm{b}}$ and $\mathrm{c}\bar{\mathrm{c}}$ jets using LHC
  Run 2 data}'',} {CMS Detector Performance Report} CMS-DP-2022-005, 2022.

\bibitem{CMS-PAS-BTV-22-001}
\href {http://cds.cern.ch/record/2866276}{{CMS Collaboration}, ``{Performance
  of heavy-flavour jet identification in boosted topologies in proton-proton
  collisions at $\sqrt{s} = 13~\mathrm{TeV}$}'',} {CMS Physics Analysis
  Summary} CMS-PAS-BTV-22-001, 2023.

\bibitem{CMS:2019ctu}
\hrefCMSnoop {}{{CMS Collaboration}, ``{Performance of missing transverse
  momentum reconstruction in proton-proton collisions at $\sqrt{s} =$ 13 TeV
  using the CMS detector}'',} \textit{ JINST} \textbf{ 14} (2019) P07004,
  \href{http://dx.doi.org/10.1088/1748-0221/14/07/P07004}{\doi{10.1088/1748-0221/14/07/P07004}},
  \href{http://www.arXiv.org/abs/1903.06078}{\texttt{arXiv:1903.06078}}.

\bibitem{CMS-PAS-JME-09-010}
\href {https://cds.cern.ch/record/1228297}{{CMS Collaboration}, ``{Performance
  of Track-Corrected Missing Transverse Energy in CMS}'',} {CMS Physics
  Analysis Summary} CMS-PAS-JME-09-010, 2009.

\bibitem{PDG2022}
\hrefCMSnoop {}{{Particle Data Group}, R.~L. Workman { et~al.}, ``Review of
  particle physics'',} \textit{ Prog. Theor. Exp. Phys.} \textbf{ 2022} (2022)
  083C01,
  \href{http://dx.doi.org/10.1093/ptep/ptac097}{\doi{10.1093/ptep/ptac097}}.

\bibitem{Brehmer:2018eca}
\hrefCMSnoop {}{J.~Brehmer, K.~Cranmer, G.~Louppe, and J.~Pavez, ``{A guide to
  constraining effective field theories with machine learning}'',} \textit{
  Phys. Rev. D} \textbf{ 98} (2018) 052004,
  \href{http://dx.doi.org/10.1103/PhysRevD.98.052004}{\doi{10.1103/PhysRevD.98.052004}},
  \href{http://www.arXiv.org/abs/1805.00020}{\texttt{arXiv:1805.00020}}.

\bibitem{NIPS2017_6449f44a}
G.~Ke\href
  {https://proceedings.neurips.cc/paper_files/paper/2017/file/6449f44a102fde848669bdd9eb6b76fa-Paper.pdf}{
  { et~al.}, ``{LightGBM}: A highly efficient gradient boosting decision
  tree'',} in \textit{ Advances in Neural Information Processing Systems 30
  (NIPS 2017)}, I.~Guyon { et~al.}, eds.
\newblock Curran Associates, Inc., 2017.

\bibitem{frazier2018tutorial}
\hrefCMSnoop {}{P.~I. Frazier, ``A tutorial on {Bayesian} optimization'',}
  2018. \href{http://www.arXiv.org/abs/1807.02811}{\texttt{arXiv:1807.02811}}.

\bibitem{pdfLHC}
\hrefCMSnoop {}{J.~Butterworth { et~al.}, ``{PDF4LHC} recommendations for {LHC}
  run {II}'',} \textit{ J. Phys. G} \textbf{ 43} (2016) 040,
  \href{http://dx.doi.org/10.1088/0954-3899/43/2/023001}{\doi{10.1088/0954-3899/43/2/023001}},
  \href{http://www.arXiv.org/abs/1510.03865}{\texttt{arXiv:1510.03865}}.

\bibitem{CMS:2021xjt}
\hrefCMSnoop {}{{CMS Collaboration}, ``{Precision luminosity measurement in
  proton-proton collisions at $\sqrt{s} =$ 13 TeV in 2015 and 2016 at CMS}'',}
  \textit{ Eur. Phys. J. C} \textbf{ 81} (2021) 800,
  \href{http://dx.doi.org/10.1140/epjc/s10052-021-09538-2}{\doi{10.1140/epjc/s10052-021-09538-2}},
  \href{http://www.arXiv.org/abs/2104.01927}{\texttt{arXiv:2104.01927}}.

\bibitem{CMS:2018elu}
\href {https://cds.cern.ch/record/2621960}{{CMS Collaboration}, ``{CMS}
  luminosity measurement for the 2017 data-taking period at {$\sqrt{s}$ = 13
  TeV}'',} CMS Physics Analysis Summary CMS-PAS-LUM-17-004, 2018.

\bibitem{CMS:2019jhq}
\href {https://cds.cern.ch/record/2676164}{{CMS Collaboration}, ``{CMS}
  luminosity measurement for the 2018 data-taking period at {$\sqrt{s} = 13$
  TeV}'',} CMS Physics Analysis Summary CMS-PAS-LUM-18-002, 2019.

\bibitem{bbb}
\hrefCMSnoop {}{R.~Barlow and C.~Beeston, ``Fitting using finite {M}onte
  {C}arlo samples'',} \textit{ Comput. Phys. Commun.} \textbf{ 77} (1993) 219,
  \href{http://dx.doi.org/10.1016/0010-4655(93)90005-W}{\doi{10.1016/0010-4655(93)90005-W}}.

\bibitem{CMS:2024onh}
\hrefCMSnoop {}{{CMS Collaboration}, ``{The CMS statistical analysis and
  combination tool: COMBINE}'',} \textit{ Comput. Softw. Big Sci.} \textbf{ 8}
  (2024) 19,
  \href{http://dx.doi.org/10.1007/s41781-024-00121-4}{\doi{10.1007/s41781-024-00121-4}},
  \href{http://www.arXiv.org/abs/2404.06614}{\texttt{arXiv:2404.06614}}.

\bibitem{Verkerke:2003ir}
\hrefCMSnoop {}{W.~Verkerke and D.~P. Kirkby, ``The {RooFit} toolkit for data
  modeling'',} in \textit{ Proc. Int. Conf. on Computing in High Energy and
  Nuclear Physics (CHEP03)}, L.~Lyons and M.~Karagoz, eds., p.~MOLT007.
\newblock 2003.
\newblock
  \href{http://www.arXiv.org/abs/physics/0306116}{\texttt{arXiv:physics/0306116}}.

\bibitem{Moneta:2010pm}
L.~Moneta\hrefCMSnoop {}{ { et~al.}, ``The {RooStats} project'',} in \textit{
  Proc. 13th Int. Workshop on Advanced Computing and Analysis Techniques in
  Physics Research}, T.~Speer { et~al.}, eds., volume ACAT2010, p.~057.
\newblock 2010.
\newblock \href{http://www.arXiv.org/abs/1009.1003}{\texttt{arXiv:1009.1003}}.
\newblock
  \href{http://dx.doi.org/10.22323/1.093.0057}{\doi{10.22323/1.093.0057}}.

\bibitem{CMS-NOTE-2011-005}
\href {https://cds.cern.ch/record/1379837}{{ATLAS and CMS Collaborations, and
  LHC Higgs Combination Group}, ``Procedure for the {LHC} {Higgs} boson search
  combination in {Summer} 2011'',} CMS Note CMS-NOTE-2011-005,
  ATL-PHYS-PUB-2011-11, 2011.

\bibitem{Thomas_1999}
\hrefCMSnoop {}{T.~Junk, ``Confidence level computation for combining searches
  with small statistics'',} \textit{ Nucl. Instrum. Meth. A} \textbf{ 434}
  (1999) 435,
  \href{http://dx.doi.org/10.1016/S0168-9002(99)00498-2}{\doi{10.1016/S0168-9002(99)00498-2}},
\href{http://www.arXiv.org/abs/hep-ex/9902006}{\texttt{arXiv:hep-ex/9902006}}.
%%CITATION = HEP-EX/9902006;%%.

\bibitem{Read_2002}
\hrefCMSnoop {}{A.~L. Read, ``Presentation of search results: The
  {CL$_{\text{s}}$} technique'',} \textit{ J. Phys. G} \textbf{ 28} (2002)
  2693,
\href{http://dx.doi.org/10.1088/0954-3899/28/10/313}{\doi{10.1088/0954-3899/28/10/313}}.
%%CITATION = JPAGA,G28,2693;%%.

\bibitem{saturated}
\hrefCMSnoop {}{R.~D. Cousins, ``Lectures on statistics in theory: {P}relude to
  statistics in practice'',} 2018.
  \href{http://www.arXiv.org/abs/1807.05996}{\texttt{arXiv:1807.05996}}.

\bibitem{Bernlochner:2022oiw}
\hrefCMSnoop {}{F.~U. Bernlochner, D.~C. Fry, S.~B. Menary, and E.~Persson,
  ``{Cover your bases: asymptotic distributions of the profile likelihood ratio
  when constraining effective field theories in high-energy physics}'',}
  \textit{ SciPost Phys. Core} \textbf{ 6} (2023) 013,
  \href{http://dx.doi.org/10.21468/SciPostPhysCore.6.1.013}{\doi{10.21468/SciPostPhysCore.6.1.013}},
  \href{http://www.arXiv.org/abs/2207.01350}{\texttt{arXiv:2207.01350}}.

\bibitem{Wilks}
\hrefCMSnoop {}{S.~S. Wilks, ``{The Large-Sample Distribution of the Likelihood
  Ratio for Testing Composite Hypotheses}'',} \textit{ The Ann. Math. Stat.}
  \textbf{ 9} (1938) 60,
  \href{http://dx.doi.org/10.1214/aoms/1177732360}{\doi{10.1214/aoms/1177732360}}.

\end{thebibliography}\endgroup
